\documentclass[%
 reprint,
superscriptaddress,
preprintnumbers,
nofootinbib,
 amsmath,amssymb,
 aps,
 prd,
]{revtex4-2}
\usepackage{graphicx}
\usepackage{dcolumn}
\usepackage{bm}
\usepackage{hyperref}
\usepackage{mathtools}

\usepackage{booktabs}
\usepackage{cleveref}

\usepackage{xcolor}

\usepackage{subfiles}
\usepackage{cancel}

\usepackage{siunitx}
\newcolumntype{d}[1]{D{.}{.}{#1}}

\usepackage{lineno}

\usepackage{subfigure}

\usepackage[normalem]{ulem}

\defcitealias{Bocquet24Ia}{SB24a}

\defcitealias{Bocquet24II}{SB24b}

\defcitealias{Vogt24}{SV24a}
\defcitealias{Vogt24b}{SV24b}

\usepackage{hyperref}
\definecolor{plasmablue}{rgb}{0.050383, 0.029803, 0.527975}
\hypersetup{
  colorlinks=true,
  linkcolor=plasmablue,
  urlcolor=plasmablue,
  citecolor=plasmablue,
  }

\newcommand{\ie}{i.\,e.\ }
\newcommand{\eg}{e.\,g.\ }

\newcommand{\dd}{\textnormal{d}}

\newcommand{\nDGP}{\ensuremath{{1/\sqrt{H_0r_{\rm c}}}}}

\newcommand{\zetahat}{\ensuremath{{\hat \zeta}}}
\newcommand{\lambdahat}{\ensuremath{{\hat \lambda}}}

\newcommand{\asz}{\ensuremath{{\zeta_0}}}
\newcommand{\bsz}{\ensuremath{{\zeta_M}}}
\newcommand{\csz}{\ensuremath{{\zeta_z}}}
\newcommand{\sigmalnzeta}{\ensuremath{{\sigma_{\ln\zeta}}}}
\newcommand{\alambda}{\ensuremath{{\lambda_0}}}
\newcommand{\blambda}{\ensuremath{{\lambda_M}}}
\newcommand{\clambda}{\ensuremath{{\lambda_z}}}
\newcommand{\sigmalnlambda}{\ensuremath{{\sigma_{\ln\lambda}}}}
\newcommand{\MWL}{\ensuremath{M_\mathrm{WL}}}
\newcommand{\bWL}{\ensuremath{{\ln M_{\mathrm{WL}_0}}}}
\newcommand{\bWLM}{\ensuremath{{M_{\mathrm{WL}_M}}}}
\newcommand{\sWLall}{\ensuremath{{\sigma_{\ln\mathrm{WL}}}}}
\newcommand{\sWL}{\ensuremath{{\ln\sigma^2_{\ln\mathrm{WL}_0}}}}
\newcommand{\sWLM}{\ensuremath{\sigma^2_{\ln\mathrm{WL}_M}}}

\newcommand{\bWLHST}{\ensuremath{{\ln M_{\mathrm{WL}_0}}}}
\newcommand{\sWLHST}{\ensuremath{{\sigma_{\ln\mathrm{WL}}}}}

\begin{document}

\preprint{DES-2025-0910}
\preprint{FERMILAB-PUB-25-0861-PPD}

\title{Constraints on the  normal branch of DGP gravity from SPT galaxy clusters with DES and HST weak-lensing mass calibration and from Planck PR4 CMB anisotropies}

\author{S.~M.~L.~Vogt}
\email{s.vogt@physik.lmu.de}
\affiliation{
University Observatory, LMU Faculty of Physics, Scheinerstra\ss e 1, 81679 Munich, Germany
}
\affiliation{
Excellence Cluster Origins, Boltzmannstr. 2, 85748 Garching, Germany
}
\affiliation{Max Planck Institute for Astrophysics, Karl-Schwarzschild-Stra\ss e 1, 85748 Garching, Germany}
\author{S.~Bocquet}
\affiliation{
University Observatory, LMU Faculty of Physics,  Scheinerstra\ss e\ 1, 81679 Munich, Germany
}
\author{C.~T.~Davies}
\affiliation{
University Observatory, LMU Faculty of Physics,  Scheinerstra\ss e 1, 81679 Munich, Germany
}
\author{J.~J.~Mohr}
\affiliation{
University Observatory, LMU Faculty of Physics, Scheinerstra\ss e 1, 81679 Munich, Germany
}
\affiliation{Max Planck Institute for Extraterrestrial Physics, Giessenbachstr. 2, 85748 Garching, Germany}
\author{F.~Schmidt}
\affiliation{Max Planck Institute for Astrophysics, Karl-Schwarzschild-Stra\ss e 1, 85748 Garching, Germany}

\author{C.-Z.~Ruan}
\affiliation{Institute of Theoretical Astrophysics, University of Oslo, 0315 Oslo, Norway}

\author{B.~Li}
\affiliation{Institute for Computational Cosmology, Department of Physics, Durham University, South Road, Durham DH1 3LE, UK}

\author{C.~Hernández-Aguayo}
\affiliation{Max Planck Institute for Astrophysics, Karl-Schwarzschild-Stra\ss e 1, 85748 Garching, Germany}

\author{S.~Grandis}
\affiliation{Universit\"at Innsbruck, Institut f\"ur Astro- und Teilchenphysik, Technikerstra\ss e 25/8, 6020 Innsbruck, Austria}
\author{L.~E.~Bleem}
\affiliation{High-Energy Physics Division, Argonne National Laboratory, 9700 South Cass Avenue, Lemont, IL 60439, USA}
\affiliation{Kavli Institute for Cosmological Physics, University of Chicago, 5640 South Ellis Avenue, Chicago, IL 60637, USA}
\author{M.~Klein}
\affiliation{
University Observatory, LMU Faculty of Physics,  Scheinerstra\ss e 1, 81679 Munich, Germany
}
\affiliation{Max Planck Institute for Extraterrestrial Physics, Giessenbachstra\ss e 2, 85748 Garching, Germany}

\author{M.~Aguena}
\affiliation{INAF-Osservatorio Astronomico di Trieste, via G. B. Tiepolo 11, I-34143 Trieste, Italy}
\affiliation{Laborat\'orio Interinstitucional de e-Astronomia - LIneA, Av. Pastor Martin Luther King Jr, 126 Del Castilho, Nova Am\'erica Offices, Torre 3000/sala 817 CEP: 20765-000, Brazil}
\author{S.~Allam}
\affiliation{Fermi National Accelerator Laboratory, P. O. Box 500, Batavia, IL 60510, USA}
\author{F.~Andrade-Oliveira}
\affiliation{Physik-Institut, University of Z\"urich, Winterthurerstrasse 190, CH-8057 Z\"urich, Switzerland}
\author{D.~Bacon}
\affiliation{Institute of Cosmology and Gravitation, University of Portsmouth, Portsmouth, PO1 3FX, UK}
\author{D.~Brooks}
\affiliation{Department of Physics \& Astronomy, University College London, Gower Street, London, WC1E 6BT, UK}
\author{R.~Camilleri}
\affiliation{School of Mathematics and Physics, University of Queensland,  Brisbane, QLD 4072, Australia}
\author{A.~Carnero~Rosell}
\affiliation{Instituto de Astrofisica de Canarias, E-38205 La Laguna, Tenerife, Spain}
\affiliation{Laborat\'orio Interinstitucional de e-Astronomia - LIneA, Av. Pastor Martin Luther King Jr, 126 Del Castilho, Nova Am\'erica Offices, Torre 3000/sala 817 CEP: 20765-000, Brazil}
\affiliation{Universidad de La Laguna, Dpto. Astrrofisica, E-38206 La Laguna, Tenerife, Spain}
\author{J.~Carretero}
\affiliation{Institut de F\'{\i}sica d'Altes Energies (IFAE), The Barcelona Institute of Science and Technology, Campus UAB, 08193 Bellaterra (Barcelona) Spain}
\author{M.~Costanzi}
\affiliation{Astronomy Unit, Department of Physics, University of Trieste, via Tiepolo 11, I-34131 Trieste, Italy}
\affiliation{INAF-Osservatorio Astronomico di Trieste, via G. B. Tiepolo 11, I-34143 Trieste, Italy}
\affiliation{Institute for Fundamental Physics of the Universe, Via Beirut 2, 34014 Trieste, Italy}
\author{L.~N.~da Costa}
\affiliation{Laborat\'orio Interinstitucional de e-Astronomia - LIneA, Av. Pastor Martin Luther King Jr, 126 Del Castilho, Nova Am\'erica Offices, Torre 3000/sala 817 CEP: 20765-000, Brazil}
\author{M.~E.~da Silva Pereira}
\affiliation{Hamburger Sternwarte, Universit\"{a}t Hamburg, Gojenbergsweg 112, 21029 Hamburg, Germany}
\author{J.~De~Vicente}
\affiliation{Centro de Investigaciones Energ\'eticas, Medioambientales y Tecnol\'ogicas (CIEMAT), Madrid, Spain}
\author{P.~Doel}
\affiliation{Department of Physics \& Astronomy, University College London, Gower Street, London, WC1E 6BT, UK}
\author{J.~Garc\'ia-Bellido}
\affiliation{Instituto de Fisica Teorica UAM/CSIC, Universidad Autonoma de Madrid, 28049 Madrid, Spain}\author{P.~Giles}
\affiliation{Department of Physics and Astronomy, Pevensey Building, University of Sussex, Brighton, BN1 9QH, UK}
\author{D.~Gruen}
\affiliation{University Observatory, LMU Faculty of Physics, Scheinerstr. 1, 81679 Munich, Germany}
\author{G.~Gutierrez}
\affiliation{Fermi National Accelerator Laboratory, P. O. Box 500, Batavia, IL 60510, USA}
\author{S.~R.~Hinton}
\affiliation{School of Mathematics and Physics, University of Queensland,  Brisbane, QLD 4072, Australia}
\author{D.~L.~Hollowood}
\affiliation{Santa Cruz Institute for Particle Physics, Santa Cruz, CA 95064, USA}
\author{D.~J.~James}
\affiliation{Center for Astrophysics $\vert$ Harvard \& Smithsonian, 60 Garden Street, Cambridge, MA 02138, USA}
\author{K.~Kuehn}
\affiliation{Australian Astronomical Optics, Macquarie University, North Ryde, NSW 2113, Australia}
\affiliation{Lowell Observatory, 1400 Mars Hill Rd, Flagstaff, AZ 86001, USA}
\author{S.~Lee}
\affiliation{Jet Propulsion Laboratory, California Institute of Technology, 4800 Oak Grove Dr., Pasadena, CA 91109, USA}
\author{J.~L.~Marshall}
\affiliation{George P. and Cynthia Woods Mitchell Institute for Fundamental Physics and Astronomy, and Department of Physics and Astronomy, Texas A\&M University, College Station, TX 77843,  USA}
\author{J. Mena-Fern{\'a}ndez}
\affiliation{Universit\'e Grenoble Alpes, CNRS, LPSC-IN2P3, 38000 Grenoble, France}
\author{F.~Menanteau}
\affiliation{Center for Astrophysical Surveys, National Center for Supercomputing Applications, 1205 West Clark St., Urbana, IL 61801, USA}
\affiliation{Department of Astronomy, University of Illinois at Urbana-Champaign, 1002 W. Green Street, Urbana, IL 61801, USA}
\author{R.~Miquel}
\affiliation{Instituci\'o Catalana de Recerca i Estudis Avan\c{c}ats, E-08010 Barcelona, Spain}
\affiliation{Institut de F\'{\i}sica d'Altes Energies (IFAE), The Barcelona Institute of Science and Technology, Campus UAB, 08193 Bellaterra (Barcelona) Spain}
\author{J.~Myles}
\affiliation{Department of Astrophysical Sciences, Princeton University, Peyton Hall, Princeton, NJ 08544, USA}
\author{A.~A.~Plazas~Malag\'on}
\affiliation{Kavli Institute for Particle Astrophysics \& Cosmology, P. O. Box 2450, Stanford University, Stanford, CA 94305, USA}
\affiliation{SLAC National Accelerator Laboratory, Menlo Park, CA 94025, USA}
\author{A.~Porredon}
\affiliation{Centro de Investigaciones Energ\'eticas, Medioambientales y Tecnol\'ogicas (CIEMAT), Madrid, Spain}
\affiliation{Ruhr University Bochum, Faculty of Physics and Astronomy, Astronomical Institute, German Centre for Cosmological Lensing, 44780 Bochum, Germany}
\author{J.~Prat}
\affiliation{Nordita, KTH Royal Institute of Technology and Stockholm University, Hannes Alfv\'ens v\"ag 12, SE-10691 Stockholm, Sweden}
\author{C.~L.~Reichardt}
\affiliation{School of Physics, University of Melbourne, Parkville, VIC 3010, Australia}
\author{A.~K.~Romer}
\affiliation{Department of Physics and Astronomy, Pevensey Building, University of Sussex, Brighton, BN1 9QH, UK}
\author{E.~Sanchez}
\affiliation{Centro de Investigaciones Energ\'eticas, Medioambientales y Tecnol\'ogicas (CIEMAT), Madrid, Spain}
\author{I.~Sevilla-Noarbe}
\affiliation{Centro de Investigaciones Energ\'eticas, Medioambientales y Tecnol\'ogicas (CIEMAT), Madrid, Spain}
\author{M.~Smith}
\affiliation{Physics Department, Lancaster University, Lancaster, LA1 4YB, UK}
\author{M.~Soares-Santos}
\affiliation{Physik-Institut, University of ZÃ¼rich, Winterthurerstrasse 190, CH-8057 ZÃ¼rich, Switzerland}
\author{E.~Suchyta}
\affiliation{Computer Science and Mathematics Division, Oak Ridge National Laboratory, Oak Ridge, TN 37831}
\author{M.~E.~C.~Swanson}
\affiliation{Center for Astrophysical Surveys, National Center for Supercomputing Applications, 1205 West Clark St., Urbana, IL 61801, USA}
\author{C.~To}
\affiliation{Department of Astronomy and Astrophysics, University of Chicago, Chicago, IL 60637, USA}
\author{V.~Vikram}
\affiliation{Argone National Laboratory, 9700 S. Cass Avenue, Lemont, IL 60439, USA}
\author{N.~Weaverdyck}
\affiliation{Berkeley Center for Cosmological Physics, Department of Physics, University of California, Berkeley, CA 94720, US}
\affiliation{Lawrence Berkeley National Laboratory, 1 Cyclotron Road, Berkeley, CA 94720, USA}

\noaffiliation
\collaboration{the SPT and DES Collaborations}
\noaffiliation


\begin{abstract}
We present constraints on the normal branch of the Dvali-Gabadadze-Porrati (nDGP) braneworld gravity model from the abundance of massive galaxy clusters. 
On scales below the nDGP crossover scale $r_{\rm c}$, the nDGP model features an effective gravity-like fifth force that alters the growth of structure, leading to an enhancement of the halo mass function (HMF) on cluster scales. The enhanced cluster abundance allows for constraints on the nDGP model using cluster samples.
We employ the SPT cluster sample, selected through the thermal Sunyaev-Zel’dovich effect (tSZE) with the South Pole Telescope (SPT) and with mass calibration using weak-lensing data from the Dark Energy Survey (DES) and the Hubble Space Telescope (HST). The cluster sample contains 1,005 clusters with redshifts $0.25 < z < 1.78$, which are confirmed with the Multi-Component Matched Filter (MCMF) algorithm using optical and near-infrared data. Weak-lensing data from DES and HST enable a robust mass measurement of the cluster sample. We use DES Year 3 data for 688 clusters with redshifts $z < 0.95$, and HST data for 39 clusters with redshifts $ 0.6 < z <1.7$. 
We account for the enhancement in the HMF through a semi-analytic correction factor to the 
standard cosmology
HMF derived from the spherical collapse model in 
the nDGP model. We then further calibrate this model using $N$-body simulations.
In addition, for the first time, we analyze the primary cosmic microwave background (CMB) temperature and polarization anisotropy measurements from Planck PR4 within the nDGP model. We obtain a competitive constraint from the joint analysis of the SPT cluster abundance with the Planck PR4 data, and report an upper bound of $\nDGP< 1.41$ at $95\,\%$ when assuming a cosmology with massive neutrinos.
\end{abstract}

\maketitle


\section{\label{sec:intro}Introduction}
Understanding the observed accelerated expansion of the Universe remains one of the most fundamental open questions in cosmology \cite{Perlmutter1999,Riess1998}.
The standard cosmological model, $\Lambda$ cold dark matter ($\Lambda$CDM), assumes gravity follows general relativity (GR) and that the accelerated expansion is sourced through a cosmological constant $\Lambda$, which is added to the Einstein-Hilbert action. This corresponds to a dark energy component with constant density that drives the acceleration of the Universe's expansion. 
An independent test of GR, and thus of our understanding of gravity, can be performed by considering modifications to GR, which are generally obtained from extensions of the Einstein-Hilbert action (for a review, see, e.g., \cite{JoyceEtal16,Koyama18,Baker19}).

One widely studied modified gravity model is the Dvali-Gabadadze-Porrati (DGP) braneworld gravity model \cite{Dvali2000}. In the DGP model, our four-dimensional Universe is embedded in a five-dimensional spacetime, and gravity can leak into the extra dimension on large scales, whereas the other fundamental forces remain four-dimensional.
Specifically, this transition happens at the so-called crossover scale $r_{\rm c}$, which serves as an extra parameter in the theory that describes the deviation from GR. On scales larger than $r_{\rm c}$ gravity becomes five-dimensional.
Conversely, GR is recovered if $r_{\rm c} \to \infty$.
However, the GR limit is nontrivial, and the gravitational dynamics and the growth of cosmic structure are modified even on scales much smaller than $r_{\rm c}$.
GR is restored only in regions with density much higher than the cosmological mean, via the Vainshtein screening mechanism which arises from non-linearities in the modified Poisson equation \cite{Vainshtein72}.
In this work, we focus on the normal branch of the DGP model (nDGP) 
with a quintessence-like dark energy component tuned to give a background expansion history consistent with $\Lambda$CDM \cite{Schmidt09b}.
Unlike the self-accelerating branch, the normal branch is stable and theoretically consistent. Due to the screening mechanism, which restores GR locally, as well as the unmodified background expansion, the growth of structure is the most stringent observational test of this scenario.

Modified gravity theories such as nDGP alter the linear and non-linear growth of structure on cosmological scales, making summary statistics of the large-scale structure powerful probes for testing these models. 
One such measure of the large-scale structure is the abundance of massive galaxy clusters \cite{Wang98,Haiman01}, which has been widely used to study standard cosmological models \cite{Vikhlinin09,Benson13,Bocquet15,Dehaan16,Bocquet19,Abbott20,Chiu23,Bocquet24II,Ghirardini24} as well as modified gravity \cite{Schmidt09a,Lombriser10,Cataneo14,Peirone17,Hagstotz19,Artis24,Vogt24} and dark matter models \cite{Mazoun24,Zelmer25}. 
To obtain cosmological constraints from the abundance of massive galaxy clusters, we need a link between the cluster observables and the underlying mass of the cluster, which is not a directly observed quantity \cite{Allen11,Pratt19}. To make an inference of the underlying cluster masses, we rely on weak lensing to inform these mass estimates.

In a previous paper \cite{Vogt24}, we constrained $f(R)$ modified gravity
using a similar dataset and methodology. A key difference between $f(R)$ and
nDGP is that the former exhibits scale-dependent growth, while the growth remains scale-independent in nDGP. Moreover, $f(R)$ gravity exhibits the chameleon screening mechanism, which is mass-dependent, while the Vainshtein screening mechanism active in nDGP operates at a certain density. This leads to quite different phenomenology in the abundance of galaxy clusters.

In this study, we use the sample of $1,005$ galaxy clusters detected with the South Pole Telescope (SPT) \cite{Carlstrom11} via the thermal Sunyaev-Zel'dovich effect (tSZE)~\cite{Sunyaev1972}. The tSZE is a spectral distortion of the cosmic microwave background (CMB) along the cluster line of sight, which is induced by the inverse Compton scattering of low-energy CMB photons with the high-energy electrons of the hot intracluster medium (ICM). 
Therefore, the tSZE is a direct tracer of the ICM and the underlying massive galaxy cluster. 
The tSZE signal is approximately redshift independent, and with sufficient angular resolution and sensitivity, clusters can be detected out to the highest redshifts where they exist. The tSZE enables the construction of a high-purity sample with well-understood completeness.
SPT cluster candidates are confirmed using the Multi-Component Matched Filter (MCMF) algorithm \cite{Klein18,Klein24,Bleem24} with optical and near-infrared data from the Dark Energy Survey (DES) \cite{flaugher15,DES16,DES18DR1} and the Wide-field Infrared Survey Explorer (WISE) \cite{WISEobservatory}.

The relation of the tSZE signal to the underlying mass must be calibrated with external data, as modeling the ICM is challenging. Here, we rely on mass measurements from weak-lensing shear data, for which the relation to the mass is well understood and can therefore be used to empirically calibrate the scaling relations between the cluster observables and the cluster mass. 
We use weak-lensing data from the DES Year 3 shape catalog for 688 clusters with redshifts $z < 0.95$ \cite{Gatti22} and targeted observation from the Hubble Space Telescope (HST) for 39 clusters in the redshift range $0.6 < z < 1.7$ \cite{Schrabback18,Schrabback21,Zohren22}. The analysis presented in this work is based on the method developed in \cite{Bocquet24Ia,Bocquet24II} (hereafter \citetalias{Bocquet24Ia,Bocquet24II}).

To account for the impact of nDGP on structure formation, we adopt a halo mass function (HMF) model which accounts for changes in spherical collapse in nDGP relative to GR \cite{Schmidt09b}. The analytical model applied here uses the critical collapse overdensity $\delta_{\rm c}$ and the virial overdensity $\Delta_{\rm vir}$ derived from a spherical collapse calculation that incorporates the Vainshtein screening. Furthermore, we calibrate the subsequent semi-analytical HMF model against the BRIDGE $N$-body simulations \cite{Harnois23,Ruan24,Davies24}, ensuring accurate predictions for the abundance of collapsed structures in nDGP.

We also perform an nDGP analysis of the primary CMB Planck PR4 data \cite{Tristram23} using the linear power spectrum in the nDGP model, which is computed from the linearized growth equation in the nDGP cosmology.
We obtain constraints on the nDGP model from Planck PR4 alone, as well as in combination with the SPT cluster dataset.
Our results are highly competitive with the constraints on the nDGP model from clustering wedge statistics of the galaxy correlation function and estimated growth rate values. This was applied to BOSS DR1 with Planck\,15 priors on $\Omega_{\rm m}$ and $A_s$ and resulted in the tightest constraint on nDGP so far \cite{Barreira16}. 

This paper is organized as follows. Section~\ref{sec:data} presents the cluster data from SPT and a short summary of the weak-lensing data from DES and HST. The nDGP model, the corresponding HMF, and its calibration with the BRIDGE simulations are outlined in Sec.~\ref{sec:MG}. The following Sec .~\ref {sec:analysis} describes the analysis methodology, including the scaling relations, weak-lensing model, and the Planck data used in this work. Section~\ref{sec:results} presents the results of this work. We conclude the paper with a summary in Sec.~\ref{sec:summary}.

\section{\label{sec:data}Data}
This section gives a brief summary of the cluster and weak-lensing data used in this work. A detailed description of the data products is presented in \citetalias{Bocquet24Ia}.

\subsection{\label{subsec:SPT} SPT cluster catalog}
The tSZE-selected galaxy cluster sample is created from the SPT-SZ, SPTpol~ECS, and SPTpol~500d surveys. The three surveys cover $5,270\,\rm deg^2$ of the southern sky \cite{Bleem15,Bleem20,Bleem24}, largely overlapping with the DES footprint. Cluster candidates are selected based on the tSZE detection significance $\zetahat$ and confirmed using optical and infrared data, which provide redshift measurements and center positions.  

In the $1,327\,\rm deg^2$ region that is outside the DES footprint ($\sim27\%$ of the total area), the candidates are selected by $\zetahat > 5$ and confirmed by targeted follow-up observations, resulting in a sample of 110 clusters, with a purity $\gtrsim95\%$ \cite{Bleem15,Bleem20}. 

In the $3,567\,\rm deg^2$ region that DES covers, clusters are confirmed using the MCMF tool \cite{Klein18,Klein24}, which also provides cluster redshifts, optical richnesses $\lambdahat$, and center positions. For clusters with $z < 1.1$ we use data from DES and for clusters with $z > 1.1$, WISE data are used. The MCMF tool confirms a cluster if the optical richness is higher than a redshift-dependent richness threshold $\hat \lambda_{\rm min}(z)$ which ensures a constant sample purity $>98\%$.  
The selection cut in \zetahat\ varies between surveys due to different depths and is given by: $\zetahat > 4.25$ for SPTpol~500d, $\zetahat > 4.5$ for SPT-SZ and $\zetahat > 5$ for SPTpol~ECS. This yields 895 confirmed clusters in the DES region \citepalias{Bocquet24II}.  

In total, the sample consists of 1,005 clusters with $z > 0.25$, where the lower redshift cut is employed due to contamination from the atmosphere and primary CMB anisotropies.

\subsection{\label{subsec:DES} DES Y3 and HST weak-lensing data}
We use (when available) weak-lensing shear profiles measured from the DES~Y3 shape catalog \cite{Gatti22} for clusters within the overlapping SPT-DES region. The profiles are computed within the radial range $0.5<r/(h^{-1}\mathrm{Mpc})<3.2\, (1+z_{\mathrm{cluster}})^{-1}$, where $z_{\mathrm{cluster}}$ is the cluster redshift, ensuring the exclusion of the central region and the two-halo regime. These weak-lensing measurements are used for clusters with $z < 0.95$, resulting in a sample of 688 SPT clusters with DES shear data \citepalias{Bocquet24Ia}. The redshift cut is based on the median redshift of DES's highest-redshift source bin.

For high-redshift clusters, we supplement the DES weak-lensing dataset with HST observations \cite{Schrabback18,Schrabback21,Zohren22}. A total of 39 clusters in the redshift range $0.6 - 1.7$ have HST shear measurements. Further details on the HST dataset and methodology can be found in \cite{Schrabback18,Schrabback21,Raihan20,Hernandes20,Zohren22,Sommer22}. In total 727 of our cluster sample of 1,005 clusters have weak-lensing information. 

Our weak-lensing analysis accounts for a wide range of systematic uncertainties, including cluster member contamination, miscentering effects, shear and photometric redshift calibration, halo mass modeling, and the influence of the large-scale structure. A detailed description of the modeling of these uncertainties is provided in \citetalias{Bocquet24Ia}. Although these models were calibrated within the $\Lambda$CDM framework, we expect that they remain essentially unchanged in the nDGP model, as significant deviations from GR are already ruled out.

\section{\label{sec:MG}\MakeLowercase{n}DGP gravity}
In this section, we briefly discuss the main aspects of the nDGP gravity model, including the modifications to GR collapse dynamics, and discuss the subsequent difference compared to the GR HMF.

In the DGP gravity model, the four-dimensional spacetime (the brane) is embedded in a five-dimensional spacetime (the bulk). While all particles are confined to the four-dimensional brane, gravity can propagate along the additional spatial dimension and leak into the five-dimensional bulk on large scales. Based on this assumption, the DGP action is given by \cite{Dvali2000}
    \begin{equation}
    \begin{split}
        \label{eq:EH_nDGP}        
        S = \int_{\rm brane} \dd^4 x \sqrt{-g} \left ( \frac{R}{16\pi G} + \mathcal{L}_m \right ) \\
        + \int_{\rm bulk} \dd^5 x \sqrt{-g^{(5)}} \left ( \frac{R^{(5)}}{16\pi G^{(5)}} \right ) \, ,
    \end{split}
    \end{equation}
where $g$ denotes the determinant of the metric tensor $g_{\mu\nu}$, $G$ is the gravitational constant, $R$ is the Ricci scalar in the brane and superscripts $(5)$ denote the quantities in the 5-dimensional bulk. Note that we use natural units with $\hbar = c = 1$. The length scale at which gravity becomes five-dimensional is called the crossover scale, which is defined as
    \begin{equation}
        \label{eq:def_rc}        
        r_{\rm c} = \frac{1}{2} \frac{G^{(5)}}{G} \, .
    \end{equation}
From matter domination onward, the modified Friedmann equation is given by \cite{Shandera13,Lombriser09}
    \begin{equation}
        \label{eq:exp_hist}        
        H(a) = H_0 \sqrt{\Omega_{\rm m} a^{-3} + \Omega_{\rm de}(a) + \Omega_{\rm rc}} \pm \sqrt{\Omega_{\rm rc}} \, .
    \end{equation}
Here $H_0$ is the expansion rate at redshift 0, $\Omega_{\rm m}$ is the matter density parameter, $\Omega_{\rm de}$ is the dark energy density parameter and $\Omega_{\rm rc} \coloneqq  1/(4H_0r_{\rm c})$. The different signs represent the two branches of the DGP theory. The self-accelerating branch ($-$ sign, sDGP) results in a late-time accelerating universe even without a dark energy component, \ie $\Omega_{\rm de} = 0$. However, sDGP suffers from ghost instabilities \cite{Luty03,Nicolis04,Koyama07} 
and is also ruled out by supernova and CMB data \cite{Fang08}. The $+$ sign refers to the normal branch nDGP and requires a dark energy component to yield an accelerating universe. In this paper, we consider the normal branch and tune the density and equation of state of the dark energy
such that the background history matches that in $\Lambda$CDM, \ie $H(a) = \sqrt{\Omega_{\rm m} a^{-3} + \Omega_{\Lambda}}$ \cite{Schmidt09b,Schmidt2010a}. It is also possible to assume an nDGP scenario without introducing an evolving dark energy component to match the $\Lambda$CDM background evolution, i.e. assuming Eq.~\eqref{eq:exp_hist} with $\Omega_{\rm de} = \Omega_{\Lambda}$ constant. 
However, in this scenario, the background evolution is significantly different from $\Lambda$CDM, and the constraint on the crossover scale $r_{\rm rc}$
is dominated by geometric probes rather than the growth of structure \cite{Lombriser09,Wyman10,Xu14}.

If we consider scales smaller than the Horizon, $H^{-1}$, as well as the crossover scale $r_{\rm c}$, which is entirely sufficient for the observables considered here (our constraints yield $r_{\rm c} > 1/(2H) \simeq 1500\, h^{-1}{\rm Mpc}$, and galaxy clusters emerge from perturbation with scales of $\sim 10$\,Mpc), nDGP can be described as an effective scalar-tensor theory with an extra scalar field, the brane-bending mode $\varphi$. This field arises from the ability of the brane to move in the extra dimension and mediates a gravity-like fifth force. Effectively, it contributes to the metric potentials by
    \begin{align}
        \label{eq:metric_potential_1}
        & \nabla^2 \Phi = 4 \pi G \delta \rho + \frac{1}{2} \nabla^2 \varphi \,, \\
        \label{eq:metric_potential_2}
        & \nabla^2 \Psi = 4 \pi G \delta \rho - \frac{1}{2} \nabla^2 \varphi \,.
    \end{align}
In the quasistatic regime, again appropriate on the sub-horizon scales considered here, the equation of motion for the brane bending mode is given by \cite{Schmidt09b}
    \begin{equation}
        \label{eq:eom_phi}
        \nabla^2 \varphi + \frac{r_{\rm c}^2}{3\beta(a)} \left [ (\nabla^2\varphi)^2 - (\nabla_i \nabla_j \varphi)^2 \right] = \frac{8\pi G}{3\beta(a)}\delta \rho \, ,
    \end{equation}
with the function
    \begin{equation}
        \label{eq:beta}        
        \beta(a) = 1 + 2H(a)r_{\rm c} \left( 1 + \frac{\dot H (a)}{3H^2(a)} \right) \, .
    \end{equation}
From the above equation, if $r_{\rm c} \to \infty$ then $\nabla^2 \varphi = 0$ and the equations for the metric potentials, Eqs.~\eqref{eq:metric_potential_1} and \eqref{eq:metric_potential_2}, return to their $\Lambda$CDM expressions. 

In low density regions, we can linearize Eqs.~\eqref{eq:metric_potential_1} and \eqref{eq:metric_potential_2} by neglecting quadratic terms in Eq.~\eqref{eq:eom_phi} and obtain the following relation for the dynamical potential:
    \begin{align}
        \label{eq:metric_potential_lin_1}
        \nabla^2 \Phi &= 4 \pi \left ( 1+\frac{1}{3\beta} \right ) G \delta \rho
        , \\
        \label{eq:metric_potential_lin_2}
        \nabla^2 (\Phi+\Psi) &= 4\pi G \delta\rho. 
    \end{align}
Note that the second relation states that the potential combination governing gravitational lensing (null geodesics) is unmodified from GR. From the first relation on the other hand, the linearized growth equation is given by
    \begin{equation}
        \label{eq:delta_lin}        
        \ddot{\delta} + 2H\dot \delta = 4 \pi \left ( 1+\frac{1}{3\beta} \right ) G \delta \rho \, ,
    \end{equation}
which differs from the $\Lambda$CDM equation by a factor $1+\frac{1}{3\beta}$. Because $\beta(a)$ depends only on time, the nDGP linear growth of structure is scale-independent. Thus the nDGP linear matter power spectrum, $P^{\rm nDGP}_{\rm L}(k)$, is given by the $\Lambda$CDM linear power spectrum with the same initial condition (\ie same primordial amplitude $A_s$) rescaled by the ratio of the growth factors squared $D^2$ derived from Eq.~\eqref{eq:delta_lin}.

Tests of gravity on solar system scales report results consistent with GR. Therefore, any deviations from GR must be suppressed on solar system scales \cite{Will14,Burrage18,Fischer24}. 
In nDGP, the nonlinearities in the field Eq.~\eqref{eq:eom_phi} suppress the fifth force in these dense environments. If $r_{\rm c} \sim H_0^{-1}$, then the suppression acts in any region with density much higher than the cosmological background. This is known as the Vainshtein screening mechanism \cite{Vainshtein72}. 

\subsection{\label{subsec:HMF} Collapse dynamics and the halo mass function}
To constrain nDGP gravity with the cluster abundance, we need a model for the nDGP halo mass function (HMF). 
In this work, we use the Sheth-Tormen HMF \cite{Sheth1999} adapted for nDGP, which is based on the model presented in Refs.~\cite {Schmidt09b,Schmidt2010a}. The Sheth-Tormen HMF is given by
    \begin{equation}
        \label{eq:ST_HMF}
        \left. \frac{\dd n}{\dd \mathrm{ln} M} \right\vert_{\mathrm{ST}} = -  \frac{1}{2}  \frac{\bar \rho_{\mathrm{m}}}{M_{\rm vir}} f(\delta_{\rm c} / \sigma)_{\mathrm{ST}}  \frac{\dd\mathrm{ln} \sigma^2}{\dd\mathrm{ln} M_{\rm vir}}  \, .
    \end{equation}
We use the following quantities calculated in the nDGP model: the variance of the linear matter power spectrum, $\sigma^2(M)$, the linearly extrapolated collapse overdensity, $\delta_{\rm c}$, and the virial overdensity, $\Delta_{\rm vir}$. Note that the Sheth-Thormen HMF uses the virial mass, $M_{\rm vir} = 4/3 \pi \Delta_{\rm vir} \bar \rho R_{\rm vir}^3$, as a mass definition. Since we are using $M_{200 \rm crit}$, we transform the above HMF to $M_{200 \rm crit}$ by performing a mass rescaling, also propagating the derivative $dM_{200 \rm crit}/dM_{\rm vir}$ \cite{Schmidt2010a}.
This model has been shown to agree well with $N$-body simulations of the nDGP model \cite{Schmidt09b}. As presented at the end of this section, we further calibrate this model with additional $N$-body simulations to improve its accuracy and utility.

As described in the previous section, the linear power spectrum in nDGP is the $\Lambda$CDM power spectrum rescaled by the growth factor ratio, and thus we have
    \begin{equation}
        \label{eq:sigma_nDGP}        
        \sigma(M, z) = \sigma^{\Lambda \rm CDM}(M, z) \frac{D^{\rm nDGP}(z)}{D^{\rm \Lambda CDM}(z)} \, .
    \end{equation}
$\delta_{\rm c}$ is computed from the full equation of motion for the density perturbation in a spherically symmetric setup,
    \begin{equation}
        \label{eq:delta_full}        
        \ddot{\delta} + 2H\dot \delta  - \frac{4}{3} \frac{\dot \delta}{1+\delta}= (1+\delta) \nabla^2 \Phi \, .
    \end{equation}
We find the initial overdensity $\delta_\mathrm{i}$ such that the collapse of a spherical top-hat overdensity occurs at the collapse redshift $z_{\rm c}$. $\delta_{\rm c}$ is computed from the linear extrapolation of the initial overdensity $\delta_\mathrm{i}$ using Eq.~\eqref{eq:delta_lin}.

The virial overdensity $\Delta_{\rm vir}$ is the mean overdensity of a spherical overdensity with radius $R_{\rm vir}$ with respect to the background density. The virial radius is defined as the radius (after turn-around) that satisfies the virial equation at a specific virial redshift $z_{\rm vir}$.
The density contrast at $z_{\rm vir}$ is then given as $1+\delta(R_{\rm vir})$ and is extrapolated to the collapse redshift $z_{\rm c}$ to define the virial overdensity:
    \begin{equation}
        \label{eq:Delta_vir}        
        \Delta_{\rm vir} = [1+\delta(R_{\rm vir})] \left ( \frac{1+ z_{\rm vir}}{1 + z_{\rm c}} \right )^3
    \end{equation}
The three computed quantities are used to calculate the Sheth-Tormen HMF in the nDGP model. In practice, we are not using the Sheth-Tormen HMF directly. Instead, we capture the deviations from the $\Lambda$CDM Sheth-Tormen HMF with the same initial conditions by computing the ratio of Sheth-Tormen HMFs in the two models 
    \begin{figure}
        \centering
        \includegraphics[width=\linewidth]{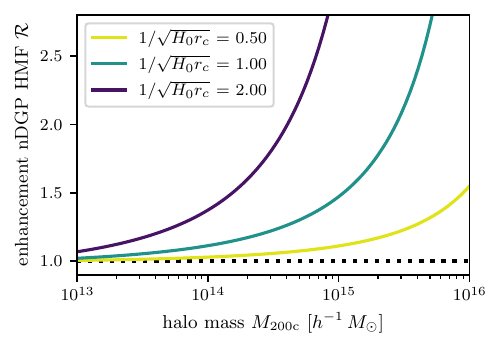}
        \vskip-0.25cm
        \caption{Enhancement of the nDGP HMF with respect to the corresponding $\Lambda$CDM cosmology from Eq.~\eqref{eq:nDGP_enhancement} in colored lines for different strengths of the nDGP model at mean cluster sample redshift $z = 0.6$. The enhancement depends on the strength of the nDGP model and grows exponentially with mass. The black dashed line shows the $\Lambda$CDM limit with no enhancement, \ie $\mathcal{R} = 1$.
        }
        \label{fig:HMF_nDGP}
    \end{figure} 
    \begin{equation}
        \label{eq:nDGP_enhancement}
        \mathcal{R} = \frac{\left. \frac{\dd n}{\dd \mathrm{ln} M} \right\vert_{\mathrm{ST},\,\mathrm{nDGP}}}{\left. \frac{\dd n}{\dd \mathrm{ln} M} \right\vert_{\mathrm{ST},\,\Lambda\mathrm{CDM}}} \, .
    \end{equation}
To obtain the HMF in nDGP gravity, we multiply this ratio by a $\Lambda$CDM HMF of our choice. 
This approach has the advantage of being independent of the Sheth-Tormen HMF, because the latter is only used to compute the relative enhancement of the $\Lambda$CDM HMF due to nDGP gravity.
In this work, we use the Tinker HMF as the $\Lambda$CDM baseline~\cite{Tinker08} 
    \begin{equation}
        \label{eq:Tinker_HMF}
        \left. \frac{\dd n}{\dd \mathrm{ln} M} \right\vert_{\mathrm{T},\,\Lambda\mathrm{CDM}} = - \frac{\bar \rho_{\mathrm{m}}}{2M} f(\sigma_{\Lambda\rm CDM})_{\mathrm{T}} \frac{\dd \mathrm{ln} \sigma_{\Lambda\rm CDM}^2}{\dd \mathrm{ln} M}\, ,
    \end{equation}
with the multiplicity function of the form
    \begin{equation}
        \label{eq:Tinker_multi_fct}
        f(\sigma_{\Lambda\rm CDM})_{\mathrm{T}} = \tilde A \left [ \left(\frac{\sigma_{\Lambda\rm CDM}}{\tilde b}\right)^{-\tilde a} +1 \right]e^{-\frac{\tilde c}{\sigma_{\Lambda\rm CDM}^2}}\, ,
    \end{equation}
where $\tilde A,\ \tilde a,\ \tilde b$ and $\tilde c$ are parameters calibrated using $N$-body simulations~\cite[see][table 2]{Tinker08} and $\sigma_{\Lambda\rm CDM}$ is the variance of the power spectrum in the corresponding GR cosmology.
The nDGP HMF is then given by
    \begin{equation}
        \label{eq:HMF_nDGP}
        \frac{\dd n}{\dd \mathrm{ln} M} =  \mathcal{R} \left. \frac{\dd n}{\dd \mathrm{ln} M} \right\vert_{\mathrm{T},\,\Lambda\mathrm{CDM}}\, .
    \end{equation}
Figure~\ref{fig:HMF_nDGP} shows the ratio predicted by Eq.~\eqref{eq:nDGP_enhancement} for different values of the nDGP parameter \nDGP\ at the mean redshift of the cluster sample $z = 0.6$. As expected from the enhanced structure formation, the HMF is increased in nDGP gravity. Moreover, the enhancement depends strongly on mass and shows a significant deviation from $\Lambda$CDM for massive halos, which is because the HMF is exponentially sensitive to $\delta_{\rm c}/\sigma(M)$ at high masses. 
We account for the effect of massive neutrinos by using the baryonic and cold dark matter only power spectrum in our calculations \cite{Ichiki12,Costanzi13} and refer to a standard cosmology with massive neutrinos as $\nu\Lambda$CDM. 

\begin{figure*}
        \centering
        \includegraphics[width=\textwidth]{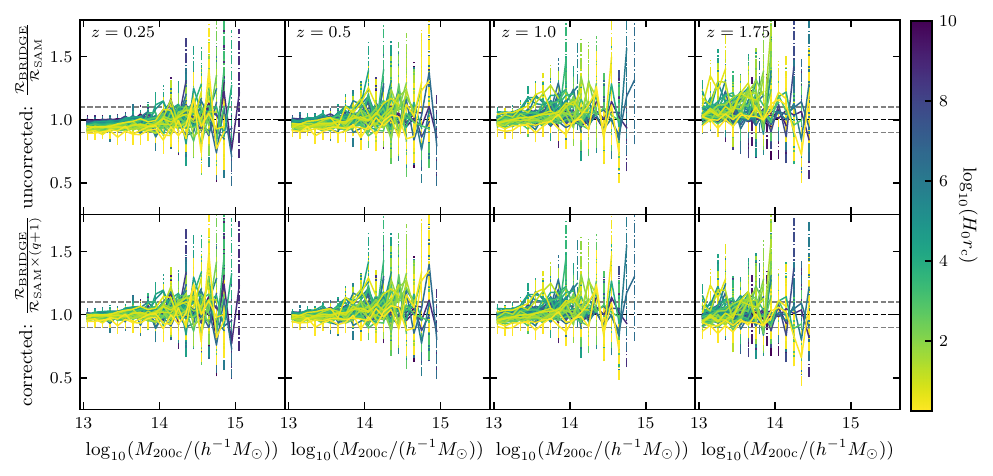}
        \vskip-0.25cm
        \caption{
        Comparison between the HMF enhancement predicted by the BRIDGE simulations, $\mathcal{R}_{\rm BRIDGE}$, and the semi-analytical model, $\mathcal{R}_{\rm SAM}$, across various redshifts. The data points are color-coded by $\log(H_0r_{rm c})$ values. Gray dashed lines indicate $\pm10\%$ deviation as a visual reference. The top four panels present the uncorrected comparison based on Eq.~\eqref{eq:nDGP_enhancement}, while the bottom four panels show the results after applying the correction factor from Eq.~\eqref{eq:nDGP_enhancement_correction}, resulting in improved agreement. Error bars reflect the jackknife covariance estimated from the BRIDGE simulation.
        }
        \label{fig:comp_BRIDGE}
    \end{figure*}

Since the Sheth-Tormen HMF is a semi-analytical model, we want to compare the results of Eq.~\eqref{eq:nDGP_enhancement} to non-linear simulations to validate these models. For this we use the BRIDGE simulations \cite{Harnois23,Ruan24,Davies24}, which cover a wide range of nDGP cosmologies by varying the parameters $\Omega_m$, $h$, $S_8^{\Lambda \rm CDM} = \sigma_{8,\Lambda \rm CDM} \sqrt{\Omega_{\rm m}/0.3}$, and $\log (H_0r_{\rm c})$, and with a box length of $L = 500\,h^{-1}\,\rm Mpc$. In addition to each nDGP simulation, a $\Lambda$CDM box was created with the same initial conditions. Note that the simulations use massless neutrinos; therefore, the calibration with these simulations does not account for the effects of massive neutrinos.
In total 49 cosmologies plus one fiducial $\Lambda$CDM cosmology are simulated, and we use the snapshots at nine evenly distributed redshifts between 0 and 2 to compute the ratio of halo numbers from the nDGP and the $\Lambda$CDM boxes, $\mathcal{R}_{\rm BRIDGE}$. The comparison for four redshifts is shown in Fig.~\ref{fig:comp_BRIDGE}, and one sees that the two ratios do not fully agree.
Based on the comparison in Fig.~\ref{fig:comp_BRIDGE}, we model the bias between the semi-analytical model and the simulations as a constant $q+1$ for each cosmology. 
We limit the correction to a constant bias, as the shot noise in the simulations
 precludes us from reliably fitting a more complex fitting function. Fitting the bias $q$ for each cosmology yields a linear trend in \nDGP\ and redshift $z$. 
Therefore, we model the correction as follows
    \begin{equation}
        \label{eq:nDGP_enhancement_correction}
        \frac{\mathcal{R}_{\rm BRIDGE}}{\mathcal{R}_{\rm SAM}} =  \tilde q \left(  \frac{1}{\sqrt{H_0r_{\rm c}}} - 0.1\right ) (z-0.8) + 1\, .
    \end{equation}
Here $\tilde q$ is the mean bias, and we 
choose pivot values in $1/\sqrt{H_0r_{\rm c}}$ and $z$ of 0.1 and 0.8, respectively. 
We separate the dependence on \nDGP\ and $z$ from the bias through the above modeling and obtain a mean bias parameter $\tilde q = 0.06$ for the correction function. Note that we assume no bias for cosmologies with $\nDGP < 0.1$, because the bias should vanish for $\nDGP = 0$ as we approach $\Lambda$CDM.
Applying the above one-parameter correction to the ratio of the semi-analytical model to the simulations reduces the variance between the semi-analytical model and the simulations by $26\%$ from $0.0073$ to $0.0058$. 
This variance is estimated from the comparison of enhancement factors shown in Fig.~\ref{fig:comp_BRIDGE} in all mass and redshift bins and is thus representative of the overall uncertainty in the amplitude of the HMF.

To summarize, the full nDGP HMF used in our analysis is given by 
    \begin{equation}
    \begin{split}
        \label{eq:HMF_nDGP_correction}
        \frac{\dd n}{\dd \mathrm{ln} M} (M, z) =& \left. \frac{\dd n}{\dd \mathrm{ln} M} (M, z)\right\vert_{\mathrm{T}} \mathcal{R} \\ 
        & \times \left [  \tilde q \left(  \frac{1}{\sqrt{H_0r_{\rm c}}} - 0.1\right ) (z-0.8) +1 \right]  \, .
    \end{split}
    \end{equation}
and we also account for an $8\%$ uncertainty in the amplitude of the HMF based on the scatter from the corrected HMF enhancement.

\section{\label{sec:analysis}Cluster Analysis Method}
Our analysis follows the state-of-the-art cluster analysis with weak-lensing informed mass calibration developed for the SPT cluster sample \citepalias{Bocquet24Ia}. This chapter briefly summarizes the cluster analysis, and we refer the reader to \citepalias{Bocquet24Ia} and \citepalias{Vogt24b} for more details.

\subsection{\label{subsec:obs-mass-rel} Observable--mass relations}
Clusters of galaxies are detected through observables such as the tSZE detection significance and optical richness, with empirical scaling relations connecting these observables to the halo mass \eg, \cite{Kaiser86,Angulo12}. A connection between the halo observable function, which is the differential cluster number in the observable space, and the HMF can be made by calibrating these relations with weak gravitational lensing measurements. This connection allows us to relate observables to the HMF, which is sensitive to cosmology, and provides constraints on cosmological parameters.

The observed tSZE detection significance, \zetahat, is linked to the intrinsic detection significance, $\zeta$ following the relation \cite{Vanderlinde10}:
    \begin{equation}
        \label{eq:zeta_to_observed}
        P(\hat\zeta | \zeta) = \mathcal{N} \left( \sqrt{\zeta^2 +3}, 1 \right)  \,,
    \end{equation}
which accounts for observational noise. The mean intrinsic tSZE detection significance, $\zeta$, is linked to the underlying mass and redshift by the following power law:
    \begin{equation}
    \begin{split}
        \label{eq:zeta_mass_rel}
        \langle \ln \zeta \rangle = \ln\asz &+ \bsz \ln\left (  \frac{M_{200\mathrm{c}}}{3 \times 10^{14}\, h^{-1} M_\odot} \right) \\
        &+ \csz\ln \left( \frac{E(z)}{E(0.6)} \right) \,.
    \end{split}
    \end{equation}
A lognormal intrinsic scatter with variance $\sigmalnzeta$ around this mean relation is assumed. To account for variation in survey depth, \asz\ and \csz\ are rescaled for each SPT field and survey, respectively \cite{Bleem15,Bleem20,Bleem24}. The normalization factor $\gamma_{\rm ECS}$ of \asz\ for the SPTpol ECS survey is difficult to determine and thus is treated as an additional free parameter in the analysis.

Similarly, the observed richness, \lambdahat, relates to the intrinsic richness, $\lambda$, through a Gaussian distribution:
    \begin{equation}
        \label{eq:richness_to_observed}
        P(\hat \lambda | \lambda) = \mathcal{N} ( \lambda,  \sqrt{\lambda} )  \,.
    \end{equation}
This approximates the Poisson sampling noise in the richness estimation. The mean intrinsic richness follows a power-law dependence on mass and redshift:
    \begin{equation}
        \label{eq:lambda_mass_rel}
        \begin{split}
        \langle \ln \lambda \rangle = \ln \alambda &+ \blambda \ln\left (  \frac{M_{200\mathrm{c}}}{3 \times 10^{14}\, h^{-1} M_\odot} \right) \\
        &+ \clambda\ln \left( \frac{1+z}{1.6} \right) \,.
        \end{split}
    \end{equation}
We assume a lognormal scatter width \sigmalnlambda\ around this relation.

For clusters with $z \leq 1.1$, richness is measured using DES, while for high-redshift ($z > 1.1$) clusters, WISE data are used. 
As these datasets have different richness measurements, we adopt separate $\lambda$--mass relations for DES and WISE clusters \citepalias{Bocquet24II}.

The scaling relation described above enables the translations of the halo observable function in the $\zeta$--$\lambda$--$z$ space to the halo mass functions and allows constraints on cosmological parameters.
However, given the lack of information on the priors of these scaling relation parameters and their scatters, we rely on additional observations to calibrate these relations empirically. As the link between the mass of a cluster and the corresponding weak-lensing signal is well understood, using weak-lensing data for this calibration is robust.

\subsection{\label{subsubsec:WL_model} Cluster weak-lensing model}
The model we adopt for the DES weak-lensing data has been studied and described in detail in \citetalias{Bocquet24Ia} and the references therein. Here, we summarize the key aspects of this approach. It is important to note that we assume any modifications arising from nDGP gravity in the mapping from cluster mass to lensing signal to be negligible. 
While in the nDGP model the lensing signal for a given mass distribution is unmodified from GR,
nDGP gravity can still change the  halo profiles. However, these modifications have been found to be minor \cite{Schmidt2010b}. Note that any variations in the cluster observables $\zetahat$ and $\lambdahat$ at fixed halo mass are accounted for by the empirical calibration of the observable--mass relations.

The weak-lensing observable is the reduced shear and is related by a projected Navarro-Frenk-White profile (NFW) \cite{Navarr01996,Bartelmann17} to a mass, \MWL, by 
    \begin{equation}
      \label{eq:g_model}
      g_\mathrm{t}(r,\MWL) = \frac{\Delta\Sigma(r,\MWL)~\Sigma_\mathrm{crit}^{-1}}{1-\Sigma(r,\MWL)~\Sigma_\mathrm{crit}^{-1}} \, .
    \end{equation}
We refer to the mass inferred from this observable as the weak-lensing mass \MWL\ of the cluster.
Since real cluster profiles deviate from a perfect NFW profile, the inferred weak-lensing mass \MWL\ is a biased and noisy estimator of the true cluster mass $M_{200 \rm c}$ \cite{becker&kravtsov11,oguri&hamana11}. To correct for this bias, we adopt a scaling relation between \MWL\ and $M_{200 \rm c}$, with the mean relation given by \cite{Grandis21}:
    \begin{small}
    \begin{equation}
        \label{eq:WL_mass_rel}
         \left\langle \mathrm{ln} \left( \frac{M_{\mathrm{WL}}}{M_0}  \right)  \right\rangle =\bWL(z) + \bWLM \mathrm{ln} \left( \frac{M_{200\mathrm{c}}}{M_0} \right) \,.
    \end{equation}
    \end{small}
Here \bWL\ is the logarithmic mass bias normalization and \bWLM\ is the mass trend in this bias at a pivot mass $M_0 = 2 \times 10^{14}\, h^{-1} M_\odot$.
We model the scatter around this relation with a lognormal distribution with a variance given by:
    \begin{equation}
        \label{eq:WL_mass_var}
       \mathrm{ln}\,\sWLall^2 =
       \sWL(z)  +\sWLM  \mathrm{ln}  \left( \frac{M_{200\mathrm{c}}}{M_0} \right)\, ,
    \end{equation}
where \sWL\ is the normalization and \sWLM\ is the mass trend of the scatter.
The parameters defining the mean scaling relation and the scatter are calibrated using hydrodynamical simulations, where the weak-lensing inferred mass is extracted and compared to the corresponding true cluster mass obtained from matched gravity-only simulations across different redshifts \cite{Grandis21,Bocquet24Ia}.
\newline
\newline 
A similar model is applied to the HST dataset.
Here, the weak-lensing mass $\MWL$ is mapped to the true halo mass with a mean relation \cite{Schrabback18},
    \begin{equation}
        \label{eq:HST_WL}
        \langle \ln\,\MWL \rangle = \bWLHST + \ln\,M_{200\mathrm{c}}\, ,
    \end{equation}
where \bWLHST is the bias between the true halo and weak-lensing mass.
The mean relation scatters around the mean with a Gaussian distribution of width \sWLHST. The two parameters are found on a cluster-by-cluster basis. We refer the reader to the original works for a more detailed explanation of the cluster lensing model employed for the HST dataset \cite{Schrabback18,Schrabback21,Zohren22,Sommer22}.

\subsection{\label{subsec:likelihood} Cluster Likelihood}
Our analysis is based on a Bayesian analysis framework, where we determine the posterior distributions of cosmological and scaling relation parameters through a cluster population model. The likelihood function used in this study follows the methodology from Refs.~\citepalias{Bocquet24Ia,Bocquet24II}. We assume that a Poisson distribution gives the multi-observable cluster abundance likelihood:
    \begin{equation}
      \label{eq:cluster_likelihood}
      \begin{split}
        \ln \mathcal L&\left(\{\zetahat,\lambdahat_i,z_i,\boldsymbol g_{\mathrm{t},i}\}_{i=1}^{N_\mathrm{cluster}}\big|\boldsymbol p\right) =\\
        & \sum_{i=1}^{N_\mathrm{cluster}} \ln\frac{\mathrm{d}^4 N(\boldsymbol p)}{\mathop{\mathrm{d}\zetahat} \mathop{\mathrm{d}\lambdahat} \mathop{\mathrm{d} \boldsymbol g_\mathrm{t}} \mathop{\mathrm{d} z}}
        \Big|_{\zetahat_i, \lambdahat_i, \boldsymbol g_{\mathrm{t},i}, z_i} \\
        &- \idotsint \mathop{\mathrm{d}\zetahat} \mathop{\mathrm{d}\lambdahat} \mathop{\mathrm{d} \boldsymbol g_\mathrm{t}} \mathop{\mathrm{d} z} 
         \frac{\mathrm{d}^4 N(\boldsymbol p)}{\mathop{\mathrm{d}\zetahat} \mathop{\mathrm{d}\lambdahat} \mathop{\mathrm{d} \boldsymbol g_\mathrm{t}} \mathop{\mathrm{d} z}} \Theta_\mathrm{s}(\zetahat,\lambdahat,z) \\
        &+\mathrm{const.}
      \end{split}
    \end{equation}
where the index $i$ runs over all observed clusters, and $\Theta_\mathrm{s}(\zetahat,\lambdahat,z)$ represents the sample selection function, see Sec.~\ref{subsec:SPT}. The differential cluster abundance entering the likelihood function 
is the differential halo observable function and
is expressed as:
    \begin{equation}
      \label{eq:dN}
      \begin{split}
        \frac{\mathrm{d}^4 N(\boldsymbol p)}{\mathop{\mathrm{d}\zetahat} \mathop{\mathrm{d}\lambdahat} \mathop{\mathrm{d} \boldsymbol g_\mathrm{t}} \mathop{\mathrm{d} z} } =&
      \int \mathop{\mathrm{d}\Omega_\mathrm{s}} \idotsint \mathop{\mathrm{d} M} \mathop{\mathrm{d}\zeta} \mathop{\mathrm{d}\lambda} \mathop{\mathrm{d} M_\mathrm{WL}} \\
      &P(\zetahat|\zeta)
      P(\lambdahat|\lambda)
      P(\boldsymbol g_\mathrm{t}|M_\mathrm{WL}, \boldsymbol p) \\
      &P(\zeta, \lambda, M_\mathrm{WL} |M,z,\boldsymbol p) \\
      &\frac{\mathrm{d}^2 N(M, z, \boldsymbol p)}{\mathop{\mathrm{d} M} \mathop{\mathrm{d} V}} \frac{\mathrm{d}^2 V(z,\boldsymbol p)}{\mathop{\mathrm{d} z} \mathop{\mathrm{d}\Omega_\mathrm{s}}}.
      \end{split}
\end{equation}
Here, $\Omega_\mathrm{s}$ represents the survey solid angle. The terms $\frac{\dd^2 N (M, z, \boldsymbol p)}{\dd M \dd z}$ and $\frac{\dd^2 V (z, \boldsymbol p)}{\dd z \dd\Omega_\mathrm{s}}$ correspond to the HMF and the differential volume element, respectively. The probabilities $P(\hat\zeta|\zeta)$, $P(\hat\lambda|\lambda)$, and $P(\boldsymbol g_\mathrm{t}|M_\mathrm{WL}, \boldsymbol p)$ describe the relationships between the observed and intrinsic quantities. The joint probability $P(\zeta, \lambda, M_\mathrm{WL} |M,z,\boldsymbol p)$ is the multi-observable--mass relation, accounting for potential correlations among the three observables. 
Our model is described by 23 astrophysical parameters and seven cosmological parameters, denoted by $\boldsymbol p$.

In the analysis, we use uniform priors for the cosmological parameters $\Omega_{\rm m}$, $\ln 10^{10}A_s$, $\Omega_\nu h^2$, and \nDGP\ as well as for all scaling relation parameters. 
We adopt Gaussian priors for the weak-lensing parameters in Eqs.~\eqref{eq:WL_mass_rel}--\eqref{eq:HST_WL}. The cluster abundance does not constrain the cosmological parameters $\Omega_{\rm b}h^2$, $n_s$ and $h$, and we apply Gaussian priors on the first two from Planck and a Gaussian prior on the latter of $h \sim \mathcal{N}(0.70, 0.05^2)$. In the joint analysis with Planck, we use uniform priors for these parameters instead.

\subsection{\label{subsec:likelihood_Plancl} Planck PR4 data and likelihood}
Primary CMB data, such as those from the Planck satellite, place tight constraints on standard cosmological parameters and can break degeneracies with modified gravity parameters. Therefore, the combination of the cluster abundance and primary CMB anisotropies can impose competitive constraints on nDGP. In this work, we use the latest Planck PR4 data to achieve such constraints, adopting \textsc{HiLLiPoP} likelihoods for the high-$l$  TT, TE and EE spectra and the \textsc{LoLLiPoP} low-$l$ EE spectrum likelihood \cite{Tristram23}. 
For the low-$l$ TT spectrum, the original Planck PR3 likelihood is used \cite{Planck2020}. The CMB power spectrum in nDGP is derived from the linearized metric potential equations~\eqref{eq:metric_potential_lin_1} and \eqref{eq:metric_potential_lin_2}, where the deviations from GR are encoded in the modification functions $\mu$ and $\gamma$ respectively. To compute the nDGP power spectrum, we implemented the two nDGP expressions for $\mu$ and $\gamma$ in the Boltzmann code \textsc{MGCAMB} \cite{Zhao08,Hojjati11,Zucca19,Wang23}.\footnote{\url{https://github.com/sfu-cosmo/MGCAMB}}

When using primary CMB data from Planck, the results on scale-independent modified gravity models change from Planck PR3 to Planck PR4 \cite{Planck2020,Ishak24,Specogna24}.
Earlier Planck analysis from Planck\,15 \cite{Planck_MG15} 
and Planck\,PR3 \cite{Planck2020} reported a $\sim 2.1\sigma$ detection of a scale-independent modified gravity model.\footnote{Note that the Planck analyses adopted a different parametrization of the modification function $\mu$ and $\gamma$, but this remains valid for nDGP gravity as shown in Appendix~\ref{app:Planck_MG}.}
This is due to the CMB lensing anomaly \cite{Calabrese08,Renzi13,Mokeddem23}. 
This anomaly refers to a systematic effect found in the Planck PR3 data, which prefers a smoother power spectrum at high $l$ than predicted from the standard cosmological model. The tension can be resolved by introducing an empirical lensing parameter $A_{\rm lens}$, which modifies the amplitude of the CMB lensing effect  and is unity for $\Lambda$CDM by definition. 
Allowing $A_{\rm lens}$ to vary improves the fit to the data and gives a value of $A_{\rm lens}$ away from unity at the $2-3\,\sigma$ level.
The empirical $A_{\rm lens}$ parameter is degenerate with the modified gravity parameter $\mu$ and, if not accounted for, the $A_{\rm{lens}} \neq 1$ anomaly can lead to a detection of scale-independent modified gravity in the analyses of Planck\,15 and Planck\,PR3 \cite{Planck_MG15,Planck2020,Specogna24}. 
Recently, the $A_{\rm lens}$ anomaly was resolved with the \texttt{LoLLiPoP} and \texttt{HiLLiPoP} likelihoods, through more advanced modeling of the systematics \cite{Tristram23}. Therefore, we decide to use this latter likelihood combination.
A detailed comparison of the nDGP constraints derived from different Planck datasets is provided in Appendix~\ref{app:Planck_MG}.

\section{\label{sec:results}Results}
In this section, we present the constraints on nDGP gravity from SPT clusters with mass calibration from DES and HST, Planck PR4 and the combination. 
Note that all reported uncertainties are provided at the $68\,\%$ credible level and upper limits at $95\,\%$ credibility.

\subsection{\label{subsec:results_cl}nDGP constraint from clusters}
\begin{figure}
        \centering
        \includegraphics[width=\linewidth]{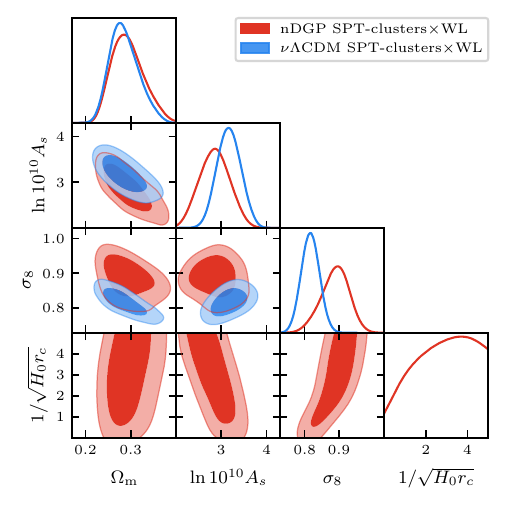}
        \vskip-0.25cm
        \caption{Posterior distribution on $\Omega_{\rm m}$, $\log_{10}A_s$, $\sigma_8$ and \nDGP\ ($68\,\%$ and $95\,\%$ credible regions) for the nDGP SPT-clusters$\times$WL analysis in red and for reference the $\nu\Lambda$CDM analysis from \citetalias{Bocquet24II} in blue. The cluster dataset alone cannot meaningfully constrain the nDGP parameter \nDGP. Compared to the $\nu\Lambda$CDM analysis, the constraint on $\Omega_{\rm m}$ remains the same while $\sigma_8$ (which is computed using the nDGP linear growth equation) is shifted to higher values due to the enhanced growth in nDGP.
        }
        \label{fig:results_cl_comp_LCDM}
    \end{figure}

Figure~\ref{fig:results_cl_comp_LCDM} shows in red the results from the SPT cluster abundance alone for the four parameters $\Omega_{\rm m}$, $A_s$, $\sigma_8$, and \nDGP.
The posterior shows that the nDGP parameter \nDGP\ cannot be meaningfully constrained by this SPT dataset alone. This is mainly due to the degeneracy between \nDGP\ and the standard cosmological parameters, such as $\Omega_{\rm m}$ and $A_s$.
These degeneracies are visible in the two-dimensional contours of Fig.~\ref{fig:results_cl_comp_LCDM}. 
A negative correlation between $A_s$ and \nDGP\ is visible which accounts for the fact that an increase in the HMF due to modified gravity can be compensated by a lower initial amplitude of fluctuations. Note that $\sigma_8$ in our definition includes the modified growth, see Eq.~\eqref{eq:sigma_nDGP}, and is a function of both $A_s$ and \nDGP. Therefore, a positive correlation emerges between \nDGP\ and $\sigma_8$.

Figure~\ref{fig:results_cl_comp_LCDM} also shows the $\nu\Lambda$CDM baseline analysis of \citetalias{Bocquet24II} in blue. In comparison, the uncertainty in $\Omega_{\rm m}$ remains almost unchanged and increases only by $7\,\%$ in the nDGP analysis. In contrast, the uncertainty in $\sigma_8$ increases by $58\,\%$, and its posterior shifts to higher values.
This shift arises because the SPT clusters constrain the amplitude of matter fluctuations at an effective redshift characteristic of the sample, here $z \sim 0.6$, and on an effective physical scale that does not correspond exactly to $8\,h^{-1}\,\mathrm{Mpc}$. Because the growth history in nDGP depends on \nDGP\ and is different than in  $\Lambda$CDM, mapping this cluster constrained amplitude to $\sigma_8(z=0)$ leads to a higher inferred present-day value of $\sigma_8$. The shift to lower values of $A_s$, discussed above, reflects the compensation of the HMF enhancement induced by modified gravity.

In summary, the current SPT cluster dataset does not provide sufficient information to constrain all three parameters simultaneously, particularly the nDGP parameter. However, ongoing galaxy cluster surveys such as SPT-3G \cite{Benson14} and the Simons Observatory \cite{Ade19} will significantly improve the sensitivity to cosmological parameters and offer the potential to constrain modified gravity models without relying on additional datasets \cite{Vogt24}.

In Fig.~\ref{fig:all_params} in Appendix~\ref{app:triangle_plot}, we present the full posterior distributions, comparing the nDGP and $\nu\Lambda$CDM analyses. The scaling relation parameters of the $\zeta$--mass and $\lambda$--mass relations show good agreement with the $\nu\Lambda$CDM baseline result. As expected, the uncertainties on these parameters are slightly larger in the nDGP analysis due to the additional degree of freedom introduced by the modified gravity model. Moreover, we do not observe any significant degeneracies between the nDGP and scaling relation parameters.

\subsection{\label{subsec:results_PR4}nDGP constraint from Planck PR4}
In this work, we also present the first constraints on the nDGP gravity model from the primary CMB Planck data.
We show the posterior distribution of $\Omega_{\rm m}$, $\sigma_8$
and \nDGP\ from Planck PR4 alone in Fig.~\ref{fig:results_combined} in green. 
We report an upper limit on the nDGP parameter \nDGP\ from Planck PR4 of 
    \begin{equation}
        \label{eq:nDGP_constaint_PR4}
        \nDGP <  1.62 \quad (95\,\% \quad \rm{limit}) \, .
    \end{equation}
This result is consistent with zero at $95\,\%$ credibility and thus we find no significant deviation from GR. 
The obtained value of $\sigma_8$ from the nDGP analysis is higher than from the $\nu\Lambda$CDM analysis due to the enhanced structure formation coming from the additional fifth force (see Sec.~\ref{sec:MG}). A positive correlation between $\sigma_8$ and \nDGP\ is seen because the CMB primarily constraints the primordial amplitude $A_s$, and, for fixed $A_s$, $\sigma_8$ depends on \nDGP\ as shown in Eq.~\eqref{eq:sigma_nDGP}.

The constraints on the nDGP model from the primary CMB Planck data depend on the underlying Planck dataset and likelihood. In Appendix~\ref{app:Planck_MG}, we present results considering different Planck datasets and analysis choices. In the main analysis, we rely on the most recent and up-to-date Planck PR4 release from Ref.~\cite{Tristram23}. 

\subsection{\label{subsec:results_cl_PR4}nDGP constraint from SPT-clusters$\times$WL+PR4} 
\begin{figure}
        \centering
        \includegraphics[width=\linewidth]{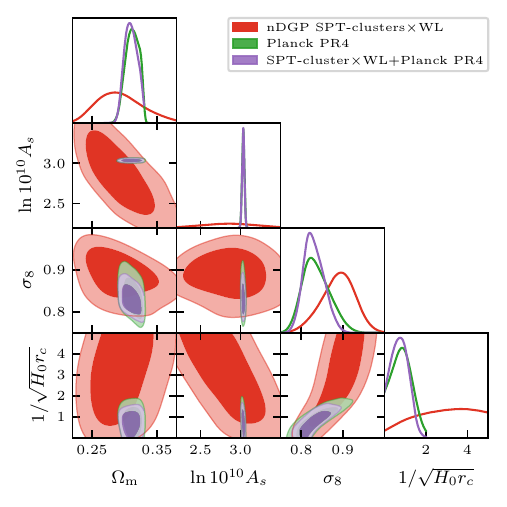}
        \vskip-0.25cm
        \caption{
        Posterior distribution on  $\Omega_{\rm m}$, $A_s$, $\sigma_8$ 
        and \nDGP\ ($68\,\%$ and $95\,\%$ credible regions) for the nDGP SPT-cluster$\times$WL analysis in red, Planck PR4 in green and the combination in purple. The joint analysis places a competitive constraint on the nDGP parameter $\nDGP < 1.41$ at $95\,\%$ credibility.
        }
        \label{fig:results_combined}
    \end{figure}

Finally, we combine both of the above considered datasets. For this, we multiply the primary CMB Planck PR4 likelihood with the SPT-cluster$\times$WL dataset likelihood, which is meaningful when the two posteriors are statistically consistent, with the additional assumption that there is no correlation between the two datasets. 
As shown in Fig.~\ref{fig:results_combined}, the combination tightens the constraints on $\Omega_{\rm m}$, $\sigma_8$ and \nDGP. While the tightening of the constraint on $\Omega_{\rm m}$ is mild, the improvement on \nDGP\ and $\sigma_8$ is much larger. From the combination, we get an upper limit on the nDGP parameter of 
    \begin{equation}
        \label{eq:nDGP_constaint}
        \nDGP < 1.41 \quad (95\,\% \quad \rm{limit}) \, ,
    \end{equation}
which is $15\,\%$ tighter than the Planck PR4 only results. 
The constraints on the other cosmological parameters can be found in Tab.~\ref{tab:results}. 
\begin{table*}
    \caption{Constraints on the cosmological parameters $\Omega_{\rm m}$, $\sigma_8$, $\sum m_\nu$ and \nDGP\ for the three datasets used in this work (mean and 68\% credible intervals, or 95\% limit). The nDGP parameter \nDGP\ and the total neutrinos mass $\sum m_\nu$ are not a meaningful constraint by the SPT-cluster$\times$WL dataset alone (missing entries (\dots) in the table), and we only quote the constraints from Planck PR4 and the combination for these parameters. For comparison to other literature results, we also quote in the last row the result from the combination with massless neutrinos.}
    \label{tab:results}
    \begin{ruledtabular}
      \begin{tabular}{lcccc}
        Dataset & $\Omega_{\rm m}$ & $\sigma_8$ & $\sum m_\nu$ [eV] &\nDGP\\
        \colrule
        SPT-clusters$\times$WL & $0.294\pm0.035$ & $0.889\pm0.040$ & \dots & \dots \\
        Planck\,PR4 & $0.312\pm0.010$ & $0.837 \pm 0.033 $ & $ < 0.26$ & $<1.62$ \\
        SPT-clusters$\times$WL + Planck\,PR4& $0.3098 \pm0.0095$ & $0.831 \pm 0.025$ &  $ < 0.24$  & $<1.41$ \\
        SPT-clusters$\times$WL + Planck\,PR4 ($\sum m_\nu = 0$\,eV)& $0.2984\pm0.0064$ & $0.842^{+0.027}_{-0.012}$ & 0 & $<1.14$ \\
        
      \end{tabular}
    \end{ruledtabular}
  \end{table*}
In comparing our results to the literature, a few considerations should to be taken into account.
First, a meaningful comparison can only be made if the nDGP model studied in the related work also assumes a $\Lambda$CDM expansion history [see discussion after Eq.~\eqref{eq:exp_hist}]. Therefore, we restrict our comparison to Refs.~\cite{Raccanelli13,Barreira16}.
In addition, our analysis allows the total neutrino mass to vary, whereas Refs.~\cite{Raccanelli13,Barreira16} assume massless neutrinos. This difference prevents a direct comparison of the results. To address this, we also perform an analysis with zero neutrino mass and obtain a tighter constraint of $\nDGP < 1.14$, which we use for the comparison below.
Ref.~\cite{Raccanelli13} uses measurements of the monopole and quadrupole of the two-point correlation function of the LRG sample from SDSS II DR7 and found $r_{\rm c} > 340$\,Mpc, which translates to $\nDGP < 3.6$ at $95\,\%$ credibility \cite{Raccanelli13}. 
Ref.~\cite{Barreira16} reported the current tightest constraints on nDGP gravity with a $\Lambda$CDM background evolution by using clustering wedges statistics of the galaxy correlation function measured from BOSS DR1 with Planck\,15 priors on $\Omega_{\rm m}$ and $A_s$. 
There, the authors reported $1/(H_0r_{\rm c}) < 0.97$ which translates to $\nDGP < 0.99$ at $95\,\%$ credibility. In comparison, we see that our constraint (when assuming massless neutrinos) is three times tighter in \nDGP\ than the one reported in Ref.~\cite{Raccanelli13} and is of the same level ($17\,\%$ less tight) as the most stringent constraint from Ref.~\cite{Barreira16}. Importantly, our analysis goes beyond simply incorporating additional priors from Planck as done by Ref.~\cite{Barreira16}, whereas we perform a self-consistent joint analysis of the cluster dataset and the primary CMB data from Planck using the explicit likelihoods for both datasets.

\section{\label{sec:summary}Summary}

In this paper, we present a cosmological analysis of the nDGP gravity model using the abundance of massive galaxy clusters detected with SPT with weak-lensing mass information from DES and HST and in combination with Planck PR4. The cluster sample consists of 1,005 clusters with redshifts $z > 0.25$, selected from the SPT-SZ, SPTpol~ECS and SPTpol~500d surveys \cite{Bleem15,Bleem20,Bleem24} and confirmed with the MCMF algorithm \cite{Klein18,Klein24} (in the DES footprint) and targeted follow-up observations. We use weak-lensing tangential shear profiles for the simultaneous weak-lensing mass calibration from DES for 688 clusters with $z < 0.95$ \cite{Gatti22} and HST measurements for 39 SPT clusters with higher redshifts $0.6 - 1.6$ \cite{Schrabback18,Schrabback21,Zohren22}. The analysis framework used here is based on the state-of-the-art cluster analysis presented in \citepalias{Bocquet24Ia,Bocquet24II}.

The nDGP gravity model alters the structure growth on cosmological scales by introducing an effective gravity-like fifth force \cite{Dvali2000,Schmidt09b,Schmidt2010a}, thereby affecting and enhancing the abundance of massive galaxy clusters compared to the standard cosmological model. Therefore, the HMF is modified, and we use a semi-analytical approach, where we rescale the $\nu\Lambda$CDM HMF with an enhancement factor to account for the increased clustering due to the fifth force \cite{Schmidt09b,Schmidt2010a}. This rescaling involves the spherical collapse quantities computed in the nDGP gravity model, including the critical overdensity $\delta_{\rm c}$ and the virial overdensity $\Delta_{\rm vir}$. 

To validate the semi-analytical model we compare it against the BRIDGE simulations \cite{Harnois23,Ruan24,Davies24}, a set of 49 $N$-body simulations for nDGP gravity with different values of the cosmological parameters $\Omega_m$, $h$, $S_8^{\rm GR} = \sigma_8^{\Lambda \rm CDM} \sqrt{\Omega_{\rm m}/0.3}$, and $\log (H_0r_{\rm c})$. We find good agreement for nDGP model realizations close to GR (\ie $\log (H_0r_{\rm c})$ large or \nDGP\ close to zero). A bias is noticeable for larger deviations from GR between the simulations and the semi-analytical HMF. We model this bias as a mass-independent factor that depends on the nDGP parameter \nDGP\ and redshift $z$, and correct our semi-analytical HMF accordingly. This correction reduces the overall scatter between the semi-analytical model and the simulations by $26\,\%$.

Using the nDGP HMF within a Bayesian analysis framework for the SPT cluster sample allows us to constrain nDGP gravity with the abundance of massive galaxy clusters. We find that SPT clusters alone cannot place a meaningful constraint on the nDGP gravity parameter \nDGP. 
This is due to the degeneracies between \nDGP\ and the two standard cosmological parameters $\Omega_{\rm m}$, $\sigma_8$. 
We conclude that an external probe is needed to break degeneracies and obtain a competitive constraint on the nDGP modified gravity model with the SPT clusters. 
Moreover, all standard cosmological parameters and the scaling relation parameters are consistent with the $\nu\Lambda$CDM analysis of Ref.~\citepalias{Bocquet24II}, as expected since our analysis does not show a statistically significant hint for the nDGP model. 

We use primary CMB data from Planck PR4 \cite{Tristram23}, which benefits from improved systematic modeling over previous releases, particularly with respect to the CMB lensing anomaly which is degenerate with the modified gravity parameter \cite{Calabrese08,Renzi13,Mokeddem23,Planck_MG15,Planck2020,Specogna24}. Using Planck PR4 data alone, we obtain an upper bound of $\nDGP < 1.62$ at 95\,\% credibility.

Combining Planck PR4 with the SPT cluster data tightens the constraint to $\nDGP < 1.41$ (95\% credible level). This result is competitive with the currently most stringent constraint from Ref.~\cite{Barreira16}. We also find that this combination improves the constraint on $\sigma_8$, while the constraint on $\Omega_{\rm m}$ is dominated by Planck (see Fig.~\ref{fig:results_combined} and Tab.~\ref{tab:results}).

Upcoming galaxy cluster surveys such as SPT-3G \cite{Benson14} or the Simons Observatory \cite{Ade19} will provide much larger cluster samples.
Paired with next-generation weak-lensing data from Euclid \cite{Laureijs11} and the Legacy Survey of Space and Time (LSST) \cite{lsst09sciencebook} at the Vera C.\ Rubin Observatory, these cluster samples will lead to tight constraints on standard cosmology and potential extensions to gravity \cite{Vogt24,Mazoun24}. 
Another avenue to tighten the constraints on modified gravity theories would be to supplement the Planck primary CMB anisotropy data we use here with the ground-based data from SPT and the Atacama Space Telescope (ACT) that probe the CMB fluctuations on smaller scales \citep{Camphuis25,Louis25}.

\begin{acknowledgments}
This research was supported by 1) the Excellence Cluster ORIGINS, which is funded by the Deutsche Forschungsgemeinschaft (DFG, German Research Foundation) under Germany's Excellence Strategy - EXC-2094-390783311, 
by 2) the Max Planck Society Faculty Fellowship program at MPE, and by 3) the Ludwig-Maximilians-Universit\"at in Munich.

BL is funded by UK STFC through Consolidate Grant ST/X001075/1.

The South Pole Telescope program is supported by
the National Science Foundation (NSF) through awards
OPP-1852617 and OPP-2332483. Partial support is
also provided by the Kavli Institute of Cosmological
Physics at the University of Chicago. Argonne National Laboratory’s work was supported by the U.S. Department of Energy, Office of High Energy Physics, under contract DE-AC02-06CH11357.

Funding for the DES Projects has been provided by the U.S. Department of Energy, the U.S. National Science Foundation, the Ministry of Science and Education of Spain, 
the Science and Technology Facilities Council of the United Kingdom, the Higher Education Funding Council for England, the National Center for Supercomputing 
Applications at the University of Illinois at Urbana-Champaign, the Kavli Institute of Cosmological Physics at the University of Chicago, 
the Center for Cosmology and Astro-Particle Physics at the Ohio State University,
the Mitchell Institute for Fundamental Physics and Astronomy at Texas A\&M University, Financiadora de Estudos e Projetos, 
Funda{\c c}{\~a}o Carlos Chagas Filho de Amparo {\`a} Pesquisa do Estado do Rio de Janeiro, Conselho Nacional de Desenvolvimento Cient{\'i}fico e Tecnol{\'o}gico and 
the Minist{\'e}rio da Ci{\^e}ncia, Tecnologia e Inova{\c c}{\~a}o, the Deutsche Forschungsgemeinschaft and the Collaborating Institutions in the Dark Energy Survey. 

The Collaborating Institutions are Argonne National Laboratory, the University of California at Santa Cruz, the University of Cambridge, Centro de Investigaciones Energ{\'e}ticas, 
Medioambientales y Tecnol{\'o}gicas-Madrid, the University of Chicago, University College London, the DES-Brazil Consortium, the University of Edinburgh, 
the Eidgen{\"o}ssische Technische Hochschule (ETH) Z{\"u}rich, 
Fermi National Accelerator Laboratory, the University of Illinois at Urbana-Champaign, the Institut de Ci{\`e}ncies de l'Espai (IEEC/CSIC), 
the Institut de F{\'i}sica d'Altes Energies, Lawrence Berkeley National Laboratory, the Ludwig-Maximilians-Universit{\"a}t M{\"u}nchen and the associated Excellence Cluster Origins, 
the University of Michigan, NSF's NOIRLab, the University of Nottingham, The Ohio State University, the University of Pennsylvania, the University of Portsmouth, 
SLAC National Accelerator Laboratory, Stanford University, the University of Sussex, Texas A\&M University, and the OzDES Membership Consortium.

Based in part on observations at Cerro Tololo Inter-American Observatory at NSF's NOIRLab (NOIRLab Prop. ID 2012B-0001; PI: J. Frieman), which is managed by the Association of Universities for Research in Astronomy (AURA) under a cooperative agreement with the National Science Foundation.

The DES data management system is supported by the National Science Foundation under Grant Numbers AST-1138766 and AST-1536171.
The DES participants from Spanish institutions are partially supported by MICINN under grants ESP2017-89838, PGC2018-094773, PGC2018-102021, SEV-2016-0588, SEV-2016-0597, and MDM-2015-0509, some of which include ERDF funds from the European Union. IFAE is partially funded by the CERCA program of the Generalitat de Catalunya.
Research leading to these results has received funding from the European Research Council under the European Union's Seventh Framework Program (FP7/2007-2013) including ERC grant agreements 240672, 291329, and 306478.
We acknowledge support from the Brazilian Instituto Nacional de Ci\^encia e Tecnologia (INCT) do e-Universo (CNPq grant 465376/2014-2).

This work is based on observations made with the NASA/ESA {\it Hubble Space Telescope}, using imaging data from the SPT follow-up GO programs 12246 (PI: C.~Stubbs), 12477 (PI: F.~W.~High), 13412 (PI: T.~Schrabback), 14252 (PI: V.~Strazzullo), 14352 (PI: J.~Hlavacek-Larrondo), and 14677 (PI: T.~Schrabback).
STScI is operated by the Association of Universities for Research in Astronomy, Inc. under NASA contract NAS 5-26555.
It is also based on observations made with ESO Telescopes at the La Silla Paranal Observatory under programs 086.A-0741 (PI: Bazin), 088.A-0796 (PI: Bazin), 088.A-0889 (PI: Mohr), 089.A-0824 (PI: Mohr), 0100.A-0204 (PI: Schrabback), 0100.A-0217 (PI: Hern\'andez-Mart\'in), 0101.A-0694 (PI: Zohren), and 0102.A-0189 (PI: Zohren).
It is also based on observations obtained at the Gemini Observatory, which is operated by the Association of Universities for Research in Astronomy, Inc., under a cooperative agreement with the NSF on behalf of the Gemini partnership: the National Science Foundation (United States), National Research Council (Canada), CONICYT (Chile), Ministerio de Ciencia, Tecnolog\'{i}a e Innovaci\'{o}n Productiva (Argentina), Minist\'{e}rio da Ci\^{e}ncia, Tecnologia e Inova\c{c}\~{a}o (Brazil), and Korea Astronomy and Space Science Institute (Republic of Korea), under programs 2014B-0338 and	2016B-0176 (PI: B.~Benson).

This manuscript has been authored by Fermi Research Alliance, LLC under Contract No. DE-AC02-07CH11359 with the U.S. Department of Energy, Office of Science, Office of High Energy Physics.

\end{acknowledgments}

\appendix

\section{\label{app:Planck_MG} Planck and modified gravity}
\begin{figure}%
        \centering
        \subfigure{\includegraphics[width=.49\linewidth]{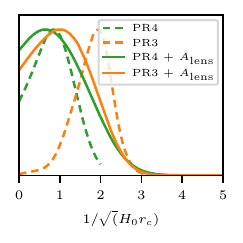} }%
        \subfigure{\includegraphics[width=.49\linewidth]{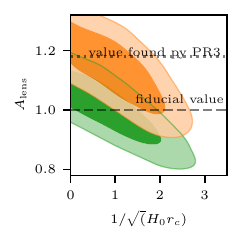} }%
        \caption{\textit{Left}: 1D posterior distribution for the nDGP parameter \nDGP\ for the different Planck datasets with and without varying the lensing parameter $A_{\rm lens}$ in solid and dashed lines, respectively. Planck PR3 indicated a $3.7\,\sigma$ detection on the nDGP model. This detection goes away if $A_{\rm lens}$ is varied as well as with the Planck PR4 analysis pipeline. \textit{Right:} Posterior distribution for \nDGP and $A_{\rm lens}$ for Planck PR3 and Planck PR4. Both datasets show a negative correlation between the two parameters, as both have a smoothing effect on the high $ l$ CMB power spectrum. For reference, the black dashed lines show the fiducial value of $A_{\rm lens}$ and the black dotted line shows the recovered value of the Planck PR3 analysis.}%
        \label{fig:A_lens}%
    \end{figure}

As mentioned in Sec.~\ref{subsec:likelihood_Plancl}, the constraint on nDGP gravity from Planck primary CMB depends on the choice of the Planck dataset and analysis pipeline. Earlier Planck analysis, such as Planck PR3, found a $\sim 2.1\,\sigma$ detection of scale-independent modified gravity away from $\Lambda$CDM \cite{Planck2020}. In this work, a different scale-independent modified gravity parametrization is used and we include massive neutrinos in our analysis. We re-analyzed the Planck PR3 data with an nDGP gravity model and found similar results with an even higher detection of the nDGP parameters, which is $3.7\,\sigma$ away from $\nu\Lambda$CDM as seen in Fig.~\ref{fig:A_lens} in blue.
This detection is related to the lensing anomaly found in the earlier Planck analyses \cite{Planck_MG15,Planck2020}.
The lensing anomaly reflects the systematic effect that the Planck data prefer a larger lensing-induced smoothing of the CMB power spectrum at high $l$ than predicted from a GR cosmology. To investigate this tension, an artificial lensing amplitude parameter $A_{\rm lens}$ was introduced, which allows for more smoothing, and the theoretical prediction in $\Lambda$CDM is $A_{\rm lens} = 1$ \cite{Calabrese08}. In Planck PR3 the data preferred $A_{\rm lens} > 1$ at the $3\,\sigma$ level in the $\Lambda$CDM analysis, leading to a tension between the data and the standard cosmological model. 

Because modified gravity models enhance lensing and thus increase lensing-induced smoothing, the effect on the CMB power spectrum is the same as having $A_{\rm lens} > 1$ in $\Lambda$CDM and $\nu\Lambda$CDM cosmologies. Therefore, earlier Planck analyses preferred modified gravity parameters away from the GR limit. Suppose $A_{\rm lens}$ is a free parameter in the analysis with a modified gravity model. In that case, the constraints on the modified gravity parameters are consistent with GR, because one accounts for the correlation between the two parameters \cite{Planck2020,Ishak24,Specogna24}. In the case of nDGP gravity, with the Planck PR3 data and pipeline we find a result consistent with $\nu\Lambda$CDM at the $1.7\,\sigma$ level. The shift in the posterior with and without varying $A_{\rm lens}$ can be seen in Fig.~\ref{fig:A_lens} on the left side for Planck PR3 in solid and dashed orange lines, respectively. 

The recent Planck PR4 analysis resolved the lensing anomaly, where $A_{\rm lens} = 1.039\pm0.052$ was reported \cite{Tristram23}. Therefore, analyzing scale-independent modified gravity models with Planck PR4 data and pipeline leads to consistent results with GR without varying the lensing parameter $A_{\rm lens}$ \cite{Ishak24,Specogna24}. In this paper, we present the constraint from Planck PR4 on the nDGP model and report $\nDGP > 1.45$ at $95\,\%$ credibility. If we also vary $A_{\rm lens}$ we find a weaker constraint of $\nDGP > 2.28$ at $95\,\%$ credible interval as seen in Fig.~\ref{fig:A_lens} on the left side in green. The constraint is weaker because $A_{\rm lens}$ already accounts for smoothing the CMB power spectrum at high $l$. 

This section explains that modified gravity and the lensing parameter $A_{\rm lens}$ have the same effect on the high-$l$ CMB power spectrum. Therefore, we expect a degeneracy between the two parameters. The right side of Fig.~\ref{fig:A_lens} shows the 2-D posterior of \nDGP\ and $A_{\rm lens}$, and as expected, a negative correlation can be seen. This is related to the fact that $A_{\rm lens} > 1$ and $\nDGP > 0$ enhance both the smoothing of the CMB lensing, and thus when we account for the smoothing from nDGP, $A_{\rm lens}$ is reduced.

\section{\label{app:triangle_plot} Full posterior results}
\begin{figure*}
        \centering
        \includegraphics[width=\textwidth]{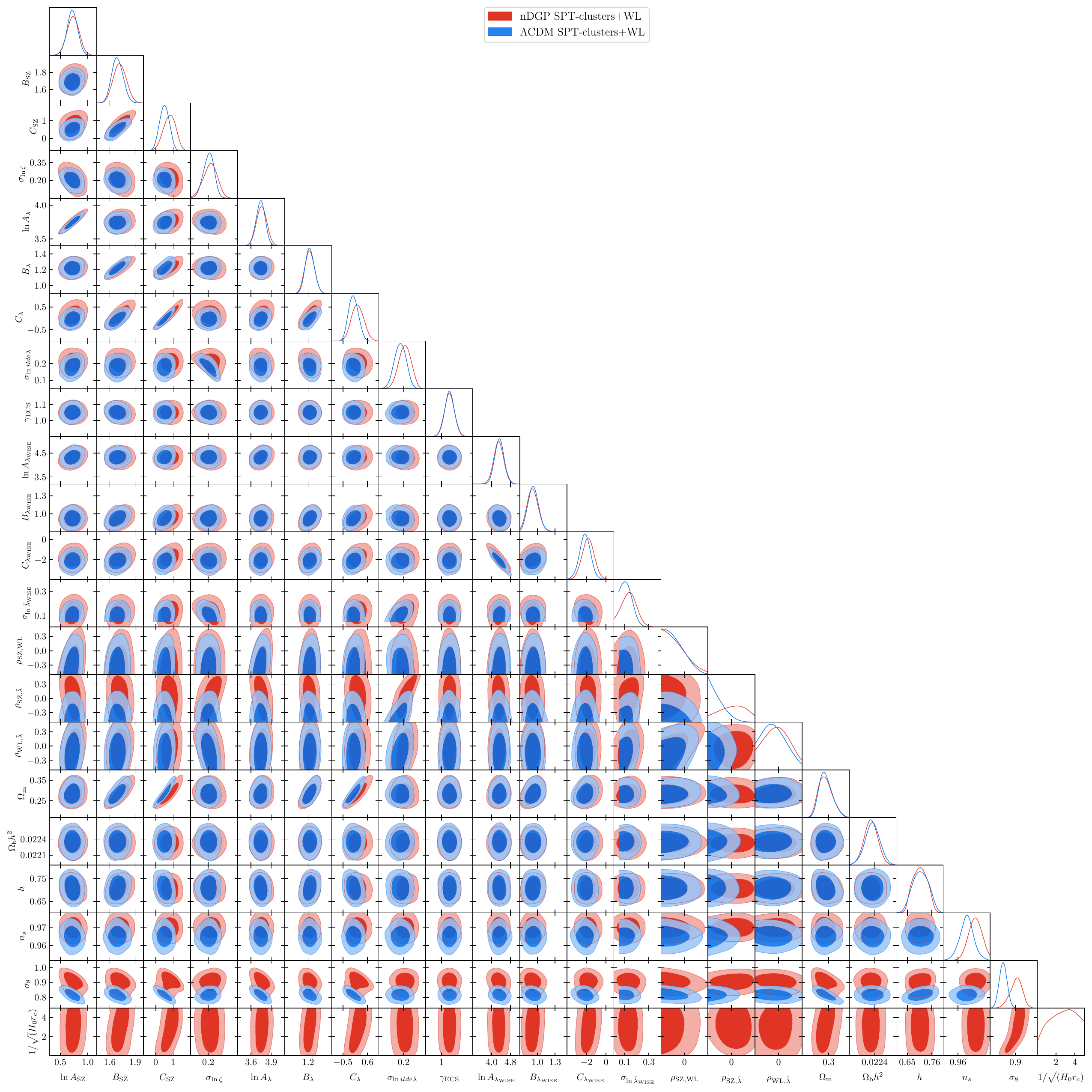}
        \vskip-0.25cm
        \caption{
        Posterior distribution for all (non prior dominated) parameters in the SPT-cluster$\times$WL analysis in the nDGP and $\nu\Lambda$CDM analyses ($68\,\%$ and $95\,\%$ credible regions). All scaling relation parameters are consistent between the analysis with wider posteriors from the nDGP analysis due to the extra degree of freedom. Only the constraint of the correlations between tSZE significance and richness changes due to a different scatter model applied to the mean richness--mass scaling relation. The shift in $n_s$ come from the different Planck prior applied to our analysis and the analysis from \citepalias{Bocquet24II}.}
        
        \label{fig:all_params}
    \end{figure*}

We present the posterior distribution for all parameters from our cluster analysis in Fig.~\ref{fig:all_params}. For comparison, we added the results from the $\nu\Lambda$CDM analysis. 

\bibliography{apssamp}

@string{june = {June}}

@article{Abbott20,
 adsurl = {https://ui.adsabs.harvard.edu/abs/2020PhRvD.102b3509A},
 archiveprefix = {arXiv},
 author = {{Abbott}, T.~M.~C. and {Aguena}, M. and {Alarcon}, A. and {Allam}, S. and {Allen}, S. and {Annis}, J. and {Avila}, S. and {Bacon}, D. and {Bechtol}, K. and {Bermeo}, A. and {Bernstein}, G.~M. and {Bertin}, E. and {Bhargava}, S. and {Bocquet}, S. and {Brooks}, D. and {Brout}, D. and {Buckley-Geer}, E. and {Burke}, D.~L. and {Carnero Rosell}, A. and {Carrasco Kind}, M. and {Carretero}, J. and {Castander}, F.~J. and {Cawthon}, R. and {Chang}, C. and {Chen}, X. and {Choi}, A. and {Costanzi}, M. and {Crocce}, M. and {da Costa}, L.~N. and {Davis}, T.~M. and {De Vicente}, J. and {DeRose}, J. and {Desai}, S. and {Diehl}, H.~T. and {Dietrich}, J.~P. and {Dodelson}, S. and {Doel}, P. and {Drlica-Wagner}, A. and {Eckert}, K. and {Eifler}, T.~F. and {Elvin-Poole}, J. and {Estrada}, J. and {Everett}, S. and {Evrard}, A.~E. and {`Farahi}, A. and {Ferrero}, I. and {Flaugher}, B. and {Fosalba}, P. and {Frieman}, J. and {Garc{\'\i}a-Bellido}, J. and {Gatti}, M. and {Gaztanaga}, E. and {Gerdes}, D.~W. and {Giannantonio}, T. and {Giles}, P. and {Grandis}, S. and {Gruen}, D. and {Gruendl}, R.~A. and {Gschwend}, J. and {Gutierrez}, G. and {Hartley}, W.~G. and {Hinton}, S.~R. and {Hollowood}, D.~L. and {Honscheid}, K. and {Hoyle}, B. and {Huterer}, D. and {James}, D.~J. and {Jarvis}, M. and {Jeltema}, T. and {Johnson}, M.~W.~G. and {Johnson}, M.~D. and {Kent}, S. and {Krause}, E. and {Kron}, R. and {Kuehn}, K. and {Kuropatkin}, N. and {Lahav}, O. and {Li}, T.~S. and {Lidman}, C. and {Lima}, M. and {Lin}, H. and {MacCrann}, N. and {Maia}, M.~A.~G. and {Mantz}, A. and {Marshall}, J.~L. and {Martini}, P. and {Mayers}, J. and {Melchior}, P. and {Mena-Fern{\'a}ndez}, J. and {Menanteau}, F. and {Miquel}, R. and {Mohr}, J.~J. and {Nichol}, R.~C. and {Nord}, B. and {Ogando}, R.~L.~C. and {Palmese}, A. and {Paz-Chinch{\'o}n}, F. and {Plazas}, A.~A. and {Prat}, J. and {Rau}, M.~M. and {Romer}, A.~K. and {Roodman}, A. and {Rooney}, P. and {Rozo}, E. and {Rykoff}, E.~S. and {Sako}, M. and {Samuroff}, S. and {S{\'a}nchez}, C. and {Sanchez}, E. and {Saro}, A. and {Scarpine}, V. and {Schubnell}, M. and {Scolnic}, D. and {Serrano}, S. and {Sevilla-Noarbe}, I. and {Sheldon}, E. and {Smith}, J. Allyn. and {Smith}, M. and {Suchyta}, E. and {Swanson}, M.~E.~C. and {Tarle}, G. and {Thomas}, D. and {To}, C. and {Troxel}, M.~A. and {Tucker}, D.~L. and {Varga}, T.~N. and {von der Linden}, A. and {Walker}, A.~R. and {Wechsler}, R.~H. and {Weller}, J. and {Wilkinson}, R.~D. and {Wu}, H. and {Yanny}, B. and {Zhang}, Y. and {Zhang}, Z. and {Zuntz}, J. and {DES Collaboration}},
 doi = {10.1103/PhysRevD.102.023509},
 eid = {023509},
 eprint = {2002.11124},
 journal = {\prd},
 month = {July},
 number = {2},
 pages = {023509},
 primaryclass = {astro-ph.CO},
 title = {{Dark Energy Survey Year 1 Results: Cosmological constraints from cluster abundances and weak lensing}},
 volume = {102},
 year = {2020}
}

@article{Ade19,
 adsurl = {https://ui.adsabs.harvard.edu/abs/2019JCAP...02..056A},
 archiveprefix = {arXiv},
 author = {{Ade}, Peter and {Aguirre}, James and {Ahmed}, Zeeshan and {Aiola}, Simone and {Ali}, Aamir and {Alonso}, David and {Alvarez}, Marcelo A. and {Arnold}, Kam and {Ashton}, Peter and {Austermann}, Jason and {Awan}, Humna and {Baccigalupi}, Carlo and {Baildon}, Taylor and {Barron}, Darcy and {Battaglia}, Nick and {Battye}, Richard and {Baxter}, Eric and {Bazarko}, Andrew and {Beall}, James A. and {Bean}, Rachel and {Beck}, Dominic and {Beckman}, Shawn and {Beringue}, Benjamin and {Bianchini}, Federico and {Boada}, Steven and {Boettger}, David and {Bond}, J. Richard and {Borrill}, Julian and {Brown}, Michael L. and {Bruno}, Sarah Marie and {Bryan}, Sean and {Calabrese}, Erminia and {Calafut}, Victoria and {Calisse}, Paolo and {Carron}, Julien and {Challinor}, Anthony and {Chesmore}, Grace and {Chinone}, Yuji and {Chluba}, Jens and {Cho}, Hsiao-Mei Sherry and {Choi}, Steve and {Coppi}, Gabriele and {Cothard}, Nicholas F. and {Coughlin}, Kevin and {Crichton}, Devin and {Crowley}, Kevin D. and {Crowley}, Kevin T. and {Cukierman}, Ari and {D'Ewart}, John M. and {D{\"u}nner}, Rolando and {de Haan}, Tijmen and {Devlin}, Mark and {Dicker}, Simon and {Didier}, Joy and {Dobbs}, Matt and {Dober}, Bradley and {Duell}, Cody J. and {Duff}, Shannon and {Duivenvoorden}, Adri and {Dunkley}, Jo and {Dusatko}, John and {Errard}, Josquin and {Fabbian}, Giulio and {Feeney}, Stephen and {Ferraro}, Simone and {Flux{\`a}}, Pedro and {Freese}, Katherine and {Frisch}, Josef C. and {Frolov}, Andrei and {Fuller}, George and {Fuzia}, Brittany and {Galitzki}, Nicholas and {Gallardo}, Patricio A. and {Tomas Galvez Ghersi}, Jose and {Gao}, Jiansong and {Gawiser}, Eric and {Gerbino}, Martina and {Gluscevic}, Vera and {Goeckner-Wald}, Neil and {Golec}, Joseph and {Gordon}, Sam and {Gralla}, Megan and {Green}, Daniel and {Grigorian}, Arpi and {Groh}, John and {Groppi}, Chris and {Guan}, Yilun and {Gudmundsson}, Jon E. and {Han}, Dongwon and {Hargrave}, Peter and {Hasegawa}, Masaya and {Hasselfield}, Matthew and {Hattori}, Makoto and {Haynes}, Victor and {Hazumi}, Masashi and {He}, Yizhou and {Healy}, Erin and {Henderson}, Shawn W. and {Hervias-Caimapo}, Carlos and {Hill}, Charles A. and {Hill}, J. Colin and {Hilton}, Gene and {Hilton}, Matt and {Hincks}, Adam D. and {Hinshaw}, Gary and {Hlo{\v{z}}ek}, Ren{\'e}e and {Ho}, Shirley and {Ho}, Shuay-Pwu Patty and {Howe}, Logan and {Huang}, Zhiqi and {Hubmayr}, Johannes and {Huffenberger}, Kevin and {Hughes}, John P. and {Ijjas}, Anna and {Ikape}, Margaret and {Irwin}, Kent and {Jaffe}, Andrew H. and {Jain}, Bhuvnesh and {Jeong}, Oliver and {Kaneko}, Daisuke and {Karpel}, Ethan D. and {Katayama}, Nobuhiko and {Keating}, Brian and {Kernasovskiy}, Sarah S. and {Keskitalo}, Reijo and {Kisner}, Theodore and {Kiuchi}, Kenji and {Klein}, Jeff and {Knowles}, Kenda and {Koopman}, Brian and {Kosowsky}, Arthur and {Krachmalnicoff}, Nicoletta and {Kuenstner}, Stephen E. and {Kuo}, Chao-Lin and {Kusaka}, Akito and {Lashner}, Jacob and {Lee}, Adrian and {Lee}, Eunseong and {Leon}, David and {Leung}, Jason S. -Y. and {Lewis}, Antony and {Li}, Yaqiong and {Li}, Zack and {Limon}, Michele and {Linder}, Eric and {Lopez-Caraballo}, Carlos and {Louis}, Thibaut and {Lowry}, Lindsay and {Lungu}, Marius and {Madhavacheril}, Mathew and {Mak}, Daisy and {Maldonado}, Felipe and {Mani}, Hamdi and {Mates}, Ben and {Matsuda}, Frederick and {Maurin}, Lo{\"\i}c and {Mauskopf}, Phil and {May}, Andrew and {McCallum}, Nialh and {McKenney}, Chris and {McMahon}, Jeff and {Meerburg}, P. Daniel and {Meyers}, Joel and {Miller}, Amber and {Mirmelstein}, Mark and {Moodley}, Kavilan and {Munchmeyer}, Moritz and {Munson}, Charles and {Naess}, Sigurd and {Nati}, Federico and {Navaroli}, Martin and {Newburgh}, Laura and {Nguyen}, Ho Nam and {Niemack}, Michael and {Nishino}, Haruki and {Orlowski-Scherer}, John and {Page}, Lyman and {Partridge}, Bruce and {Peloton}, Julien and {Perrotta}, Francesca and {Piccirillo}, Lucio and {Pisano}, Giampaolo and {Poletti}, Davide and {Puddu}, Roberto and {Puglisi}, Giuseppe and {Raum}, Chris and {Reichardt}, Christian L. and {Remazeilles}, Mathieu and {Rephaeli}, Yoel and {Riechers}, Dominik and {Rojas}, Felipe and {Roy}, Anirban and {Sadeh}, Sharon and {Sakurai}, Yuki and {Salatino}, Maria and {Sathyanarayana Rao}, Mayuri and {Schaan}, Emmanuel and {Schmittfull}, Marcel and {Sehgal}, Neelima and {Seibert}, Joseph and {Seljak}, Uros and {Sherwin}, Blake and {Shimon}, Meir and {Sierra}, Carlos and {Sievers}, Jonathan and {Sikhosana}, Precious and {Silva-Feaver}, Maximiliano and {Simon}, Sara M. and {Sinclair}, Adrian and {Siritanasak}, Praween and {Smith}, Kendrick and {Smith}, Stephen R. and {Spergel}, David and {Staggs}, Suzanne T. and {Stein}, George and {Stevens}, Jason R. and {Stompor}, Radek and {Suzuki}, Aritoki and {Tajima}, Osamu and {Takakura}, Satoru and {Teply}, Grant and {Thomas}, Daniel B. and {Thorne}, Ben and {Thornton}, Robert and {Trac}, Hy and {Tsai}, Calvin and {Tucker}, Carole and {Ullom}, Joel and {Vagnozzi}, Sunny and {van Engelen}, Alexander and {Van Lanen}, Jeff and {Van Winkle}, Daniel D. and {Vavagiakis}, Eve M. and {Verg{\`e}s}, Clara and {Vissers}, Michael and {Wagoner}, Kasey and {Walker}, Samantha and {Ward}, Jon and {Westbrook}, Ben and {Whitehorn}, Nathan and {Williams}, Jason and {Williams}, Joel and {Wollack}, Edward J. and {Xu}, Zhilei and {Yu}, Byeonghee and {Yu}, Cyndia and {Zago}, Fernando and {Zhang}, Hezi and {Zhu}, Ningfeng and {Simons Observatory Collaboration}},
 doi = {10.1088/1475-7516/2019/02/056},
 eid = {056},
 eprint = {1808.07445},
 journal = {\jcap},
 month = {February},
 number = {2},
 pages = {056},
 primaryclass = {astro-ph.CO},
 title = {{The Simons Observatory: science goals and forecasts}},
 volume = {2019},
 year = {2019}
}

@article{Allen11,
 adsurl = {https://ui.adsabs.harvard.edu/abs/2011ARA&A..49..409A},
 archiveprefix = {arXiv},
 author = {{Allen}, Steven W. and {Evrard}, August E. and {Mantz}, Adam B.},
 doi = {10.1146/annurev-astro-081710-102514},
 eprint = {1103.4829},
 journal = {\araa},
 month = {September},
 number = {1},
 pages = {409-470},
 primaryclass = {astro-ph.CO},
 title = {{Cosmological Parameters from Observations of Galaxy Clusters}},
 volume = {49},
 year = {2011}
}

@article{Angulo12,
 adsurl = {https://ui.adsabs.harvard.edu/abs/2012MNRAS.426.2046A},
 archiveprefix = {arXiv},
 author = {{Angulo}, R.~E. and {Springel}, V. and {White}, S.~D.~M. and {Jenkins}, A. and {Baugh}, C.~M. and {Frenk}, C.~S.},
 doi = {10.1111/j.1365-2966.2012.21830.x},
 eprint = {1203.3216},
 journal = {\mnras},
 month = {November},
 number = {3},
 pages = {2046-2062},
 primaryclass = {astro-ph.CO},
 title = {{Scaling relations for galaxy clusters in the Millennium-XXL simulation}},
 volume = {426},
 year = {2012}
}

@article{Artis24,
 adsurl = {https://ui.adsabs.harvard.edu/abs/2024A&A...691A.301A},
 archiveprefix = {arXiv},
 author = {{Artis}, E. and {Ghirardini}, V. and {Bulbul}, E. and {Grandis}, S. and {Garrel}, C. and {Clerc}, N. and {Seppi}, R. and {Comparat}, J. and {Cataneo}, M. and {Bahar}, Y.~E. and {Balzer}, F. and {Chiu}, I. and {Gruen}, D. and {Kleinebreil}, F. and {Kluge}, M. and {Krippendorf}, S. and {Li}, X. and {Liu}, A. and {Merloni}, A. and {Miyatake}, H. and {Miyazaki}, S. and {Nandra}, K. and {Okabe}, N. and {Pacaud}, F. and {Predehl}, P. and {Ramos-Ceja}, M.~E. and {Reiprich}, T.~H. and {Sanders}, J.~S. and {Schrabback}, T. and {Zelmer}, S. and {Zhang}, X.},
 doi = {10.1051/0004-6361/202449587},
 eid = {A301},
 eprint = {2402.08459},
 journal = {\aap},
 month = {November},
 pages = {A301},
 primaryclass = {astro-ph.CO},
 title = {{The SRG/eROSITA All-Sky Survey: Constraints on f (R) gravity from cluster abundances}},
 volume = {691},
 year = {2024}
}

@article{Baker19,
 adsurl = {https://ui.adsabs.harvard.edu/abs/2019arXiv190803430B},
 archiveprefix = {arXiv},
 author = {{Baker}, Tessa and {Barreira}, Alexandre and {Desmond}, Harry and {Ferreira}, Pedro and {Jain}, Bhuvnesh and {Koyama}, Kazuya and {Li}, Baojiu and {Lombriser}, Lucas and {Nicola}, Andrina and {Sakstein}, Jeremy and {Schmidt}, Fabian},
 doi = {10.48550/arXiv.1908.03430},
 eid = {arXiv:1908.03430},
 eprint = {1908.03430},
 journal = {arXiv e-prints},
 month = {August},
 pages = {arXiv:1908.03430},
 primaryclass = {astro-ph.CO},
 title = {{The Novel Probes Project -- Tests of Gravity on Astrophysical Scales}},
 year = {2019}
}

@article{Barreira16,
 adsurl = {https://ui.adsabs.harvard.edu/abs/2016PhRvD..94h4022B},
 archiveprefix = {arXiv},
 author = {{Barreira}, Alexandre and {S{\'a}nchez}, Ariel G. and {Schmidt}, Fabian},
 doi = {10.1103/PhysRevD.94.084022},
 eid = {084022},
 eprint = {1605.03965},
 journal = {\prd},
 month = {October},
 number = {8},
 pages = {084022},
 primaryclass = {astro-ph.CO},
 title = {{Validating estimates of the growth rate of structure with modified gravity simulations}},
 volume = {94},
 year = {2016}
}

@article{Bartelmann17,
 adsurl = {https://ui.adsabs.harvard.edu/abs/2017SchpJ..1232440B},
 archiveprefix = {arXiv},
 author = {{Bartelmann}, Matthias and {Maturi}, Matteo},
 doi = {10.4249/scholarpedia.32440},
 eprint = {1612.06535},
 journal = {Scholarpedia},
 month = {January},
 number = {1},
 pages = {32440},
 primaryclass = {astro-ph.CO},
 title = {{Weak gravitational lensing}},
 volume = {12},
 year = {2017}
}

@article{becker&kravtsov11,
 adsurl = {https://ui.adsabs.harvard.edu/abs/2011ApJ...740...25B},
 archiveprefix = {arXiv},
 author = {{Becker}, Matthew R. and {Kravtsov}, Andrey V.},
 doi = {10.1088/0004-637X/740/1/25},
 eid = {25},
 eprint = {1011.1681},
 journal = {\apj},
 month = {October},
 number = {1},
 pages = {25},
 primaryclass = {astro-ph.CO},
 title = {{On the Accuracy of Weak-lensing Cluster Mass Reconstructions}},
 volume = {740},
 year = {2011}
}

@article{Benson13,
 adsurl = {https://ui.adsabs.harvard.edu/abs/2013ApJ...763..147B},
 archiveprefix = {arXiv},
 author = {{Benson}, B.~A. and {de Haan}, T. and {Dudley}, J.~P. and {Reichardt}, C.~L. and {Aird}, K.~A. and {Andersson}, K. and {Armstrong}, R. and {Ashby}, M.~L.~N. and {Bautz}, M. and {Bayliss}, M. and {Bazin}, G. and {Bleem}, L.~E. and {Brodwin}, M. and {Carlstrom}, J.~E. and {Chang}, C.~L. and {Cho}, H.~M. and {Clocchiatti}, A. and {Crawford}, T.~M. and {Crites}, A.~T. and {Desai}, S. and {Dobbs}, M.~A. and {Foley}, R.~J. and {Forman}, W.~R. and {George}, E.~M. and {Gladders}, M.~D. and {Gonzalez}, A.~H. and {Halverson}, N.~W. and {Harrington}, N. and {High}, F.~W. and {Holder}, G.~P. and {Holzapfel}, W.~L. and {Hoover}, S. and {Hrubes}, J.~D. and {Jones}, C. and {Joy}, M. and {Keisler}, R. and {Knox}, L. and {Lee}, A.~T. and {Leitch}, E.~M. and {Liu}, J. and {Lueker}, M. and {Luong-Van}, D. and {Mantz}, A. and {Marrone}, D.~P. and {McDonald}, M. and {McMahon}, J.~J. and {Mehl}, J. and {Meyer}, S.~S. and {Mocanu}, L. and {Mohr}, J.~J. and {Montroy}, T.~E. and {Murray}, S.~S. and {Natoli}, T. and {Padin}, S. and {Plagge}, T. and {Pryke}, C. and {Rest}, A. and {Ruel}, J. and {Ruhl}, J.~E. and {Saliwanchik}, B.~R. and {Saro}, A. and {Sayre}, J.~T. and {Schaffer}, K.~K. and {Shaw}, L. and {Shirokoff}, E. and {Song}, J. and {Spieler}, H.~G. and {Stalder}, B. and {Staniszewski}, Z. and {Stark}, A.~A. and {Story}, K. and {Stubbs}, C.~W. and {Suhada}, R. and {van Engelen}, A. and {Vanderlinde}, K. and {Vieira}, J.~D. and {Vikhlinin}, A. and {Williamson}, R. and {Zahn}, O. and {Zenteno}, A.},
 doi = {10.1088/0004-637X/763/2/147},
 eid = {147},
 eprint = {1112.5435},
 journal = {\apj},
 month = {February},
 number = {2},
 pages = {147},
 primaryclass = {astro-ph.CO},
 title = {{Cosmological Constraints from Sunyaev-Zel'dovich-selected Clusters with X-Ray Observations in the First 178 deg$^{2}$ of the South Pole Telescope Survey}},
 volume = {763},
 year = {2013}
}

@article{Benson14,
 adsurl = {https://ui.adsabs.harvard.edu/abs/2014SPIE.9153E..1PB},
 archiveprefix = {arXiv},
 author = {{Benson}, B.~A. and {Ade}, P.~A.~R. and {Ahmed}, Z. and {Allen}, S.~W. and {Arnold}, K. and {Austermann}, J.~E. and {Bender}, A.~N. and {Bleem}, L.~E. and {Carlstrom}, J.~E. and {Chang}, C.~L. and {Cho}, H.~M. and {Cliche}, J.~F. and {Crawford}, T.~M. and {Cukierman}, A. and {de Haan}, T. and {Dobbs}, M.~A. and {Dutcher}, D. and {Everett}, W. and {Gilbert}, A. and {Halverson}, N.~W. and {Hanson}, D. and {Harrington}, N.~L. and {Hattori}, K. and {Henning}, J.~W. and {Hilton}, G.~C. and {Holder}, G.~P. and {Holzapfel}, W.~L. and {Irwin}, K.~D. and {Keisler}, R. and {Knox}, L. and {Kubik}, D. and {Kuo}, C.~L. and {Lee}, A.~T. and {Leitch}, E.~M. and {Li}, D. and {McDonald}, M. and {Meyer}, S.~S. and {Montgomery}, J. and {Myers}, M. and {Natoli}, T. and {Nguyen}, H. and {Novosad}, V. and {Padin}, S. and {Pan}, Z. and {Pearson}, J. and {Reichardt}, C. and {Ruhl}, J.~E. and {Saliwanchik}, B.~R. and {Simard}, G. and {Smecher}, G. and {Sayre}, J.~T. and {Shirokoff}, E. and {Stark}, A.~A. and {Story}, K. and {Suzuki}, A. and {Thompson}, K.~L. and {Tucker}, C. and {Vanderlinde}, K. and {Vieira}, J.~D. and {Vikhlinin}, A. and {Wang}, G. and {Yefremenko}, V. and {Yoon}, K.~W.},
 booktitle = {Millimeter, Submillimeter, and Far-Infrared Detectors and Instrumentation for Astronomy VII},
 doi = {10.1117/12.2057305},
 editor = {{Holland}, Wayne S. and {Zmuidzinas}, Jonas},
 eid = {91531P},
 eprint = {1407.2973},
 month = {July},
 pages = {91531P},
 primaryclass = {astro-ph.IM},
 series = {Society of Photo-Optical Instrumentation Engineers (SPIE) Conference Series},
 title = {{SPT-3G: a next-generation cosmic microwave background polarization experiment on the South Pole telescope}},
 volume = {9153},
 year = {2014}
}

@article{Bleem15,
 adsurl = {https://ui.adsabs.harvard.edu/abs/2015ApJS..216...27B},
 archiveprefix = {arXiv},
 author = {{Bleem}, L.~E. and {Stalder}, B. and {de Haan}, T. and {Aird}, K.~A. and {Allen}, S.~W. and {Applegate}, D.~E. and {Ashby}, M.~L.~N. and {Bautz}, M. and {Bayliss}, M. and {Benson}, B.~A. and {Bocquet}, S. and {Brodwin}, M. and {Carlstrom}, J.~E. and {Chang}, C.~L. and {Chiu}, I. and {Cho}, H.~M. and {Clocchiatti}, A. and {Crawford}, T.~M. and {Crites}, A.~T. and {Desai}, S. and {Dietrich}, J.~P. and {Dobbs}, M.~A. and {Foley}, R.~J. and {Forman}, W.~R. and {George}, E.~M. and {Gladders}, M.~D. and {Gonzalez}, A.~H. and {Halverson}, N.~W. and {Hennig}, C. and {Hoekstra}, H. and {Holder}, G.~P. and {Holzapfel}, W.~L. and {Hrubes}, J.~D. and {Jones}, C. and {Keisler}, R. and {Knox}, L. and {Lee}, A.~T. and {Leitch}, E.~M. and {Liu}, J. and {Lueker}, M. and {Luong-Van}, D. and {Mantz}, A. and {Marrone}, D.~P. and {McDonald}, M. and {McMahon}, J.~J. and {Meyer}, S.~S. and {Mocanu}, L. and {Mohr}, J.~J. and {Murray}, S.~S. and {Padin}, S. and {Pryke}, C. and {Reichardt}, C.~L. and {Rest}, A. and {Ruel}, J. and {Ruhl}, J.~E. and {Saliwanchik}, B.~R. and {Saro}, A. and {Sayre}, J.~T. and {Schaffer}, K.~K. and {Schrabback}, T. and {Shirokoff}, E. and {Song}, J. and {Spieler}, H.~G. and {Stanford}, S.~A. and {Staniszewski}, Z. and {Stark}, A.~A. and {Story}, K.~T. and {Stubbs}, C.~W. and {Vanderlinde}, K. and {Vieira}, J.~D. and {Vikhlinin}, A. and {Williamson}, R. and {Zahn}, O. and {Zenteno}, A.},
 doi = {10.1088/0067-0049/216/2/27},
 eid = {27},
 eprint = {1409.0850},
 journal = {\apjs},
 month = {February},
 number = {2},
 pages = {27},
 primaryclass = {astro-ph.CO},
 title = {{Galaxy Clusters Discovered via the Sunyaev-Zel'dovich Effect in the 2500-Square-Degree SPT-SZ Survey}},
 volume = {216},
 year = {2015}
}

@article{Bleem20,
 adsurl = {https://ui.adsabs.harvard.edu/abs/2020ApJS..247...25B},
 archiveprefix = {arXiv},
 author = {{Bleem}, L.~E. and {Bocquet}, S. and {Stalder}, B. and {Gladders}, M.~D. and {Ade}, P.~A.~R. and {Allen}, S.~W. and {Anderson}, A.~J. and {Annis}, J. and {Ashby}, M.~L.~N. and {Austermann}, J.~E. and {Avila}, S. and {Avva}, J.~S. and {Bayliss}, M. and {Beall}, J.~A. and {Bechtol}, K. and {Bender}, A.~N. and {Benson}, B.~A. and {Bertin}, E. and {Bianchini}, F. and {Blake}, C. and {Brodwin}, M. and {Brooks}, D. and {Buckley-Geer}, E. and {Burke}, D.~L. and {Carlstrom}, J.~E. and {Rosell}, A. Carnero and {Carrasco Kind}, M. and {Carretero}, J. and {Chang}, C.~L. and {Chiang}, H.~C. and {Citron}, R. and {Moran}, C. Corbett and {Costanzi}, M. and {Crawford}, T.~M. and {Crites}, A.~T. and {da Costa}, L.~N. and {de Haan}, T. and {De Vicente}, J. and {Desai}, S. and {Diehl}, H.~T. and {Dietrich}, J.~P. and {Dobbs}, M.~A. and {Eifler}, T.~F. and {Everett}, W. and {Flaugher}, B. and {Floyd}, B. and {Frieman}, J. and {Gallicchio}, J. and {Garc{\'\i}a-Bellido}, J. and {George}, E.~M. and {Gerdes}, D.~W. and {Gilbert}, A. and {Gruen}, D. and {Gruendl}, R.~A. and {Gschwend}, J. and {Gupta}, N. and {Gutierrez}, G. and {Halverson}, N.~W. and {Harrington}, N. and {Henning}, J.~W. and {Heymans}, C. and {Holder}, G.~P. and {Hollowood}, D.~L. and {Holzapfel}, W.~L. and {Honscheid}, K. and {Hrubes}, J.~D. and {Huang}, N. and {Hubmayr}, J. and {Irwin}, K.~D. and {James}, D.~J. and {Jeltema}, T. and {Joudaki}, S. and {Khullar}, G. and {Klein}, M. and {Knox}, L. and {Kuropatkin}, N. and {Lee}, A.~T. and {Li}, D. and {Lidman}, C. and {Lowitz}, A. and {MacCrann}, N. and {Mahler}, G. and {Maia}, M.~A.~G. and {Marshall}, J.~L. and {McDonald}, M. and {McMahon}, J.~J. and {Melchior}, P. and {Menanteau}, F. and {Meyer}, S.~S. and {Miquel}, R. and {Mocanu}, L.~M. and {Mohr}, J.~J. and {Montgomery}, J. and {Nadolski}, A. and {Natoli}, T. and {Nibarger}, J.~P. and {Noble}, G. and {Novosad}, V. and {Padin}, S. and {Palmese}, A. and {Parkinson}, D. and {Patil}, S. and {Paz-Chinch{\'o}n}, F. and {Plazas}, A.~A. and {Pryke}, C. and {Ramachandra}, N.~S. and {Reichardt}, C.~L. and {Remolina Gonz{\'a}lez}, J.~D. and {Romer}, A.~K. and {Roodman}, A. and {Ruhl}, J.~E. and {Rykoff}, E.~S. and {Saliwanchik}, B.~R. and {Sanchez}, E. and {Saro}, A. and {Sayre}, J.~T. and {Schaffer}, K.~K. and {Schrabback}, T. and {Serrano}, S. and {Sharon}, K. and {Sievers}, C. and {Smecher}, G. and {Smith}, M. and {Soares-Santos}, M. and {Stark}, A.~A. and {Story}, K.~T. and {Suchyta}, E. and {Tarle}, G. and {Tucker}, C. and {Vanderlinde}, K. and {Veach}, T. and {Vieira}, J.~D. and {Wang}, G. and {Weller}, J. and {Whitehorn}, N. and {Wu}, W.~L.~K. and {Yefremenko}, V. and {Zhang}, Y.},
 doi = {10.3847/1538-4365/ab6993},
 eid = {25},
 eprint = {1910.04121},
 journal = {\apjs},
 month = {March},
 number = {1},
 pages = {25},
 primaryclass = {astro-ph.CO},
 title = {{The SPTpol Extended Cluster Survey}},
 volume = {247},
 year = {2020}
}

@article{Bleem24,
 adsurl = {https://ui.adsabs.harvard.edu/abs/2024OJAp....7E..13B},
 archiveprefix = {arXiv},
 author = {{Bleem}, L.~E. and {Klein}, M. and {Abbot}, T.~M.~C. and {Ade}, P.~A.~R. and {Aguena}, M. and {Alves}, O. and {Anderson}, A.~J. and {Andrade-Oliveira}, F. and {Ansarinejad}, B. and {Archipley}, M. and {Ashby}, M.~L.~N. and {Austermann}, J.~E. and {Bacon}, D. and {Beall}, J.~A. and {Bender}, A.~N. and {Benson}, B.~A. and {Bianchini}, F. and {Bocquet}, S. and {Brooks}, D. and {Burke}, D.~L. and {Calzadilla}, M. and {Carlstrom}, J.~E. and {Carnero Rosell}, A. and {Carretero}, J. and {Chang}, C.~L. and {Chaubal}, P. and {Chiang}, H.~C. and {Chou}, T-L. and {Citron}, R. and {Corbett Moran}, C. and {Costanzi}, M. and {Constanzi}, M. and {Crawford}, T.~M. and {Crites}, A.~T. and {da Costa}, L.~N. and {de Haan}, T. and {De Vicente}, J. and {Desai}, S. and {Dobbs}, M.~A. and {Doel}, P. and {Everett}, W. and {Ferrero}, I. and {Flaugher}, B. and {Floyd}, B. and {Friedel}, D. and {Frieman}, J. and {Gallicchio}, J. and {Garc'ia-Bellido}, J. and {Gatti}, M. and {George}, E.~M. and {Giannini}, G. and {Grandis}, S. and {Gruen}, D. and {Gruendl}, R.~A. and {Gupta}, N. and {Gutierrez}, G. and {Halverson}, N.~W. and {Hinton}, S.~R. and {Hinton}, S.~R. and {Holder}, G.~P. and {Hollowood}, D.~L. and {Holzapfel}, W.~L. and {Honscheid}, K. and {Hrubes}, J.~D. and {Huang}, N. and {Hubmayr}, J. and {Irwin}, K.~D. and {Mena-Fern{\'a}ndez}, J. and {James}, D.~J. and {K{\'e}ruzor{\'e}}, F. and {Knox}, L. and {Kuehn}, K. and {Lahav}, O. and {Lee}, A.~T. and {Lee}, S. and {Li}, D. and {Lowitz}, A. and {Marshal}, J.~L. and {McDonald}, M. and {McMahon}, J.~J. and {Menanteau}, F. and {Meyer}, S.~S. and {Miquel}, R. and {Mohr}, J.~J. and {Montgomery}, J. and {Myles}, J. and {Natoli}, T. and {Nibarger}, J.~P. and {Noble}, G.~I. and {Novosad}, V. and {Ogando}, R.~L.~C. and {Padin}, S. and {Patil}, S. and {Pereira}, M.~E.~S. and {Pieres}, A. and {Plazas Malag'on}, A.~A. and {Pryke}, C. and {Reichardt}, C.~L. and {Rodr'iguez-Monroy}, M. and {Romer}, A.~K. and {Ruhl}, J.~E. and {Saliwanchik}, B.~R. and {Salvati}, L. and {Sanchez}, E. and {Saro}, A. and {Schaffer}, K.~K. and {Schrabback}, T. and {Sevilla-Noarbe}, I. and {Sievers}, C. and {Smecher}, G. and {Smith}, M. and {Somboonpanyakul}, T. and {Stalder}, B. and {Stark}, A.~A. and {Suchyta}, E. and {Swanson}, M.~E.~C. and {Tarle}, G. and {To}, C. and {Tucker}, C. and {Veach}, T. and {Vieira}, J.~D. and {Vincenzi}, M. and {Wang}, G. and {Weller}, J. and {Whitehorn}, N. and {Wiseman}, P. and {Wu}, W.~L.~K. and {Yefremenko}, V. and {Zebrowski}, J.~A. and {Zhang}, Y.},
 doi = {10.21105/astro.2311.07512},
 eid = {13},
 eprint = {2311.07512},
 journal = {The Open Journal of Astrophysics},
 month = {February},
 pages = {13},
 primaryclass = {astro-ph.CO},
 title = {{Galaxy Clusters Discovered via the Thermal Sunyaev-Zel'dovich Effect in the 500-square-degree SPTpol Survey}},
 volume = {7},
 year = {2024}
}

@article{Bocquet15,
 adsurl = {https://ui.adsabs.harvard.edu/abs/2015ApJ...799..214B},
 archiveprefix = {arXiv},
 author = {{Bocquet}, S. and {Saro}, A. and {Mohr}, J.~J. and {Aird}, K.~A. and {Ashby}, M.~L.~N. and {Bautz}, M. and {Bayliss}, M. and {Bazin}, G. and {Benson}, B.~A. and {Bleem}, L.~E. and {Brodwin}, M. and {Carlstrom}, J.~E. and {Chang}, C.~L. and {Chiu}, I. and {Cho}, H.~M. and {Clocchiatti}, A. and {Crawford}, T.~M. and {Crites}, A.~T. and {Desai}, S. and {de Haan}, T. and {Dietrich}, J.~P. and {Dobbs}, M.~A. and {Foley}, R.~J. and {Forman}, W.~R. and {Gangkofner}, D. and {George}, E.~M. and {Gladders}, M.~D. and {Gonzalez}, A.~H. and {Halverson}, N.~W. and {Hennig}, C. and {Hlavacek-Larrondo}, J. and {Holder}, G.~P. and {Holzapfel}, W.~L. and {Hrubes}, J.~D. and {Jones}, C. and {Keisler}, R. and {Knox}, L. and {Lee}, A.~T. and {Leitch}, E.~M. and {Liu}, J. and {Lueker}, M. and {Luong-Van}, D. and {Marrone}, D.~P. and {McDonald}, M. and {McMahon}, J.~J. and {Meyer}, S.~S. and {Mocanu}, L. and {Murray}, S.~S. and {Padin}, S. and {Pryke}, C. and {Reichardt}, C.~L. and {Rest}, A. and {Ruel}, J. and {Ruhl}, J.~E. and {Saliwanchik}, B.~R. and {Sayre}, J.~T. and {Schaffer}, K.~K. and {Shirokoff}, E. and {Spieler}, H.~G. and {Stalder}, B. and {Stanford}, S.~A. and {Staniszewski}, Z. and {Stark}, A.~A. and {Story}, K. and {Stubbs}, C.~W. and {Vanderlinde}, K. and {Vieira}, J.~D. and {Vikhlinin}, A. and {Williamson}, R. and {Zahn}, O. and {Zenteno}, A.},
 doi = {10.1088/0004-637X/799/2/214},
 eid = {214},
 eprint = {1407.2942},
 journal = {\apj},
 month = {February},
 number = {2},
 pages = {214},
 primaryclass = {astro-ph.CO},
 title = {{Mass Calibration and Cosmological Analysis of the SPT-SZ Galaxy Cluster Sample Using Velocity Dispersion {\ensuremath{\sigma}}$_{ v }$ and X-Ray Y $_{X}$ Measurements}},
 volume = {799},
 year = {2015}
}

@article{Bocquet19,
 adsurl = {https://ui.adsabs.harvard.edu/abs/2019ApJ...878...55B},
 archiveprefix = {arXiv},
 author = {{Bocquet}, S. and {Dietrich}, J.~P. and {Schrabback}, T. and {Bleem}, L.~E. and {Klein}, M. and {Allen}, S.~W. and {Applegate}, D.~E. and {Ashby}, M.~L.~N. and {Bautz}, M. and {Bayliss}, M. and {Benson}, B.~A. and {Brodwin}, M. and {Bulbul}, E. and {Canning}, R.~E.~A. and {Capasso}, R. and {Carlstrom}, J.~E. and {Chang}, C.~L. and {Chiu}, I. and {Cho}, H. -M. and {Clocchiatti}, A. and {Crawford}, T.~M. and {Crites}, A.~T. and {de Haan}, T. and {Desai}, S. and {Dobbs}, M.~A. and {Foley}, R.~J. and {Forman}, W.~R. and {Garmire}, G.~P. and {George}, E.~M. and {Gladders}, M.~D. and {Gonzalez}, A.~H. and {Grandis}, S. and {Gupta}, N. and {Halverson}, N.~W. and {Hlavacek-Larrondo}, J. and {Hoekstra}, H. and {Holder}, G.~P. and {Holzapfel}, W.~L. and {Hou}, Z. and {Hrubes}, J.~D. and {Huang}, N. and {Jones}, C. and {Khullar}, G. and {Knox}, L. and {Kraft}, R. and {Lee}, A.~T. and {von der Linden}, A. and {Luong-Van}, D. and {Mantz}, A. and {Marrone}, D.~P. and {McDonald}, M. and {McMahon}, J.~J. and {Meyer}, S.~S. and {Mocanu}, L.~M. and {Mohr}, J.~J. and {Morris}, R.~G. and {Padin}, S. and {Patil}, S. and {Pryke}, C. and {Rapetti}, D. and {Reichardt}, C.~L. and {Rest}, A. and {Ruhl}, J.~E. and {Saliwanchik}, B.~R. and {Saro}, A. and {Sayre}, J.~T. and {Schaffer}, K.~K. and {Shirokoff}, E. and {Stalder}, B. and {Stanford}, S.~A. and {Staniszewski}, Z. and {Stark}, A.~A. and {Story}, K.~T. and {Strazzullo}, V. and {Stubbs}, C.~W. and {Vanderlinde}, K. and {Vieira}, J.~D. and {Vikhlinin}, A. and {Williamson}, R. and {Zenteno}, A.},
 doi = {10.3847/1538-4357/ab1f10},
 eid = {55},
 eprint = {1812.01679},
 journal = {\apj},
 month = {June},
 number = {1},
 pages = {55},
 primaryclass = {astro-ph.CO},
 title = {{Cluster Cosmology Constraints from the 2500 deg$^{2}$ SPT-SZ Survey: Inclusion of Weak Gravitational Lensing Data from Magellan and the Hubble Space Telescope}},
 volume = {878},
 year = {2019}
}

@article{Bocquet24Ia,
 adsurl = {https://ui.adsabs.harvard.edu/abs/2024PhRvD.110h3509B},
 archiveprefix = {arXiv},
 author = {{Bocquet}, S. and {Grandis}, S. and {Bleem}, L.~E. and {Klein}, M. and {Mohr}, J.~J. and {Aguena}, M. and {Alarcon}, A. and {Allam}, S. and {Allen}, S.~W. and {Alves}, O. and {Amon}, A. and {Ansarinejad}, B. and {Bacon}, D. and {Bayliss}, M. and {Bechtol}, K. and {Becker}, M.~R. and {Benson}, B.~A. and {Bernstein}, G.~M. and {Brodwin}, M. and {Brooks}, D. and {Campos}, A. and {Canning}, R.~E.~A. and {Carlstrom}, J.~E. and {Carnero Rosell}, A. and {Carrasco Kind}, M. and {Carretero}, J. and {Cawthon}, R. and {Chang}, C. and {Chen}, R. and {Choi}, A. and {Cordero}, J. and {Costanzi}, M. and {da Costa}, L.~N. and {Pereira}, M.~E.~S. and {Davis}, C. and {DeRose}, J. and {Desai}, S. and {de Haan}, T. and {De Vicente}, J. and {Diehl}, H.~T. and {Dodelson}, S. and {Doel}, P. and {Doux}, C. and {Drlica-Wagner}, A. and {Eckert}, K. and {Elvin-Poole}, J. and {Everett}, S. and {Ferrero}, I. and {Fert{\'e}}, A. and {Flores}, A.~M. and {Frieman}, J. and {Garc{\'\i}a-Bellido}, J. and {Gatti}, M. and {Giannini}, G. and {Gladders}, M.~D. and {Gruen}, D. and {Gruendl}, R.~A. and {Harrison}, I. and {Hartley}, W.~G. and {Herner}, K. and {Hinton}, S.~R. and {Hollowood}, D.~L. and {Holzapfel}, W.~L. and {Honscheid}, K. and {Huang}, N. and {Huff}, E.~M. and {James}, D.~J. and {Jarvis}, M. and {Khullar}, G. and {Kim}, K. and {Kraft}, R. and {Kuehn}, K. and {Kuropatkin}, N. and {K{\'e}ruzor{\'e}}, F. and {Lee}, S. and {Leget}, P. -F. and {MacCrann}, N. and {Mahler}, G. and {Mantz}, A. and {Marshall}, J.~L. and {McCullough}, J. and {McDonald}, M. and {Mena-Fern{\'a}ndez}, J. and {Miquel}, R. and {Myles}, J. and {Navarro-Alsina}, A. and {Ogando}, R.~L.~C. and {Palmese}, A. and {Pandey}, S. and {Pieres}, A. and {Plazas Malag{\'o}n}, A.~A. and {Prat}, J. and {Raveri}, M. and {Reichardt}, C.~L. and {Roberson}, J. and {Rollins}, R.~P. and {Romer}, A.~K. and {Romero}, C. and {Roodman}, A. and {Ross}, A.~J. and {Rykoff}, E.~S. and {Salvati}, L. and {S{\'a}nchez}, C. and {Sanchez}, E. and {Sanchez Cid}, D. and {Saro}, A. and {Schrabback}, T. and {Schubnell}, M. and {Secco}, L.~F. and {Sevilla-Noarbe}, I. and {Sharon}, K. and {Sheldon}, E. and {Shin}, T. and {Smith}, M. and {Somboonpanyakul}, T. and {Stalder}, B. and {Stark}, A.~A. and {Strazzullo}, V. and {Suchyta}, E. and {Swanson}, M.~E.~C. and {Tarle}, G. and {To}, C. and {Troxel}, M.~A. and {Tutusaus}, I. and {Varga}, T.~N. and {von der Linden}, A. and {Weaverdyck}, N. and {Weller}, J. and {Wiseman}, P. and {Yanny}, B. and {Yin}, B. and {Young}, M. and {Zhang}, Y. and {Zuntz}, J. and {(The DES} and {SPT Collaborations)}},
 doi = {10.1103/PhysRevD.110.083509},
 eid = {083509},
 eprint = {2310.12213},
 journal = {\prd},
 month = {October},
 number = {8},
 pages = {083509},
 primaryclass = {astro-ph.CO},
 title = {{SPT clusters with DES and HST weak lensing. I. Cluster lensing and Bayesian population modeling of multiwavelength cluster datasets}},
 volume = {110},
 year = {2024}
}

@article{Bocquet24II,
 adsurl = {https://ui.adsabs.harvard.edu/abs/2024PhRvD.110h3510B},
 archiveprefix = {arXiv},
 author = {{Bocquet}, S. and {Grandis}, S. and {Bleem}, L.~E. and {Klein}, M. and {Mohr}, J.~J. and {Schrabback}, T. and {Abbott}, T.~M.~C. and {Ade}, P.~A.~R. and {Aguena}, M. and {Alarcon}, A. and {Allam}, S. and {Allen}, S.~W. and {Alves}, O. and {Amon}, A. and {Anderson}, A.~J. and {Annis}, J. and {Ansarinejad}, B. and {Austermann}, J.~E. and {Avila}, S. and {Bacon}, D. and {Bayliss}, M. and {Beall}, J.~A. and {Bechtol}, K. and {Becker}, M.~R. and {Bender}, A.~N. and {Benson}, B.~A. and {Bernstein}, G.~M. and {Bhargava}, S. and {Bianchini}, F. and {Brodwin}, M. and {Brooks}, D. and {Bryant}, L. and {Campos}, A. and {Canning}, R.~E.~A. and {Carlstrom}, J.~E. and {Carnero Rosell}, A. and {Carrasco Kind}, M. and {Carretero}, J. and {Castander}, F.~J. and {Cawthon}, R. and {Chang}, C.~L. and {Chang}, C. and {Chaubal}, P. and {Chen}, R. and {Chiang}, H.~C. and {Choi}, A. and {Chou}, T. -L. and {Citron}, R. and {Corbett Moran}, C. and {Cordero}, J. and {Costanzi}, M. and {Crawford}, T.~M. and {Crites}, A.~T. and {da Costa}, L.~N. and {Pereira}, M.~E.~S. and {Davis}, C. and {Davis}, T.~M. and {DeRose}, J. and {Desai}, S. and {de Haan}, T. and {Diehl}, H.~T. and {Dobbs}, M.~A. and {Dodelson}, S. and {Doux}, C. and {Drlica-Wagner}, A. and {Eckert}, K. and {Elvin-Poole}, J. and {Everett}, S. and {Everett}, W. and {Ferrero}, I. and {Fert{\'e}}, A. and {Flores}, A.~M. and {Frieman}, J. and {Gallicchio}, J. and {Garc{\'\i}a-Bellido}, J. and {Gatti}, M. and {George}, E.~M. and {Giannini}, G. and {Gladders}, M.~D. and {Gruen}, D. and {Gruendl}, R.~A. and {Gupta}, N. and {Gutierrez}, G. and {Halverson}, N.~W. and {Harrison}, I. and {Hartley}, W.~G. and {Herner}, K. and {Hinton}, S.~R. and {Holder}, G.~P. and {Hollowood}, D.~L. and {Holzapfel}, W.~L. and {Honscheid}, K. and {Hrubes}, J.~D. and {Huang}, N. and {Hubmayr}, J. and {Huff}, E.~M. and {Huterer}, D. and {Irwin}, K.~D. and {James}, D.~J. and {Jarvis}, M. and {Khullar}, G. and {Kim}, K. and {Knox}, L. and {Kraft}, R. and {Krause}, E. and {Kuehn}, K. and {Kuropatkin}, N. and {K{\'e}ruzor{\'e}}, F. and {Lahav}, O. and {Lee}, A.~T. and {Leget}, P. -F. and {Li}, D. and {Lin}, H. and {Lowitz}, A. and {MacCrann}, N. and {Mahler}, G. and {Mantz}, A. and {Marshall}, J.~L. and {McCullough}, J. and {McDonald}, M. and {McMahon}, J.~J. and {Mena-Fern{\'a}ndez}, J. and {Menanteau}, F. and {Meyer}, S.~S. and {Miquel}, R. and {Montgomery}, J. and {Myles}, J. and {Natoli}, T. and {Navarro-Alsina}, A. and {Nibarger}, J.~P. and {Noble}, G.~I. and {Novosad}, V. and {Ogando}, R.~L.~C. and {Omori}, Y. and {Padin}, S. and {Pandey}, S. and {Paschos}, P. and {Patil}, S. and {Pieres}, A. and {Plazas Malag{\'o}n}, A.~A. and {Porredon}, A. and {Prat}, J. and {Pryke}, C. and {Raveri}, M. and {Reichardt}, C.~L. and {Roberson}, J. and {Rollins}, R.~P. and {Romero}, C. and {Roodman}, A. and {Ruhl}, J.~E. and {Rykoff}, E.~S. and {Saliwanchik}, B.~R. and {Salvati}, L. and {S{\'a}nchez}, C. and {Sanchez}, E. and {Sanchez Cid}, D. and {Saro}, A. and {Schaffer}, K.~K. and {Secco}, L.~F. and {Sevilla-Noarbe}, I. and {Sharon}, K. and {Sheldon}, E. and {Shin}, T. and {Sievers}, C. and {Smecher}, G. and {Smith}, M. and {Somboonpanyakul}, T. and {Sommer}, M. and {Stalder}, B. and {Stark}, A.~A. and {Stephen}, J. and {Strazzullo}, V. and {Suchyta}, E. and {Tarle}, G. and {To}, C. and {Troxel}, M.~A. and {Tucker}, C. and {Tutusaus}, I. and {Varga}, T.~N. and {Veach}, T. and {Vieira}, J.~D. and {Vikhlinin}, A. and {von der Linden}, A. and {Wang}, G. and {Weaverdyck}, N. and {Weller}, J. and {Whitehorn}, N. and {Wu}, W.~L.~K. and {Yanny}, B. and {Yefremenko}, V. and {Yin}, B. and {Young}, M. and {Zebrowski}, J.~A. and {Zhang}, Y. and {Zohren}, H. and {Zuntz}, J. and {(SPT} and {DES Collaborations)}},
 doi = {10.1103/PhysRevD.110.083510},
 eid = {083510},
 eprint = {2401.02075},
 journal = {\prd},
 month = {October},
 number = {8},
 pages = {083510},
 primaryclass = {astro-ph.CO},
 title = {{SPT clusters with DES and HST weak lensing. II. Cosmological constraints from the abundance of massive halos}},
 volume = {110},
 year = {2024}
}

@article{Burrage18,
 adsurl = {https://ui.adsabs.harvard.edu/abs/2018LRR....21....1B},
 archiveprefix = {arXiv},
 author = {{Burrage}, Clare and {Sakstein}, Jeremy},
 doi = {10.1007/s41114-018-0011-x},
 eid = {1},
 eprint = {1709.09071},
 journal = {Living Reviews in Relativity},
 month = {March},
 number = {1},
 pages = {1},
 primaryclass = {astro-ph.CO},
 title = {{Tests of chameleon gravity}},
 volume = {21},
 year = {2018}
}

@article{Calabrese08,
 adsurl = {https://ui.adsabs.harvard.edu/abs/2008PhRvD..77l3531C},
 archiveprefix = {arXiv},
 author = {{Calabrese}, Erminia and {Slosar}, An{\v{z}}e and {Melchiorri}, Alessandro and {Smoot}, George F. and {Zahn}, Oliver},
 doi = {10.1103/PhysRevD.77.123531},
 eid = {123531},
 eprint = {0803.2309},
 journal = {\prd},
 month = {June},
 number = {12},
 pages = {123531},
 primaryclass = {astro-ph},
 title = {{Cosmic microwave weak lensing data as a test for the dark universe}},
 volume = {77},
 year = {2008}
}

@article{Camphuis25,
 adsurl = {https://ui.adsabs.harvard.edu/abs/2026PhRvD.113h3504C},
 archiveprefix = {arXiv},
 author = {{Camphuis}, E. and {Quan}, W. and {Balkenhol}, L. and {Khalife}, A.~R. and {Ge}, F. and {Guidi}, F. and {Huang}, N. and {Lynch}, G.~P. and {Omori}, Y. and {Trendafilova}, C. and {Anderson}, A.~J. and {Ansarinejad}, B. and {Archipley}, M. and {Barry}, P.~S. and {Benabed}, K. and {Bender}, A.~N. and {Benson}, B.~A. and {Bianchini}, F. and {Bleem}, L.~E. and {Bouchet}, F.~R. and {Bryant}, L. and {Campitiello}, M.~G. and {Carlstrom}, J.~E. and {Chang}, C.~L. and {Chaubal}, P. and {Chichura}, P.~M. and {Chokshi}, A. and {Chou}, T.-L. and {Coerver}, A. and {Crawford}, T.~M. and {Daley}, C. and {de Haan}, T. and {Dibert}, K.~R. and {Dobbs}, M.~A. and {Doohan}, M. and {Doussot}, A. and {Dutcher}, D. and {Everett}, W. and {Feng}, C. and {Ferguson}, K.~R. and {Fichman}, K. and {Foster}, A. and {Galli}, S. and {Gambrel}, A.~E. and {Gardner}, R.~W. and {Goeckner-Wald}, N. and {Gualtieri}, R. and {Guns}, S. and {Halverson}, N.~W. and {Hivon}, E. and {Holder}, G.~P. and {Holzapfel}, W.~L. and {Hood}, J.~C. and {Hryciuk}, A. and {K{\'e}ruzor{\'e}}, F. and {Knox}, L. and {Korman}, M. and {Kornoelje}, K. and {Kuo}, C.-L. and {Levy}, K. and {Lowitz}, A.~E. and {Lu}, C. and {Maniyar}, A. and {Martsen}, E.~S. and {Menanteau}, F. and {Millea}, M. and {Montgomery}, J. and {Nakato}, Y. and {Natoli}, T. and {Noble}, G.~I. and {Ouellette}, A. and {Pan}, Z. and {Paschos}, P. and {Phadke}, K.~A. and {Pollak}, A.~W. and {Prabhu}, K. and {Raghunathan}, S. and {Rahimi}, M. and {Rahlin}, A. and {Reichardt}, C.~L. and {Rouble}, M. and {Ruhl}, J.~E. and {Schiappucci}, E. and {Simpson}, A. and {Sobrin}, J.~A. and {Stark}, A.~A. and {Stephen}, J. and {Tandoi}, C. and {Thorne}, B. and {Umilta}, C. and {Vieira}, J.~D. and {Vitrier}, A. and {Wan}, Y. and {Whitehorn}, N. and {Wu}, W.~L.~K. and {Young}, M.~R. and {Zebrowski}, J.~A. and {SPT-3G Collaboration}},
 doi = {10.1103/7wt3-9v2y},
 eid = {083504},
 eprint = {2506.20707},
 journal = {\prd},
 month = {April},
 number = {8},
 pages = {083504},
 primaryclass = {astro-ph.CO},
 title = {{SPT-3G D1: CMB temperature and polarization power spectra and cosmology from 2019 and 2020 observations of the SPT-3G main field}},
 volume = {113},
 year = {2026}
}

@article{Carlstrom11,
 adsurl = {https://ui.adsabs.harvard.edu/abs/2011PASP..123..568C},
 archiveprefix = {arXiv},
 author = {{Carlstrom}, J.~E. and {Ade}, P.~A.~R. and {Aird}, K.~A. and {Benson}, B.~A. and {Bleem}, L.~E. and {Busetti}, S. and {Chang}, C.~L. and {Chauvin}, E. and {Cho}, H. -M. and {Crawford}, T.~M. and {Crites}, A.~T. and {Dobbs}, M.~A. and {Halverson}, N.~W. and {Heimsath}, S. and {Holzapfel}, W.~L. and {Hrubes}, J.~D. and {Joy}, M. and {Keisler}, R. and {Lanting}, T.~M. and {Lee}, A.~T. and {Leitch}, E.~M. and {Leong}, J. and {Lu}, W. and {Lueker}, M. and {Luong-Van}, D. and {McMahon}, J.~J. and {Mehl}, J. and {Meyer}, S.~S. and {Mohr}, J.~J. and {Montroy}, T.~E. and {Padin}, S. and {Plagge}, T. and {Pryke}, C. and {Ruhl}, J.~E. and {Schaffer}, K.~K. and {Schwan}, D. and {Shirokoff}, E. and {Spieler}, H.~G. and {Staniszewski}, Z. and {Stark}, A.~A. and {Tucker}, C. and {Vanderlinde}, K. and {Vieira}, J.~D. and {Williamson}, R.},
 doi = {10.1086/659879},
 eprint = {0907.4445},
 journal = {\pasp},
 month = {May},
 number = {903},
 pages = {568},
 primaryclass = {astro-ph.IM},
 title = {{The 10 Meter South Pole Telescope}},
 volume = {123},
 year = {2011}
}

@article{Cataneo14,
 adsurl = {https://ui.adsabs.harvard.edu/abs/2015PhRvD..92d4009C},
 archiveprefix = {arXiv},
 author = {{Cataneo}, Matteo and {Rapetti}, David and {Schmidt}, Fabian and {Mantz}, Adam B. and {Allen}, Steven W. and {Applegate}, Douglas E. and {Kelly}, Patrick L. and {von der Linden}, Anja and {Morris}, R. Glenn},
 doi = {10.1103/PhysRevD.92.044009},
 eid = {044009},
 eprint = {1412.0133},
 journal = {\prd},
 month = {August},
 number = {4},
 pages = {044009},
 primaryclass = {astro-ph.CO},
 title = {{New constraints on f(R) gravity from clusters of galaxies}},
 volume = {92},
 year = {2015}
}

@article{Chiu23,
 adsurl = {https://ui.adsabs.harvard.edu/abs/2023MNRAS.522.1601C},
 archiveprefix = {arXiv},
 author = {{Chiu}, I. -Non and {Klein}, Matthias and {Mohr}, Joseph and {Bocquet}, Sebastian},
 doi = {10.1093/mnras/stad957},
 eprint = {2207.12429},
 journal = {\mnras},
 month = {June},
 number = {2},
 pages = {1601-1642},
 primaryclass = {astro-ph.CO},
 title = {{Cosmological constraints from galaxy clusters and groups in the eROSITA final equatorial depth survey}},
 volume = {522},
 year = {2023}
}

@article{Costanzi13,
 adsurl = {https://ui.adsabs.harvard.edu/abs/2013JCAP...12..012C},
 archiveprefix = {arXiv},
 author = {{Costanzi}, Matteo and {Villaescusa-Navarro}, Francisco and {Viel}, Matteo and {Xia}, Jun-Qing and {Borgani}, Stefano and {Castorina}, Emanuele and {Sefusatti}, Emiliano},
 doi = {10.1088/1475-7516/2013/12/012},
 eid = {012},
 eprint = {1311.1514},
 journal = {\jcap},
 month = {December},
 number = {12},
 pages = {012},
 primaryclass = {astro-ph.CO},
 title = {{Cosmology with massive neutrinos III: the halo mass function and an application to galaxy clusters}},
 volume = {2013},
 year = {2013}
}

@article{Davies24,
 adsurl = {https://ui.adsabs.harvard.edu/abs/2024MNRAS.533.3546D},
 archiveprefix = {arXiv},
 author = {{Davies}, Christopher T. and {Harnois-D{\'e}raps}, Joachim and {Li}, Baojiu and {Giblin}, Benjamin and {Hern{\'a}ndez-Aguayo}, C{\'e}sar and {Paillas}, Enrique},
 doi = {10.1093/mnras/stae1966},
 eprint = {2406.11958},
 journal = {\mnras},
 month = {September},
 number = {3},
 pages = {3546-3569},
 primaryclass = {astro-ph.CO},
 title = {{Constraining modified gravity with weak-lensing peaks}},
 volume = {533},
 year = {2024}
}

@article{Dehaan16,
 adsurl = {https://ui.adsabs.harvard.edu/abs/2016ApJ...832...95D},
 archiveprefix = {arXiv},
 author = {{de Haan}, T. and {Benson}, B.~A. and {Bleem}, L.~E. and {Allen}, S.~W. and {Applegate}, D.~E. and {Ashby}, M.~L.~N. and {Bautz}, M. and {Bayliss}, M. and {Bocquet}, S. and {Brodwin}, M. and {Carlstrom}, J.~E. and {Chang}, C.~L. and {Chiu}, I. and {Cho}, H. -M. and {Clocchiatti}, A. and {Crawford}, T.~M. and {Crites}, A.~T. and {Desai}, S. and {Dietrich}, J.~P. and {Dobbs}, M.~A. and {Doucouliagos}, A.~N. and {Foley}, R.~J. and {Forman}, W.~R. and {Garmire}, G.~P. and {George}, E.~M. and {Gladders}, M.~D. and {Gonzalez}, A.~H. and {Gupta}, N. and {Halverson}, N.~W. and {Hlavacek-Larrondo}, J. and {Hoekstra}, H. and {Holder}, G.~P. and {Holzapfel}, W.~L. and {Hou}, Z. and {Hrubes}, J.~D. and {Huang}, N. and {Jones}, C. and {Keisler}, R. and {Knox}, L. and {Lee}, A.~T. and {Leitch}, E.~M. and {von der Linden}, A. and {Luong-Van}, D. and {Mantz}, A. and {Marrone}, D.~P. and {McDonald}, M. and {McMahon}, J.~J. and {Meyer}, S.~S. and {Mocanu}, L.~M. and {Mohr}, J.~J. and {Murray}, S.~S. and {Padin}, S. and {Pryke}, C. and {Rapetti}, D. and {Reichardt}, C.~L. and {Rest}, A. and {Ruel}, J. and {Ruhl}, J.~E. and {Saliwanchik}, B.~R. and {Saro}, A. and {Sayre}, J.~T. and {Schaffer}, K.~K. and {Schrabback}, T. and {Shirokoff}, E. and {Song}, J. and {Spieler}, H.~G. and {Stalder}, B. and {Stanford}, S.~A. and {Staniszewski}, Z. and {Stark}, A.~A. and {Story}, K.~T. and {Stubbs}, C.~W. and {Vanderlinde}, K. and {Vieira}, J.~D. and {Vikhlinin}, A. and {Williamson}, R. and {Zenteno}, A.},
 doi = {10.3847/0004-637X/832/1/95},
 eid = {95},
 eprint = {1603.06522},
 journal = {\apj},
 month = {November},
 number = {1},
 pages = {95},
 primaryclass = {astro-ph.CO},
 title = {{Cosmological Constraints from Galaxy Clusters in the 2500 Square-degree SPT-SZ Survey}},
 volume = {832},
 year = {2016}
}

@article{DES16,
 adsurl = {https://ui.adsabs.harvard.edu/abs/2016MNRAS.460.1270D},
 archiveprefix = {arXiv},
 author = {{Dark Energy Survey Collaboration} and {Abbott}, T. and {Abdalla}, F.~B. and {Aleksi{\'c}}, J. and {Allam}, S. and {Amara}, A. and {Bacon}, D. and {Balbinot}, E. and {Banerji}, M. and {Bechtol}, K. and {Benoit-L{\'e}vy}, A. and {Bernstein}, G.~M. and {Bertin}, E. and {Blazek}, J. and {Bonnett}, C. and {Bridle}, S. and {Brooks}, D. and {Brunner}, R.~J. and {Buckley-Geer}, E. and {Burke}, D.~L. and {Caminha}, G.~B. and {Capozzi}, D. and {Carlsen}, J. and {Carnero-Rosell}, A. and {Carollo}, M. and {Carrasco-Kind}, M. and {Carretero}, J. and {Castander}, F.~J. and {Clerkin}, L. and {Collett}, T. and {Conselice}, C. and {Crocce}, M. and {Cunha}, C.~E. and {D'Andrea}, C.~B. and {da Costa}, L.~N. and {Davis}, T.~M. and {Desai}, S. and {Diehl}, H.~T. and {Dietrich}, J.~P. and {Dodelson}, S. and {Doel}, P. and {Drlica-Wagner}, A. and {Estrada}, J. and {Etherington}, J. and {Evrard}, A.~E. and {Fabbri}, J. and {Finley}, D.~A. and {Flaugher}, B. and {Foley}, R.~J. and {Fosalba}, P. and {Frieman}, J. and {Garc{\'\i}a-Bellido}, J. and {Gaztanaga}, E. and {Gerdes}, D.~W. and {Giannantonio}, T. and {Goldstein}, D.~A. and {Gruen}, D. and {Gruendl}, R.~A. and {Guarnieri}, P. and {Gutierrez}, G. and {Hartley}, W. and {Honscheid}, K. and {Jain}, B. and {James}, D.~J. and {Jeltema}, T. and {Jouvel}, S. and {Kessler}, R. and {King}, A. and {Kirk}, D. and {Kron}, R. and {Kuehn}, K. and {Kuropatkin}, N. and {Lahav}, O. and {Li}, T.~S. and {Lima}, M. and {Lin}, H. and {Maia}, M.~A.~G. and {Makler}, M. and {Manera}, M. and {Maraston}, C. and {Marshall}, J.~L. and {Martini}, P. and {McMahon}, R.~G. and {Melchior}, P. and {Merson}, A. and {Miller}, C.~J. and {Miquel}, R. and {Mohr}, J.~J. and {Morice-Atkinson}, X. and {Naidoo}, K. and {Neilsen}, E. and {Nichol}, R.~C. and {Nord}, B. and {Ogando}, R. and {Ostrovski}, F. and {Palmese}, A. and {Papadopoulos}, A. and {Peiris}, H.~V. and {Peoples}, J. and {Percival}, W.~J. and {Plazas}, A.~A. and {Reed}, S.~L. and {Refregier}, A. and {Romer}, A.~K. and {Roodman}, A. and {Ross}, A. and {Rozo}, E. and {Rykoff}, E.~S. and {Sadeh}, I. and {Sako}, M. and {S{\'a}nchez}, C. and {Sanchez}, E. and {Santiago}, B. and {Scarpine}, V. and {Schubnell}, M. and {Sevilla-Noarbe}, I. and {Sheldon}, E. and {Smith}, M. and {Smith}, R.~C. and {Soares-Santos}, M. and {Sobreira}, F. and {Soumagnac}, M. and {Suchyta}, E. and {Sullivan}, M. and {Swanson}, M. and {Tarle}, G. and {Thaler}, J. and {Thomas}, D. and {Thomas}, R.~C. and {Tucker}, D. and {Vieira}, J.~D. and {Vikram}, V. and {Walker}, A.~R. and {Wechsler}, R.~H. and {Weller}, J. and {Wester}, W. and {Whiteway}, L. and {Wilcox}, H. and {Yanny}, B. and {Zhang}, Y. and {Zuntz}, J.},
 doi = {10.1093/mnras/stw641},
 eprint = {1601.00329},
 journal = {\mnras},
 month = {August},
 number = {2},
 pages = {1270-1299},
 primaryclass = {astro-ph.CO},
 title = {{The Dark Energy Survey: more than dark energy - an overview}},
 volume = {460},
 year = {2016}
}

@article{DES18DR1,
 adsurl = {https://ui.adsabs.harvard.edu/abs/2018ApJS..239...18A},
 archiveprefix = {arXiv},
 author = {{Abbott}, T.~M.~C. and et al.},
 doi = {10.3847/1538-4365/aae9f0},
 eid = {18},
 eprint = {1801.03181},
 journal = {\apjs},
 month = {December},
 number = {2},
 pages = {18},
 primaryclass = {astro-ph.IM},
 title = {{The Dark Energy Survey: Data Release 1}},
 volume = {239},
 year = {2018}
}

@article{Dvali2000,
 adsurl = {https://ui.adsabs.harvard.edu/abs/2000PhLB..485..208D},
 archiveprefix = {arXiv},
 author = {{Dvali}, G. and {Gabadadze}, G. and {Porrati}, M.},
 doi = {10.1016/S0370-2693(00)00669-9},
 eprint = {hep-th/0005016},
 journal = {Physics Letters B},
 month = {July},
 number = {1-3},
 pages = {208-214},
 primaryclass = {hep-th},
 title = {{4D gravity on a brane in 5D Minkowski space}},
 volume = {485},
 year = {2000}
}

@article{Fang08,
 adsurl = {https://ui.adsabs.harvard.edu/abs/2008PhRvD..78j3509F},
 archiveprefix = {arXiv},
 author = {{Fang}, Wenjuan and {Wang}, Sheng and {Hu}, Wayne and {Haiman}, Zolt{\'a}n and {Hui}, Lam and {May}, Morgan},
 doi = {10.1103/PhysRevD.78.103509},
 eid = {103509},
 eprint = {0808.2208},
 journal = {\prd},
 month = {November},
 number = {10},
 pages = {103509},
 primaryclass = {astro-ph},
 title = {{Challenges to the DGP model from horizon-scale growth and geometry}},
 volume = {78},
 year = {2008}
}

@article{Fischer24,
 adsurl = {https://ui.adsabs.harvard.edu/abs/2024Univ...10..297F},
 archiveprefix = {arXiv},
 author = {{Fischer}, Hauke and {K{\"a}ding}, Christian and {Pitschmann}, Mario},
 doi = {10.3390/universe10070297},
 eid = {297},
 eprint = {2405.14638},
 journal = {Universe},
 month = {July},
 number = {7},
 pages = {297},
 primaryclass = {gr-qc},
 title = {{Screened Scalar Fields in the Laboratory and the Solar System}},
 volume = {10},
 year = {2024}
}

@article{flaugher15,
 adsurl = {https://ui.adsabs.harvard.edu/abs/2015AJ....150..150F},
 archiveprefix = {arXiv},
 author = {{Flaugher}, B. and {Diehl}, H.~T. and {Honscheid}, K. and {Abbott}, T.~M.~C. and {Alvarez}, O. and {Angstadt}, R. and {Annis}, J.~T. and {Antonik}, M. and {Ballester}, O. and {Beaufore}, L. and {Bernstein}, G.~M. and {Bernstein}, R.~A. and {Bigelow}, B. and {Bonati}, M. and {Boprie}, D. and {Brooks}, D. and {Buckley-Geer}, E.~J. and {Campa}, J. and {Cardiel-Sas}, L. and {Castander}, F.~J. and {Castilla}, J. and {Cease}, H. and {Cela-Ruiz}, J.~M. and {Chappa}, S. and {Chi}, E. and {Cooper}, C. and {da Costa}, L.~N. and {Dede}, E. and {Derylo}, G. and {DePoy}, D.~L. and {de Vicente}, J. and {Doel}, P. and {Drlica-Wagner}, A. and {Eiting}, J. and {Elliott}, A.~E. and {Emes}, J. and {Estrada}, J. and {Fausti Neto}, A. and {Finley}, D.~A. and {Flores}, R. and {Frieman}, J. and {Gerdes}, D. and {Gladders}, M.~D. and {Gregory}, B. and {Gutierrez}, G.~R. and {Hao}, J. and {Holland}, S.~E. and {Holm}, S. and {Huffman}, D. and {Jackson}, C. and {James}, D.~J. and {Jonas}, M. and {Karcher}, A. and {Karliner}, I. and {Kent}, S. and {Kessler}, R. and {Kozlovsky}, M. and {Kron}, R.~G. and {Kubik}, D. and {Kuehn}, K. and {Kuhlmann}, S. and {Kuk}, K. and {Lahav}, O. and {Lathrop}, A. and {Lee}, J. and {Levi}, M.~E. and {Lewis}, P. and {Li}, T.~S. and {Mandrichenko}, I. and {Marshall}, J.~L. and {Martinez}, G. and {Merritt}, K.~W. and {Miquel}, R. and {Mu{\~n}oz}, F. and {Neilsen}, E.~H. and {Nichol}, R.~C. and {Nord}, B. and {Ogando}, R. and {Olsen}, J. and {Palaio}, N. and {Patton}, K. and {Peoples}, J. and {Plazas}, A.~A. and {Rauch}, J. and {Reil}, K. and {Rheault}, J. -P. and {Roe}, N.~A. and {Rogers}, H. and {Roodman}, A. and {Sanchez}, E. and {Scarpine}, V. and {Schindler}, R.~H. and {Schmidt}, R. and {Schmitt}, R. and {Schubnell}, M. and {Schultz}, K. and {Schurter}, P. and {Scott}, L. and {Serrano}, S. and {Shaw}, T.~M. and {Smith}, R.~C. and {Soares-Santos}, M. and {Stefanik}, A. and {Stuermer}, W. and {Suchyta}, E. and {Sypniewski}, A. and {Tarle}, G. and {Thaler}, J. and {Tighe}, R. and {Tran}, C. and {Tucker}, D. and {Walker}, A.~R. and {Wang}, G. and {Watson}, M. and {Weaverdyck}, C. and {Wester}, W. and {Woods}, R. and {Yanny}, B. and {DES Collaboration}},
 doi = {10.1088/0004-6256/150/5/150},
 eid = {150},
 eprint = {1504.02900},
 journal = {\aj},
 month = {November},
 number = {5},
 pages = {150},
 primaryclass = {astro-ph.IM},
 title = {{The Dark Energy Camera}},
 volume = {150},
 year = {2015}
}

@article{Gatti22,
 adsurl = {https://ui.adsabs.harvard.edu/abs/2022MNRAS.510.1223G},
 archiveprefix = {arXiv},
 author = {{Gatti}, M. and {Giannini}, G. and {Bernstein}, G.~M. and {Alarcon}, A. and {Myles}, J. and {Amon}, A. and {Cawthon}, R. and {Troxel}, M. and {DeRose}, J. and {Everett}, S. and {Ross}, A.~J. and {Rykoff}, E.~S. and {Elvin-Poole}, J. and {Cordero}, J. and {Harrison}, I. and {Sanchez}, C. and {Prat}, J. and {Gruen}, D. and {Lin}, H. and {Crocce}, M. and {Rozo}, E. and {Abbott}, T.~M.~C. and {Aguena}, M. and {Allam}, S. and {Annis}, J. and {Avila}, S. and {Bacon}, D. and {Bertin}, E. and {Brooks}, D. and {Burke}, D.~L. and {Rosell}, A. Carnero and {Kind}, M. Carrasco and {Carretero}, J. and {Castander}, F.~J. and {Choi}, A. and {Conselice}, C. and {Costanzi}, M. and {Crocce}, M. and {da Costa}, L.~N. and {Pereira}, M.~E.~S. and {Dawson}, K. and {Desai}, S. and {Diehl}, H.~T. and {Eckert}, K. and {Eifler}, T.~F. and {Evrard}, A.~E. and {Ferrero}, I. and {Flaugher}, B. and {Fosalba}, P. and {Frieman}, J. and {Garc{\'\i}a-Bellido}, J. and {Gaztanaga}, E. and {Giannantonio}, T. and {Gruendl}, R.~A. and {Gschwend}, J. and {Hinton}, S.~R. and {Hollowood}, D.~L. and {Honscheid}, K. and {Hoyle}, B. and {Huterer}, D. and {James}, D.~J. and {Kuehn}, K. and {Kuropatkin}, N. and {Lahav}, O. and {Lima}, M. and {MacCrann}, N. and {Maia}, M.~A.~G. and {March}, M. and {Marshall}, J.~L. and {Melchior}, P. and {Menanteau}, F. and {Miquel}, R. and {Mohr}, J.~J. and {Morgan}, R. and {Ogando}, R.~L.~C. and {Palmese}, A. and {Paz-Chinch{\'o}n}, F. and {Percival}, W.~J. and {Plazas}, A.~A. and {Rodriguez-Monroy}, M. and {Roodman}, A. and {Rossi}, G. and {Samuroff}, S. and {Sanchez}, E. and {Scarpine}, V. and {Secco}, L.~F. and {Serrano}, S. and {Sevilla-Noarbe}, I. and {Smith}, M. and {Soares-Santos}, M. and {Suchyta}, E. and {Swanson}, M.~E.~C. and {Tarle}, G. and {Thomas}, D. and {To}, C. and {Varga}, T.~N. and {Weller}, J. and {Wilkinson}, R.~D. and {Wilkinson}, R.~D. and {DES Collaboration}},
 doi = {10.1093/mnras/stab3311},
 eprint = {2012.08569},
 journal = {\mnras},
 month = {February},
 number = {1},
 pages = {1223-1247},
 primaryclass = {astro-ph.CO},
 title = {{Dark Energy Survey Year 3 Results: clustering redshifts - calibration of the weak lensing source redshift distributions with redMaGiC and BOSS/eBOSS}},
 volume = {510},
 year = {2022}
}

@article{Ghirardini24,
 adsurl = {https://ui.adsabs.harvard.edu/abs/2024A&A...689A.298G},
 archiveprefix = {arXiv},
 author = {{Ghirardini}, V. and {Bulbul}, E. and {Artis}, E. and {Clerc}, N. and {Garrel}, C. and {Grandis}, S. and {Kluge}, M. and {Liu}, A. and {Bahar}, Y.~E. and {Balzer}, F. and {Chiu}, I. and {Comparat}, J. and {Gruen}, D. and {Kleinebreil}, F. and {Krippendorf}, S. and {Merloni}, A. and {Nandra}, K. and {Okabe}, N. and {Pacaud}, F. and {Predehl}, P. and {Ramos-Ceja}, M.~E. and {Reiprich}, T.~H. and {Sanders}, J.~S. and {Schrabback}, T. and {Seppi}, R. and {Zelmer}, S. and {Zhang}, X. and {Bornemann}, W. and {Brunner}, H. and {Burwitz}, V. and {Coutinho}, D. and {Dennerl}, K. and {Freyberg}, M. and {Friedrich}, S. and {Gaida}, R. and {Gueguen}, A. and {Haberl}, F. and {Kink}, W. and {Lamer}, G. and {Li}, X. and {Liu}, T. and {Maitra}, C. and {Meidinger}, N. and {Mueller}, S. and {Miyatake}, H. and {Miyazaki}, S. and {Robrade}, J. and {Schwope}, A. and {Stewart}, I.},
 doi = {10.1051/0004-6361/202348852},
 eid = {A298},
 eprint = {2402.08458},
 journal = {\aap},
 month = {September},
 pages = {A298},
 primaryclass = {astro-ph.CO},
 title = {{The SRG/eROSITA all-sky survey: Cosmology constraints from cluster abundances in the western Galactic hemisphere}},
 volume = {689},
 year = {2024}
}

@article{Grandis21,
 adsurl = {https://ui.adsabs.harvard.edu/abs/2021MNRAS.507.5671G},
 archiveprefix = {arXiv},
 author = {{Grandis}, Sebastian and {Bocquet}, Sebastian and {Mohr}, Joseph J. and {Klein}, Matthias and {Dolag}, Klaus},
 doi = {10.1093/mnras/stab2414},
 eprint = {2103.16212},
 journal = {\mnras},
 month = {November},
 number = {4},
 pages = {5671-5689},
 primaryclass = {astro-ph.CO},
 title = {{Calibration of bias and scatter involved in cluster mass measurements using optical weak gravitational lensing}},
 volume = {507},
 year = {2021}
}

@article{Hagstotz19,
 adsurl = {https://ui.adsabs.harvard.edu/abs/2019MNRAS.486.3927H},
 archiveprefix = {arXiv},
 author = {{Hagstotz}, Steffen and {Costanzi}, Matteo and {Baldi}, Marco and {Weller}, Jochen},
 doi = {10.1093/mnras/stz1051},
 eprint = {1806.07400},
 journal = {\mnras},
 month = {July},
 number = {3},
 pages = {3927-3941},
 primaryclass = {astro-ph.CO},
 title = {{Joint halo-mass function for modified gravity and massive neutrinos - I. Simulations and cosmological forecasts}},
 volume = {486},
 year = {2019}
}

@article{Haiman01,
 adsurl = {https://ui.adsabs.harvard.edu/abs/2001ApJ...553..545H},
 archiveprefix = {arXiv},
 author = {{Haiman}, Zolt{\'a}n and {Mohr}, Joseph J. and {Holder}, Gilbert P.},
 doi = {10.1086/320939},
 eprint = {astro-ph/0002336},
 journal = {\apj},
 month = {June},
 number = {2},
 pages = {545-561},
 primaryclass = {astro-ph},
 title = {{Constraints on Cosmological Parameters from Future Galaxy Cluster Surveys}},
 volume = {553},
 year = {2001}
}

@article{Harnois23,
 adsurl = {https://ui.adsabs.harvard.edu/abs/2023MNRAS.525.6336H},
 archiveprefix = {arXiv},
 author = {{Harnois-D{\'e}raps}, Joachim and {Hernandez-Aguayo}, Cesar and {Cuesta-Lazaro}, Carolina and {Arnold}, Christian and {Li}, Baojiu and {Davies}, Christopher T. and {Cai}, Yan-Chuan},
 doi = {10.1093/mnras/stad2700},
 eprint = {2211.05779},
 journal = {\mnras},
 month = {November},
 number = {4},
 pages = {6336-6358},
 primaryclass = {astro-ph.CO},
 title = {{MGLENS: Modified gravity weak lensing simulations for emulation-based cosmological inference}},
 volume = {525},
 year = {2023}
}

@article{Hernandes20,
 adsurl = {https://ui.adsabs.harvard.edu/abs/2020A&A...640A.117H},
 archiveprefix = {arXiv},
 author = {{Hern{\'a}ndez-Mart{\'\i}n}, B. and {Schrabback}, T. and {Hoekstra}, H. and {Martinet}, N. and {Hlavacek-Larrondo}, J. and {Bleem}, L.~E. and {Gladders}, M.~D. and {Stalder}, B. and {Stark}, A.~A. and {Bayliss}, M.},
 doi = {10.1051/0004-6361/202037844},
 eid = {A117},
 eprint = {2007.00386},
 journal = {\aap},
 month = {August},
 pages = {A117},
 primaryclass = {astro-ph.CO},
 title = {{Constraining the masses of high-redshift clusters with weak lensing: Revised shape calibration testing for the impact of stronger shears and increased blending}},
 volume = {640},
 year = {2020}
}

@article{Hojjati11,
 adsurl = {https://ui.adsabs.harvard.edu/abs/2011JCAP...08..005H},
 archiveprefix = {arXiv},
 author = {{Hojjati}, Alireza and {Pogosian}, Levon and {Zhao}, Gong-Bo},
 doi = {10.1088/1475-7516/2011/08/005},
 eid = {005},
 eprint = {1106.4543},
 journal = {\jcap},
 month = {August},
 number = {8},
 pages = {005},
 primaryclass = {astro-ph.CO},
 title = {{Testing gravity with CAMB and CosmoMC}},
 volume = {2011},
 year = {2011}
}

@article{Ichiki12,
 adsurl = {https://ui.adsabs.harvard.edu/abs/2012PhRvD..85f3521I},
 archiveprefix = {arXiv},
 author = {{Ichiki}, Kiyotomo and {Takada}, Masahiro},
 doi = {10.1103/PhysRevD.85.063521},
 eid = {063521},
 eprint = {1108.4688},
 journal = {\prd},
 month = {March},
 number = {6},
 pages = {063521},
 primaryclass = {astro-ph.CO},
 title = {{Impact of massive neutrinos on the abundance of massive clusters}},
 volume = {85},
 year = {2012}
}

@article{Ishak24,
 adsurl = {https://ui.adsabs.harvard.edu/abs/2025JCAP...09..053I},
 archiveprefix = {arXiv},
 author = {{Ishak}, M. and {Pan}, J. and {Calderon}, R. and {Lodha}, K. and {Valogiannis}, G. and {Aviles}, A. and {Niz}, G. and {Yi}, L. and {Zheng}, C. and {Garcia-Quintero}, C. and {de Mattia}, A. and {Medina-Varela}, L. and {Cervantes-Cota}, J.~L. and {Andrade}, U. and {Huterer}, D. and {Noriega}, H.~E. and {Zhao}, G. and {Shafieloo}, A. and {Fang}, W. and {Ahlen}, S. and {Bianchi}, D. and {Brooks}, D. and {Burtin}, E. and {Chaussidon}, E. and {Claybaugh}, T. and {Cole}, S. and {de la Macorra}, A. and {Dey}, A. and {Fanning}, K. and {Ferraro}, S. and {Font-Ribera}, A. and {Forero-Romero}, J.~E. and {Gazta{\~n}aga}, E. and {Gil-Mar{\'\i}n}, H. and {Gontcho A. Gontcho}, S. and {Gutierrez}, G. and {Hahn}, C. and {Honscheid}, K. and {Howlett}, C. and {Juneau}, S. and {Kirkby}, D. and {Kisner}, T. and {Kremin}, A. and {Landriau}, M. and {Le Guillou}, L. and {Leauthaud}, A. and {Levi}, M.~E. and {Meisner}, A. and {Miquel}, R. and {Moustakas}, J. and {Newman}, J.~A. and {Palanque-Delabrouille}, N. and {Percival}, W.~J. and {Poppett}, C. and {Prada}, F. and {P{\'e}rez-R{\`a}fols}, I. and {Ross}, A.~J. and {Rossi}, G. and {Sanchez}, E. and {Schlegel}, D. and {Schubnell}, M. and {Seo}, H. and {Sprayberry}, D. and {Tarl{\'e}}, G. and {Vargas-Maga{\~n}a}, M. and {Weaver}, B.~A. and {Wechsler}, R.~H. and {Y{\`e}che}, C. and {Zarrouk}, P. and {Zhou}, R. and {Zou}, H.},
 doi = {10.1088/1475-7516/2025/09/053},
 eid = {053},
 eprint = {2411.12026},
 journal = {\jcap},
 month = {September},
 number = {9},
 pages = {053},
 primaryclass = {astro-ph.CO},
 title = {{Modified gravity constraints from the full shape modeling of clustering measurements from DESI 2024}},
 volume = {2025},
 year = {2025}
}

@article{JoyceEtal16,
 adsurl = {https://ui.adsabs.harvard.edu/abs/2016ARNPS..66...95J},
 archiveprefix = {arXiv},
 author = {{Joyce}, Austin and {Lombriser}, Lucas and {Schmidt}, Fabian},
 doi = {10.1146/annurev-nucl-102115-044553},
 eprint = {1601.06133},
 journal = {Annual Review of Nuclear and Particle Science},
 month = {October},
 number = {1},
 pages = {95-122},
 primaryclass = {astro-ph.CO},
 title = {{Dark Energy Versus Modified Gravity}},
 volume = {66},
 year = {2016}
}

@article{Kaiser86,
 adsurl = {https://ui.adsabs.harvard.edu/abs/1986MNRAS.222..323K},
 author = {{Kaiser}, N.},
 doi = {10.1093/mnras/222.2.323},
 journal = {\mnras},
 month = {September},
 pages = {323-345},
 title = {{Evolution and clustering of rich clusters.}},
 volume = {222},
 year = {1986}
}

@article{Klein18,
 adsurl = {https://ui.adsabs.harvard.edu/abs/2018MNRAS.474.3324K},
 archiveprefix = {arXiv},
 author = {{Klein}, M. and {Mohr}, J.~J. and {Desai}, S. and {Israel}, H. and {Allam}, S. and {Benoit-L{\'e}vy}, A. and {Brooks}, D. and {Buckley-Geer}, E. and {Carnero Rosell}, A. and {Carrasco Kind}, M. and {Cunha}, C.~E. and {da Costa}, L.~N. and {Dietrich}, J.~P. and {Eifler}, T.~F. and {Evrard}, A.~E. and {Frieman}, J. and {Gruen}, D. and {Gruendl}, R.~A. and {Gutierrez}, G. and {Honscheid}, K. and {James}, D.~J. and {Kuehn}, K. and {Lima}, M. and {Maia}, M.~A.~G. and {March}, M. and {Melchior}, P. and {Menanteau}, F. and {Miquel}, R. and {Plazas}, A.~A. and {Reil}, K. and {Romer}, A.~K. and {Sanchez}, E. and {Santiago}, B. and {Scarpine}, V. and {Schubnell}, M. and {Sevilla-Noarbe}, I. and {Smith}, M. and {Soares-Santos}, M. and {Sobreira}, F. and {Suchyta}, E. and {Swanson}, M.~E.~C. and {Tarle}, G. and {DES Collaboration}},
 doi = {10.1093/mnras/stx2929},
 eprint = {1706.06577},
 journal = {\mnras},
 month = {March},
 number = {3},
 pages = {3324-3343},
 primaryclass = {astro-ph.CO},
 title = {{A multicomponent matched filter cluster confirmation tool for eROSITA: initial application to the RASS and DES-SV data sets}},
 volume = {474},
 year = {2018}
}

@article{Klein24,
 adsurl = {https://ui.adsabs.harvard.edu/abs/2024MNRAS.531.3973K},
 author = {{Klein}, M. and {Mohr}, J.~J. and {Bocquet}, S. and {Aguena}, M. and {Allen}, S.~W. and {Alves}, O. and {Ansarinejad}, B. and {Ashby}, M.~L.~N. and {Bacon}, D. and {Bayliss}, M. and {Benson}, B.~A. and {Bleem}, L.~E. and {Brodwin}, M. and {Brooks}, D. and {Bulbul}, E. and {Burke}, D.~L. and {Canning}, R.~E.~A. and {Carlstrom}, J.~E. and {Rosell}, A. Carnero and {Carretero}, J. and {Chang}, C.~L. and {Conselice}, C. and {Costanzi}, M. and {Crites}, A.~T. and {da Costa}, L.~N. and {Pereira}, M.~E.~S. and {Davis}, T.~M. and {De Vicente}, J. and {Desai}, S. and {de Haan}, T. and {Dobbs}, M.~A. and {Doel}, P. and {Ferrero}, I. and {Flores}, A.~M. and {Frieman}, J. and {George}, E.~M. and {Giannini}, G. and {Gladders}, M.~D. and {Gonzalez}, A.~H. and {Grandis}, S. and {Gruen}, D. and {Gruendl}, R.~A. and {Gutierrez}, G. and {Halverson}, N.~W. and {Hinton}, S.~R. and {Holder}, G.~P. and {Hollowood}, D.~L. and {Holzapfel}, W.~L. and {Honscheid}, K. and {Hrubes}, J.~D. and {Huang}, N. and {James}, D.~J. and {Khullar}, G. and {Kim}, K. and {Knox}, L. and {Kraft}, R. and {K{\'e}ruzor{\'e}}, F. and {Lee}, A.~T. and {Luong-Van}, D. and {Mahler}, G. and {Mantz}, A. and {Marrone}, D.~P. and {Marshall}, J.~L. and {McDonald}, M. and {McMahon}, J.~J. and {Mena-Fern{\'a}ndez}, J. and {Menanteau}, F. and {Meyer}, S.~S. and {Miquel}, R. and {Myles}, J. and {Padin}, S. and {Pieres}, A. and {Plazas Malag{\'o}n}, A.~A. and {Pryke}, C. and {Reichardt}, C.~L. and {Reil}, K. and {Roberson}, J. and {Romer}, A.~K. and {Romero}, C. and {Ruhl}, J.~E. and {Saliwanchik}, B.~R. and {Salvati}, L. and {Sanchez}, E. and {Saro}, A. and {Schaffer}, K.~K. and {Schrabback}, T. and {Schubnell}, M. and {Sevilla-Noarbe}, I. and {Sharon}, K. and {Shirokoff}, E. and {Smith}, M. and {Somboonpanyakul}, T. and {Stalder}, B. and {Stanford}, S.~A. and {Stark}, A.~A. and {Strazzullo}, V. and {Suchyta}, E. and {Swanson}, M.~E.~C. and {Tarle}, G. and {To}, C. and {Vanderlinde}, K. and {Vieira}, J.~D. and {von der Linden}, A. and {Weaverdyck}, N. and {Williamson}, R. and {Wiseman}, P. and {Young}, M.},
 doi = {10.1093/mnras/stae1359},
 journal = {\mnras},
 month = {July},
 number = {4},
 pages = {3973-3990},
 title = {{SPT-SZ MCMF: an extension of the SPT-SZ catalogue over the DES region}},
 volume = {531},
 year = {2024}
}

@article{Koyama07,
 adsurl = {https://ui.adsabs.harvard.edu/abs/2007CQGra..24R.231K},
 archiveprefix = {arXiv},
 author = {{Koyama}, Kazuya},
 doi = {10.1088/0264-9381/24/24/R01},
 eprint = {0709.2399},
 journal = {Classical and Quantum Gravity},
 month = {December},
 number = {24},
 pages = {R231-R253},
 primaryclass = {hep-th},
 title = {{TOPICAL REVIEW:  Ghosts in the self-accelerating universe}},
 volume = {24},
 year = {2007}
}

@article{Koyama18,
 adsurl = {https://ui.adsabs.harvard.edu/abs/2018IJMPD..2748001K},
 author = {{Koyama}, Kazuya},
 doi = {10.1142/S0218271818480012},
 eid = {1848001},
 journal = {International Journal of Modern Physics D},
 month = {January},
 number = {15},
 pages = {1848001},
 title = {{Gravity beyond general relativity}},
 volume = {27},
 year = {2018}
}

@article{Laureijs11,
 adsurl = {https://ui.adsabs.harvard.edu/abs/2011arXiv1110.3193L},
 archiveprefix = {arXiv},
 author = {{Laureijs}, R. and {Amiaux}, J. and {Arduini}, S. and {Augu{\`e}res}, J. -L. and {Brinchmann}, J. and {Cole}, R. and {Cropper}, M. and {Dabin}, C. and {Duvet}, L. and {Ealet}, A. and {Garilli}, B. and {Gondoin}, P. and {Guzzo}, L. and {Hoar}, J. and {Hoekstra}, H. and {Holmes}, R. and {Kitching}, T. and {Maciaszek}, T. and {Mellier}, Y. and {Pasian}, F. and {Percival}, W. and {Rhodes}, J. and {Saavedra Criado}, G. and {Sauvage}, M. and {Scaramella}, R. and {Valenziano}, L. and {Warren}, S. and {Bender}, R. and {Castander}, F. and {Cimatti}, A. and {Le F{\`e}vre}, O. and {Kurki-Suonio}, H. and {Levi}, M. and {Lilje}, P. and {Meylan}, G. and {Nichol}, R. and {Pedersen}, K. and {Popa}, V. and {Rebolo Lopez}, R. and {Rix}, H. -W. and {Rottgering}, H. and {Zeilinger}, W. and {Grupp}, F. and {Hudelot}, P. and {Massey}, R. and {Meneghetti}, M. and {Miller}, L. and {Paltani}, S. and {Paulin-Henriksson}, S. and {Pires}, S. and {Saxton}, C. and {Schrabback}, T. and {Seidel}, G. and {Walsh}, J. and {Aghanim}, N. and {Amendola}, L. and {Bartlett}, J. and {Baccigalupi}, C. and {Beaulieu}, J. -P. and {Benabed}, K. and {Cuby}, J. -G. and {Elbaz}, D. and {Fosalba}, P. and {Gavazzi}, G. and {Helmi}, A. and {Hook}, I. and {Irwin}, M. and {Kneib}, J. -P. and {Kunz}, M. and {Mannucci}, F. and {Moscardini}, L. and {Tao}, C. and {Teyssier}, R. and {Weller}, J. and {Zamorani}, G. and {Zapatero Osorio}, M.~R. and {Boulade}, O. and {Foumond}, J.~J. and {Di Giorgio}, A. and {Guttridge}, P. and {James}, A. and {Kemp}, M. and {Martignac}, J. and {Spencer}, A. and {Walton}, D. and {Bl{\"u}mchen}, T. and {Bonoli}, C. and {Bortoletto}, F. and {Cerna}, C. and {Corcione}, L. and {Fabron}, C. and {Jahnke}, K. and {Ligori}, S. and {Madrid}, F. and {Martin}, L. and {Morgante}, G. and {Pamplona}, T. and {Prieto}, E. and {Riva}, M. and {Toledo}, R. and {Trifoglio}, M. and {Zerbi}, F. and {Abdalla}, F. and {Douspis}, M. and {Grenet}, C. and {Borgani}, S. and {Bouwens}, R. and {Courbin}, F. and {Delouis}, J. -M. and {Dubath}, P. and {Fontana}, A. and {Frailis}, M. and {Grazian}, A. and {Koppenh{\"o}fer}, J. and {Mansutti}, O. and {Melchior}, M. and {Mignoli}, M. and {Mohr}, J. and {Neissner}, C. and {Noddle}, K. and {Poncet}, M. and {Scodeggio}, M. and {Serrano}, S. and {Shane}, N. and {Starck}, J. -L. and {Surace}, C. and {Taylor}, A. and {Verdoes-Kleijn}, G. and {Vuerli}, C. and {Williams}, O.~R. and {Zacchei}, A. and {Altieri}, B. and {Escudero Sanz}, I. and {Kohley}, R. and {Oosterbroek}, T. and {Astier}, P. and {Bacon}, D. and {Bardelli}, S. and {Baugh}, C. and {Bellagamba}, F. and {Benoist}, C. and {Bianchi}, D. and {Biviano}, A. and {Branchini}, E. and {Carbone}, C. and {Cardone}, V. and {Clements}, D. and {Colombi}, S. and {Conselice}, C. and {Cresci}, G. and {Deacon}, N. and {Dunlop}, J. and {Fedeli}, C. and {Fontanot}, F. and {Franzetti}, P. and {Giocoli}, C. and {Garcia-Bellido}, J. and {Gow}, J. and {Heavens}, A. and {Hewett}, P. and {Heymans}, C. and {Holland}, A. and {Huang}, Z. and {Ilbert}, O. and {Joachimi}, B. and {Jennins}, E. and {Kerins}, E. and {Kiessling}, A. and {Kirk}, D. and {Kotak}, R. and {Krause}, O. and {Lahav}, O. and {van Leeuwen}, F. and {Lesgourgues}, J. and {Lombardi}, M. and {Magliocchetti}, M. and {Maguire}, K. and {Majerotto}, E. and {Maoli}, R. and {Marulli}, F. and {Maurogordato}, S. and {McCracken}, H. and {McLure}, R. and {Melchiorri}, A. and {Merson}, A. and {Moresco}, M. and {Nonino}, M. and {Norberg}, P. and {Peacock}, J. and {Pello}, R. and {Penny}, M. and {Pettorino}, V. and {Di Porto}, C. and {Pozzetti}, L. and {Quercellini}, C. and {Radovich}, M. and {Rassat}, A. and {Roche}, N. and {Ronayette}, S. and {Rossetti}, E. and {Sartoris}, B. and {Schneider}, P. and {Semboloni}, E. and {Serjeant}, S. and {Simpson}, F. and {Skordis}, C. and {Smadja}, G. and {Smartt}, S. and {Spano}, P. and {Spiro}, S. and {Sullivan}, M. and {Tilquin}, A. and {Trotta}, R. and {Verde}, L. and {Wang}, Y. and {Williger}, G. and {Zhao}, G. and {Zoubian}, J. and {Zucca}, E.},
 doi = {10.48550/arXiv.1110.3193},
 eid = {arXiv:1110.3193},
 eprint = {1110.3193},
 journal = {arXiv e-prints},
 month = {October},
 pages = {arXiv:1110.3193},
 primaryclass = {astro-ph.CO},
 title = {{Euclid Definition Study Report}},
 year = {2011}
}

@article{Lombriser09,
 adsurl = {https://ui.adsabs.harvard.edu/abs/2009PhRvD..80f3536L},
 archiveprefix = {arXiv},
 author = {{Lombriser}, Lucas and {Hu}, Wayne and {Fang}, Wenjuan and {Seljak}, Uro{\v{s}}},
 doi = {10.1103/PhysRevD.80.063536},
 eid = {063536},
 eprint = {0905.1112},
 journal = {\prd},
 month = {September},
 number = {6},
 pages = {063536},
 primaryclass = {astro-ph.CO},
 title = {{Cosmological constraints on DGP braneworld gravity with brane tension}},
 volume = {80},
 year = {2009}
}

@article{Lombriser10,
 adsurl = {https://ui.adsabs.harvard.edu/abs/2012PhRvD..85l4038L},
 archiveprefix = {arXiv},
 author = {{Lombriser}, Lucas and {Slosar}, An{\v{z}}e and {Seljak}, Uro{\v{s}} and {Hu}, Wayne},
 doi = {10.1103/PhysRevD.85.124038},
 eid = {124038},
 eprint = {1003.3009},
 journal = {\prd},
 month = {June},
 number = {12},
 pages = {124038},
 primaryclass = {astro-ph.CO},
 title = {{Constraints on f(R) gravity from probing the large-scale structure}},
 volume = {85},
 year = {2012}
}

@article{Louis25,
 adsurl = {https://ui.adsabs.harvard.edu/abs/2025JCAP...11..062L},
 archiveprefix = {arXiv},
 author = {{Louis}, Thibaut and {La Posta}, Adrien and {Atkins}, Zachary and {Jense}, Hidde T. and {Abril-Cabezas}, Irene and {Addison}, Graeme E. and {Ade}, Peter A.~R. and {Aiola}, Simone and {Alford}, Tommy and {Alonso}, David and {Amiri}, Mandana and {An}, Rui and {Austermann}, Jason E. and {Barbavara}, Eleonora and {Battaglia}, Nicholas and {Battistelli}, Elia Stefano and {Beall}, James A. and {Bean}, Rachel and {Beheshti}, Ali and {Beringue}, Benjamin and {Bhandarkar}, Tanay and {Biermann}, Emily and {Bolliet}, Boris and {Bond}, J. Richard and {Calabrese}, Erminia and {Capalbo}, Valentina and {Carrero}, Felipe and {Chen}, Shi-Fan and {Chesmore}, Grace and {Cho}, Hsiao-mei and {Choi}, Steve K. and {Clark}, Susan E. and {Cothard}, Nicholas F. and {Coughlin}, Kevin and {Coulton}, William and {Crichton}, Devin and {Crowley}, Kevin T. and {Darwish}, Omar and {Devlin}, Mark J. and {Dicker}, Simon and {Duell}, Cody J. and {Duff}, Shannon M. and {Duivenvoorden}, Adriaan J. and {Dunkley}, Jo and {Dunner}, Rolando and {Embil Villagra}, Carmen and {Fankhanel}, Max and {Farren}, Gerrit S. and {Ferraro}, Simone and {Foster}, Allen and {Freundt}, Rodrigo and {Fuzia}, Brittany and {Gallardo}, Patricio A. and {Garrido}, Xavier and {Gerbino}, Martina and {Giardiello}, Serena and {Gill}, Ajay and {Givans}, Jahmour and {Gluscevic}, Vera and {Goldstein}, Samuel and {Golec}, Joseph E. and {Gong}, Yulin and {Guan}, Yilun and {Halpern}, Mark and {Harrison}, Ian and {Hasselfield}, Matthew and {Healy}, Erin and {Henderson}, Shawn and {Hensley}, Brandon and {Herv{\'\i}as-Caimapo}, Carlos and {Hill}, J. Colin and {Hilton}, Gene C. and {Hilton}, Matt and {Hincks}, Adam D. and {Hlo{\v{z}}ek}, Ren{\'e}e and {Ho}, Shuay-Pwu Patty and {Hood}, John and {Hornecker}, Erika and {Huber}, Zachary B. and {Hubmayr}, Johannes and {Huffenberger}, Kevin M. and {Hughes}, John P. and {Ikape}, Margaret and {Irwin}, Kent and {Isopi}, Giovanni and {Joshi}, Neha and {Keller}, Ben and {Kim}, Joshua and {Knowles}, Kenda and {Koopman}, Brian J. and {Kosowsky}, Arthur and {Kramer}, Darby and {Kusiak}, Aleksandra and {Lagu{\"e}}, Alex and {Lakey}, Victoria and {Lee}, Eunseong and {Li}, Yaqiong and {Li}, Zack and {Limon}, Michele and {Lokken}, Martine and {Lungu}, Marius and {MacCrann}, Niall and {MacInnis}, Amanda and {Madhavacheril}, Mathew S. and {Maldonado}, Diego and {Maldonado}, Felipe and {Mallaby-Kay}, Maya and {Marques}, Gabriela A. and {van Marrewijk}, Joshiwa and {McCarthy}, Fiona and {McMahon}, Jeff and {Mehta}, Yogesh and {Menanteau}, Felipe and {Moodley}, Kavilan and {Morris}, Thomas W. and {Mroczkowski}, Tony and {Naess}, Sigurd and {Namikawa}, Toshiya and {Nati}, Federico and {Nerval}, Simran K. and {Newburgh}, Laura and {Nicola}, Andrina and {Niemack}, Michael D. and {Nolta}, Michael R. and {Orlowski-Scherer}, John and {Pagano}, Luca and {Page}, Lyman A. and {Pandey}, Shivam and {Partridge}, Bruce and {Perez Sarmiento}, Karen and {Prince}, Heather and {Puddu}, Roberto and {Qu}, Frank J. and {Ragavan}, Damien C. and {Ried Guachalla}, Bernardita and {Rogers}, Keir K. and {Rojas}, Felipe and {Sakuma}, Tai and {Schaan}, Emmanuel and {Schmitt}, Benjamin L. and {Sehgal}, Neelima and {Shaikh}, Shabbir and {Sherwin}, Blake D. and {Sierra}, Carlos and {Sievers}, Jon and {Sif{\'o}n}, Crist{\'o}bal and {Simon}, Sara and {Sonka}, Rita and {Spergel}, David N. and {Staggs}, Suzanne T. and {Storer}, Emilie and {Surrao}, Kristen and {Switzer}, Eric R. and {Tampier}, Niklas and {Thornton}, Robert and {Trac}, Hy and {Tucker}, Carole and {Ullom}, Joel and {Vale}, Leila R. and {Van Engelen}, Alexander and {Van Lanen}, Jeff and {Vargas}, Cristian and {Vavagiakis}, Eve M. and {Wagoner}, Kasey and {Wang}, Yuhan and {Wenzl}, Lukas and {Wollack}, Edward J. and {Zheng}, Kaiwen and {The Atacama Cosmology Telescope collaboration}},
 doi = {10.1088/1475-7516/2025/11/062},
 eid = {062},
 eprint = {2503.14452},
 journal = {\jcap},
 month = {November},
 number = {11},
 pages = {062},
 primaryclass = {astro-ph.CO},
 title = {{The Atacama Cosmology Telescope: DR6 power spectra, likelihoods and {\ensuremath{\Lambda}}CDM parameters}},
 volume = {2025},
 year = {2025}
}

@article{lsst09sciencebook,
 adsurl = {https://ui.adsabs.harvard.edu/abs/2009arXiv0912.0201L},
 archiveprefix = {arXiv},
 author = {{LSST Science Collaboration} and {Abell}, Paul A. and {Allison}, Julius and {Anderson}, Scott F. and {Andrew}, John R. and {Angel}, J. Roger P. and {Armus}, Lee and {Arnett}, David and {Asztalos}, S.~J. and {Axelrod}, Tim S. and {Bailey}, Stephen and {Ballantyne}, D.~R. and {Bankert}, Justin R. and {Barkhouse}, Wayne A. and {Barr}, Jeffrey D. and {Barrientos}, L. Felipe and {Barth}, Aaron J. and {Bartlett}, James G. and {Becker}, Andrew C. and {Becla}, Jacek and {Beers}, Timothy C. and {Bernstein}, Joseph P. and {Biswas}, Rahul and {Blanton}, Michael R. and {Bloom}, Joshua S. and {Bochanski}, John J. and {Boeshaar}, Pat and {Borne}, Kirk D. and {Bradac}, Marusa and {Brandt}, W.~N. and {Bridge}, Carrie R. and {Brown}, Michael E. and {Brunner}, Robert J. and {Bullock}, James S. and {Burgasser}, Adam J. and {Burge}, James H. and {Burke}, David L. and {Cargile}, Phillip A. and {Chandrasekharan}, Srinivasan and {Chartas}, George and {Chesley}, Steven R. and {Chu}, You-Hua and {Cinabro}, David and {Claire}, Mark W. and {Claver}, Charles F. and {Clowe}, Douglas and {Connolly}, A.~J. and {Cook}, Kem H. and {Cooke}, Jeff and {Cooray}, Asantha and {Covey}, Kevin R. and {Culliton}, Christopher S. and {de Jong}, Roelof and {de Vries}, Willem H. and {Debattista}, Victor P. and {Delgado}, Francisco and {Dell'Antonio}, Ian P. and {Dhital}, Saurav and {Di Stefano}, Rosanne and {Dickinson}, Mark and {Dilday}, Benjamin and {Djorgovski}, S.~G. and {Dobler}, Gregory and {Donalek}, Ciro and {Dubois-Felsmann}, Gregory and {Durech}, Josef and {Eliasdottir}, Ardis and {Eracleous}, Michael and {Eyer}, Laurent and {Falco}, Emilio E. and {Fan}, Xiaohui and {Fassnacht}, Christopher D. and {Ferguson}, Harry C. and {Fernandez}, Yanga R. and {Fields}, Brian D. and {Finkbeiner}, Douglas and {Figueroa}, Eduardo E. and {Fox}, Derek B. and {Francke}, Harold and {Frank}, James S. and {Frieman}, Josh and {Fromenteau}, Sebastien and {Furqan}, Muhammad and {Galaz}, Gaspar and {Gal-Yam}, A. and {Garnavich}, Peter and {Gawiser}, Eric and {Geary}, John and {Gee}, Perry and {Gibson}, Robert R. and {Gilmore}, Kirk and {Grace}, Emily A. and {Green}, Richard F. and {Gressler}, William J. and {Grillmair}, Carl J. and {Habib}, Salman and {Haggerty}, J.~S. and {Hamuy}, Mario and {Harris}, Alan W. and {Hawley}, Suzanne L. and {Heavens}, Alan F. and {Hebb}, Leslie and {Henry}, Todd J. and {Hileman}, Edward and {Hilton}, Eric J. and {Hoadley}, Keri and {Holberg}, J.~B. and {Holman}, Matt J. and {Howell}, Steve B. and {Infante}, Leopoldo and {Ivezic}, Zeljko and {Jacoby}, Suzanne H. and {Jain}, Bhuvnesh and {R} and {Jedicke} and {Jee}, M. James and {Garrett Jernigan}, J. and {Jha}, Saurabh W. and {Johnston}, Kathryn V. and {Jones}, R. Lynne and {Juric}, Mario and {Kaasalainen}, Mikko and {Styliani} and {Kafka} and {Kahn}, Steven M. and {Kaib}, Nathan A. and {Kalirai}, Jason and {Kantor}, Jeff and {Kasliwal}, Mansi M. and {Keeton}, Charles R. and {Kessler}, Richard and {Knezevic}, Zoran and {Kowalski}, Adam and {Krabbendam}, Victor L. and {Krughoff}, K. Simon and {Kulkarni}, Shrinivas and {Kuhlman}, Stephen and {Lacy}, Mark and {Lepine}, Sebastien and {Liang}, Ming and {Lien}, Amy and {Lira}, Paulina and {Long}, Knox S. and {Lorenz}, Suzanne and {Lotz}, Jennifer M. and {Lupton}, R.~H. and {Lutz}, Julie and {Macri}, Lucas M. and {Mahabal}, Ashish A. and {Mandelbaum}, Rachel and {Marshall}, Phil and {May}, Morgan and {McGehee}, Peregrine M. and {Meadows}, Brian T. and {Meert}, Alan and {Milani}, Andrea and {Miller}, Christopher J. and {Miller}, Michelle and {Mills}, David and {Minniti}, Dante and {Monet}, David and {Mukadam}, Anjum S. and {Nakar}, Ehud and {Neill}, Douglas R. and {Newman}, Jeffrey A. and {Nikolaev}, Sergei and {Nordby}, Martin and {O'Connor}, Paul and {Oguri}, Masamune and {Oliver}, John and {Olivier}, Scot S. and {Olsen}, Julia K. and {Olsen}, Knut and {Olszewski}, Edward W. and {Oluseyi}, Hakeem and {Padilla}, Nelson D. and {Parker}, Alex and {Pepper}, Joshua and {Peterson}, John R. and {Petry}, Catherine and {Pinto}, Philip A. and {Pizagno}, James L. and {Popescu}, Bogdan and {Prsa}, Andrej and {Radcka}, Veljko and {Raddick}, M. Jordan and {Rasmussen}, Andrew and {Rau}, Arne and {Rho}, Jeonghee and {Rhoads}, James E. and {Richards}, Gordon T. and {Ridgway}, Stephen T. and {Robertson}, Brant E. and {Roskar}, Rok and {Saha}, Abhijit and {Sarajedini}, Ata and {Scannapieco}, Evan and {Schalk}, Terry and {Schindler}, Rafe and {Schmidt}, Samuel},
 doi = {10.48550/arXiv.0912.0201},
 eid = {arXiv:0912.0201},
 eprint = {0912.0201},
 journal = {arXiv e-prints},
 month = {December},
 pages = {arXiv:0912.0201},
 primaryclass = {astro-ph.IM},
 title = {{LSST Science Book, Version 2.0}},
 year = {2009}
}

@article{Luty03,
 adsurl = {https://ui.adsabs.harvard.edu/abs/2003JHEP...09..029L},
 archiveprefix = {arXiv},
 author = {{Luty}, Markus A. and {Porrati}, Massimo and {Rattazzi}, Riccardo},
 doi = {10.1088/1126-6708/2003/09/029},
 eid = {029},
 eprint = {hep-th/0303116},
 journal = {Journal of High Energy Physics},
 month = {September},
 number = {9},
 pages = {029},
 primaryclass = {hep-th},
 title = {{Strong interactions and stability in the DGP model}},
 volume = {2003},
 year = {2003}
}

@article{Mokeddem23,
 adsurl = {https://ui.adsabs.harvard.edu/abs/2023JCAP...01..017M},
 archiveprefix = {arXiv},
 author = {{Mokeddem}, Rahima and {Hip{\'o}lito-Ricaldi}, Wiliam S. and {Bernui}, Armando},
 doi = {10.1088/1475-7516/2023/01/017},
 eid = {017},
 eprint = {2209.11660},
 journal = {\jcap},
 month = {January},
 number = {1},
 pages = {017},
 primaryclass = {astro-ph.CO},
 title = {{Excess of lensing amplitude in the Planck CMB power spectrum}},
 volume = {2023},
 year = {2023}
}

@phdthesis{Renzi13,
 adsurl = {https://ui.adsabs.harvard.edu/abs/2013PhDT.......373R},
 author = {{Renzi}, Alessandro},
 month = {January},
 school = {SISSA, International School for Advanced Studies, Italy},
 title = {{Primordial non-Gaussianity with Planck}},
 year = {2013}
}

\end{document}